\documentclass[]{interact}
\usepackage{multirow}
\usepackage{lscape}
\usepackage[table,xcdraw]{xcolor}
\usepackage{miller}
\usepackage{nicefrac}
\usepackage{epstopdf}
\usepackage[caption=false]{subfig}
\usepackage[numbers,sort&compress]{natbib}
\bibpunct[, ]{[}{]}{,}{n}{,}{,}
\renewcommand\bibfont{\fontsize{10}{12}\selectfont}
\usepackage{comment}

\theoremstyle{plain}

\theoremstyle{definition}

\theoremstyle{remark}

\begin{document}

\articletype{REVIEW ARTICLE}

\title{A review of uranium-based thin films}

\author{
\name{R. Springell\textsuperscript{a}\thanks{Corresponding author email: phrss@bristol.ac.uk, lottie.harding@bristol.ac.uk}, E. Lawrence Bright\textsuperscript{a,b}, D. A. Chaney\textsuperscript{a,b}, L. M. Harding\textsuperscript{a}, C. Bell\textsuperscript{a}, R. C. C. Ward\textsuperscript{c} and G. H. Lander\textsuperscript{a,d}}
\affil{\textsuperscript{a} School of Physics, University of Bristol, Tyndall Avenue, Bristol BS8 1TL.
\textsuperscript{b} European Synchrotron Radiation Facility, 38040 Grenoble, France.
\textsuperscript{c} Clarendon Laboratory, Oxford Physics, Parks Road, Oxford OX1 3PU, UK.
\textsuperscript{d} European Commission, Joint Research Centre (JRC), Directorate for Nuclear Safety and Security, Postfach 2340, D-76125 Karlsruhe, Germany.}}

\maketitle

\begin{abstract}
Thin films based on silicon and transition-metal elements dominate the semi-conducting industry and are ubiquitous in all modern devices.
Films have also been produced in the rare-earth series of elements for both research and specialized applications.
Thin films of uranium and uranium dioxide were fabricated in the 1960s and 1970s, but there was little sustained effort until the early 2000s. Significant  programmes started at Oxford University (transferring to Bristol University in 2011), and Los Alamos National Laboratory (LANL) in New Mexico, USA.
In this review we cover the work that has been published over the last $\sim$20 years with these materials.
Important breakthroughs occurred with the fabrication of epitaxial thin films of initially uranium metal and UO$_2$, but more recently of many other uranium compounds and alloys.
These have led to a number of different experiments that are reviewed, as well as some important trends.
The interaction with the substrate leads to differing strain and hence changes in properties.
An important advantage is that epitaxial films can often be made of materials that are impossible to produce as bulk single crystals.
Examples are U$_3$O$_8$, U$_2$N$_3$ and alloys of U-Mo, which form in a modified {\it bcc} structure.
Epitaxial films may also be used in applied research.
They represent excellent surfaces, and it is at the surfaces that most of the important reactions occur in the nuclear fuel cycle.
For example, the fuel-cladding interactions, and the dissolution of fuel by water in the long-term storage of spent fuel.
To conclude, we discuss possible future prospects, examples include bilayers containing uranium for spintronics, and superlattices that could be used in heterostructures.
Such applications will require a more detailed knowledge of the interface interactions in these systems, and this is an important direction for future research.
\end{abstract}

\begin{keywords}
Uranium, actinides, thin films, epitaxy
\end{keywords}

 \tableofcontents

\section{Introduction} \label{s:intro}

Thin films are ubiquitous in modern technology. They form the basis of the semiconductor industry: from light emitting diodes to the millions of transistors in every single computer central processing unit. Digital memory technologies are similarly underpinned by thin films, from the spin-valve heterostructures in hard drive read heads to ferroelectric random-access memory. More recently, the rapid developments of several high profile quantum computing architectures are based on thin films of superconducting aluminium. It is not unreasonable to argue that thin films have transformed the technology of the late 20th century, and continue to do so to this day. At the same time, for many years thin films have provided researchers with key insights into fundamental condensed matter physics. Examples in this area include the discovery of the integer and fractional quantum Hall effects in GaAs heterostructures \cite{vonKlitzing1986} and oscillatory exchange coupling in magnetic/non-magnetic multilayers \cite{Fert2008,Grunberg2008}. These discoveries earned Nobel prizes, but there are a host of other novel effects in thin film layers and heterostructures.

Sitting at the bottom of the periodic table, the actinides are defined by the presence of 5$f$ electrons which give rise to a plethora of weird and wonderful physical properties \cite{Moore2009} that vary significantly across the series as the nature of the 5$f$ electrons changes from largely itinerant in Th and U, to almost fully localized in Am and beyond, with the notoriously complex Pu separating the two sub-series. The elements up to and including Pu exhibit a vast range of isomorphs, whereas Am and beyond crystalise into a double hexagonal close-packed ({\it dhcp}) structure and behave akin to the heavier rare-earth metals. Likewise, superconductivity gives way to magnetism across the series. Plutonium and uranium also display two properties unique to single element materials: negative thermal expansion in Pu and ambient pressure charge-density modulations in U. The phenomena found in actinide containing compounds are no less fascinating and include heavy-fermion behavior \cite{Stewart2017,Mydosh2011,Pfleiderer2009}, spin fluctuation states, large spin-orbit coupling, Jahn-Teller distortions, quadrupolar ordering \cite{Santini2009} piezoelectricity, and magnetorestriction \cite{Lander1990}, to name but a few.

\begin{figure}[htb]
\centering
\includegraphics[width=1.0\linewidth]{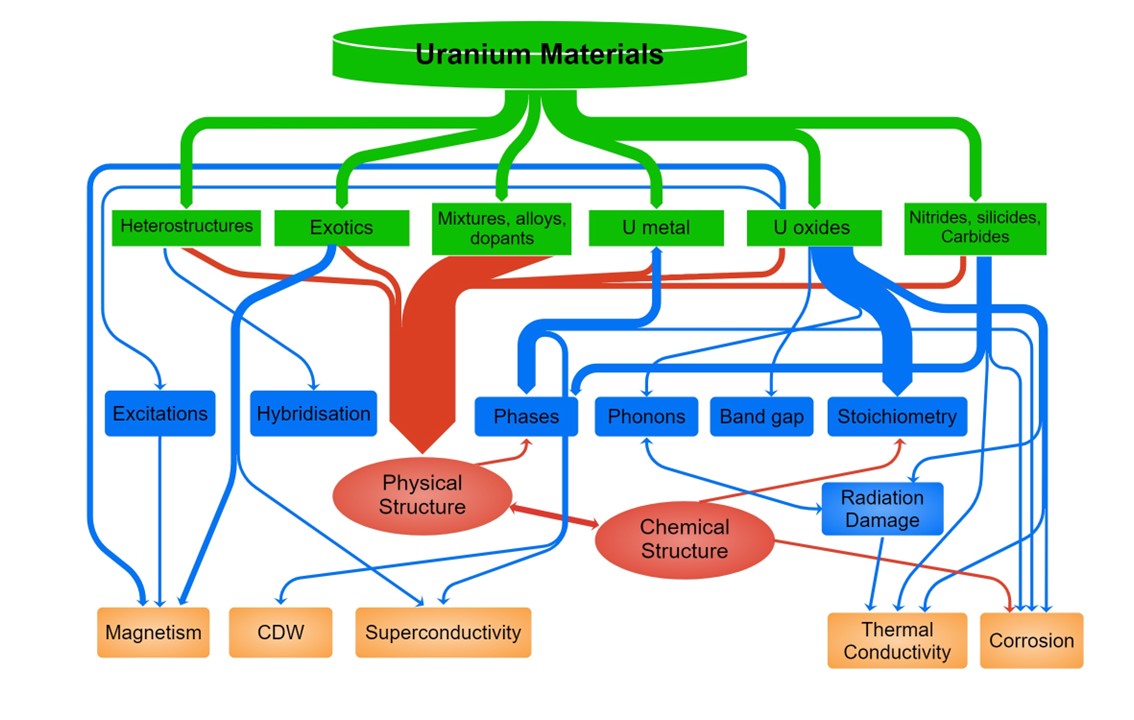}
\caption{Image showing range of uranium-based materials; properties and phenomena that have been investigated in thin film form. \label{Octopus}}
\end{figure}

The marriage of thin films and actinides provides a vast parameter space of experimental and theoretical exploration and there are some distinct areas of study where real advances have been made, and there are still many exciting opportunities for the future; in fundamental research and investigations on applied nuclear fuel and waste materials, driven by a renewed global appetite for advanced fuel materials for modern 21st century nuclear reactor fleets. Fig.\ref{Octopus} highlights the range and connectivity of the materials, properties and phenomena that can be found in the current body of literature. This approach offers some practical experimental and theoretical advantages over more traditional bulk materials, as well as opening new scientific avenues for study.

These sample systems have macroscopic surface areas (typically of the order 1\,cm$^{2}$), with typical masses of 100’s of micrograms. This provides a basic advantage for active work in that many facilities and institutions become accessible that otherwise would be restricted, and transport, handling, and storage of these samples is significantly easier than for their bulk counterparts. For example, a typical 1000\,\AA\,film of UO$_{2}$ would have an activity of only $\sim$1.5\,Bq and would contain a comparable uranium mass to that found in the human body ($\sim$100\,$\mu$g).

Sample synthesis is rapid compared to typical bulk methods, not by mass, but by sample type. This means that one can study a range of compositions quickly, which is ideal for surveys of phase diagrams, alloys, additives etc. These techniques are good for controlling growth parameters in situ, which means that one can mimic surfaces and interfaces of relevant nuclear materials, or design multilayers/interfaces to explore fundamental 2D physics of 5$f$ states. One can engineer the structure, composition, phase, and stoichiometry, one can grow amorphous materials, polycrystalline materials, controlling the grain size, single crystals: modifying the crystal quality, strain fields etc. It allows us to experimentally model many complex effects, (such as radiation damage) in a much simpler way by more careful control of variables and removal of complexity from bulk systems. This connects much more directly with theoretical modelling in nuclear materials, where these idealised systems are able to feed into computational models and vice versa.

Epitaxial matching to a substrate crystal allows fine control of the crystalline growth. This raises the question of whether different allotropes of the films can be stabilised by using different substrates, which act as "guiding templates". As we shall see, this use of substrates to apply strain and alter the properties has already been done for $\alpha$-uranium metal. Some success, also with uranium, has been achieved in preparing a hexagonal-close packed ({\it hcp}) structure using templating. Notably here the {\it hcp} structure does not exist in the bulk. If the interface between film and substrate is fully understood there is the possibility that new allotropes of bulk materials can be produced with unknown properties. The first challenge, of course, is to find the right substrate and method to prepare an epitaxial sample, but this search for new structures could clearly open exciting scientific possibilities.

Broadly, the power of thin films for fundamental physics or materials science can be categorised in terms of (a) surface and dimensional effects; (b) proximity effects; (c) strain effects. Some of these are already being exploited. The interface between the film and the substrate provides not only flexibility in a geometric sense, but also provides a pathway for electronic interactions. For example, the actinide elements have a large spin-orbit parameter, which is roughly dependent on Z$^4$, where Z is the atomic number, and this parameter is important for spintronic applications. Preparing and exploiting such samples clearly is a considerable challenge, but understanding the interfaces is the primary one.

New physics can be expected, such as topological ground states. Such states have already been predicted in 2012 \cite{Zhang2012} for Pu and Am compounds. This comes from the fact that the 5$f$ states in Pu are almost exactly at the boundary between itinerant (before Pu) and localized (after Pu), but this condition can also be found for uranium compounds, e$.$g$.$ UNiSn, \cite{Ivanov2019} and UTe$_2$ \cite{Ran2019}. Initially, of course, the effort on actinide films has been confined almost entirely to using the elements Th and U, which can be handled without difficulty in most laboratories. However, in the longer term, facilities that can handle actinides up to at least Cm can be envisaged. There are huge advantages working with small quantities of these elements, and much that could be discovered, especially about Pu, an element that has six allotropes in the solid state before it melts \cite{Puhandbook}. In this respect it is worth noting that the only transuranium epitaxial films produced so far are those of NpO$_2$ and PuO$_2$ at Los Alamos National Laboratory \cite{Scott2014}. Some of the science deduced from experiments on these samples are discussed below.

The 5$f$ electrons are also the key features of many heavy-fermion compounds, including ferromagnetic superconductors containing uranium \cite{Pfleiderer2009,Aoki2019,Stewart2017}, where it is commonly assumed that it is the hybridization of the conduction and 5$f$ states that lead to the peculiar properties of these systems. In the famous case of URu$_2$Si$_2$ the physics of the system remains unresolved, despite a vast amount of both theory and experiment \cite{Mydosh2011} since the discovery of superconductivity in this material in 1986. More recently, UTe$_2$ has attracted wide attention as a possible topological triplet superconductor \cite{Ran2019}. The prize for the highest superconducting T$_c$ (18 K) of any heavy-fermion material still goes to PuCoGa$_5$, which is mainly a mystery 20 years after its discovery \cite{Sarrao2002}. All the studies referenced above have been performed on bulk samples. Usually, but not all, the experiments reported used single crystals. If epitaxial films of these materials, especially the heavy fermions, were available, further experiments could be easily envisaged.

The basic research motivation for producing actinide thin films is thus abundantly clear. But there is another motivation, equally important. Beyond their undoubted fundamental interest, understanding actinide compounds has substantial importance due to the property that all are radioactive and from the 15 elements, one typically finds 7 fissile isotopes. From this subset, fissile uranium-235 in the chemical form of UO$_2$ powers the large majority of the world’s operational nuclear reactors, generating $\sim$10\% of world power, which is more than a quarter of the world’s low carbon electricity production \cite{IAEA2021}.

Although much is understood about UO$_2$ \cite{Hurley2022}, there are still questions to be answered, particularly relating to the surface and interface reactions and properties of UO$_2$. For example, the synthesis of fuel/clad interfaces opens up the possibility of designing experiments to test pellet-clad-interaction, and uranium metal/oxide interfaces can be used to investigate the behaviour of stored metal wastes. UO$_2$ surfaces can be used to investigate interactions with aqueous environments, simulating 'leakers' (split fuel pins during operation, giving rise to high temperature water and steam exposure) and intermediate longer term spent fuel storage scenarios. Doped UO$_2$ systems could pave the way for studies of modified fuel types to improve thermal conductivity, to improve structural degradation during operation, or to improve end-of-life behaviour.

In addition, there is intense interest in developing alternative actinide compounds to fuel reactors in a safer and more efficient way. These so-called “advanced technology fuels (ATF)” include uranium silicides, nitrides, metallic uranium alloys, thorium compounds, as well as other more exotic fuel designs. A campaign of any new fuel composition in bulk form, and then proceeding studies on radiation behaviour, thermal properties, or interaction with coolant/storage media, are understandably, intensive operations. Using thin films can shortcut many of the typical hurdles and provide a great deal of supporting information in a much shorter time. It is possible for example, to synthesise a new fuel design, simulate corrosion behaviour in long-term storage to assess its feasibility, without ever having to embark on a full in-reactor fuel performance review. 

Hopefully, it should now be clear that thin films, and particularly epitaxial films, have an important and irreplaceable role to play in advancing our understanding of the actinides and their compounds both from a fundamental aspect as well as those compounds of great practical importance to meeting our ongoing energy needs in a decarbonising world. In this review we will cover the growth methods and considerations for various uranium-based films, detail many of the key experiments conducted to date and what they have taught us already, before laying out a roadmap for the future, highlighting the scientific areas we feel hold the most promise and would benefit most from a thin film approach.

\subsection{Early efforts (before $\sim$ 2000) on uranium-based films} \label{s:earlyefforts}

Probably the first recorded use of thin films was by Steeb \cite{Steeb1961} who demonstrated in 1961 that vapour deposition onto heated substrates such as MgO produced an epitaxial film of UO$_2$ with a thickness of $\sim100\,\mbox{\AA}$ that could be further oxidized to U$_4$O$_9$. Further work on the structure of UO$_{2+x}$ was done by electron microscopy at Stuttgart by Steeb {\it et al.}. This was followed by work with electron microscopy by Navinsek \cite{Navinsek1971} and Nasu {\it et al.} \cite{Nasu1972} using different substrates - the best being identified as LiF and NaF. They also observed fission tracks after the samples were irradiated in a reactor. These efforts seem to have reduced once suitable bulk single-crystal samples were produced and the quantitative study of the structure of UO$_{2+x}$ using neutron diffraction was demonstrated. The use of neutrons allowed the positions of the light oxygen atoms to be deduced, and became a major tool in characterizing such systems \cite{Willis1963}.

For uranium metal, the first production of thin films was reported by T. Gouder in 1993 \cite{Gouder1993} in Karlsruhe who deposited monolayers of uranium onto various substrates to explore localisation effects in the uranium overlayer. The first effort to produce epitaxial thin films was reported by Molodtsov {\it et al.} in 1998 \cite{Molodtsov1998} in Dresden. The main objective of their work was to measure resonant photoemission from the surface of uranium \cite{Molodtsov1998,Molodtsov2001}, and scanning tunneling spectroscopy \cite{Berbil-Bautista2004}. As discussed later, in Section \ref{s:hcpgrowth}, a key difficulty in the interpretation of these studies is that there was no X-ray characterization of the samples, as it was not possible to cap the samples and remove them from the preparation chamber. Nonetheless, interest in different structural forms of uranium, as well as the surface layers, was stimulated by these experiments. Theoretically Hao {\it et al.} \cite{Hao1993}, had earlier predicted that the surface 5$f$ states in the {\it bcc} form would be more localized than in the $\alpha$ (orthorhombic) form. Later Stojic {\it et al.}\cite{Stojic2003} predicted that such localization, and ordered magnetism, would even occur at the surface of $\alpha$-U, but no evidence for this has been found.

In a series of experiments, a group in first Darmstadt and then Mainz in Germany grew epitaxial films of the heavy fermion compounds UPd$_2$Al$_3$ and UNi$_2$Al$_3$, which have hexagonal symmetry, with molecular-beam techniques and used heated (111) oriented LaAlO$_3$ (LAO) substrates \cite{Huth1994}. Their interest was in transport measurements, as both systems show antiferromagnetic order with superconductivity at lower temperatures, but with the material remaining antiferromagnetic. Tunneling spectroscopy was used to demonstrate the crucial role of the antiferromagnetic fluctuations in inducing the superconductivity in UPd$_2$Al$_3$ \cite{Jourdan1999}, and measurements of the optical conductivity were also made \cite{Dressel2002}. Rather similar measurements were made on epitaxial films of UNi$_2$Al$_3$ \cite{Jourdan2004,Foerster2007}. At the same time one of the thin films of UPd$_2$Al$_3$ was used in a series of synchrotron experiments to show how the coherence of the x-ray beam, together with the large absorption at the uranium $M_4$ edge, allows information to be obtained on the spatial position of the scattering volume \cite{Bernhoeft1998}. This technique is also discussed in Section \ref{s:AF_UO2}, for more recent experiments on UO$_2$ epitaxial films.

In the 1990's there was also considerable interest into whether memory systems could be based on the magneto-optical Kerr effect (MOKE), and many different systems were studied. This effect requires a bulk ferromagnetic signal. Samples consisting of multilayers of amorphous UAs and elemental Co were produced and the MOKE measurements showed that the uranium had a magnetic moment at room temperature \cite{Fumagalli1993}. A more detailed experiment later took place \cite{Kernavanois2004} to measure the XMCD signal at the uranium $M_4$ edge in a sample of the form [UAs$_{80}$/Co$_{20}$]$_{12}$. The XMCD data confirmed a moment of $\sim 0.80$ $\mu_{\mathrm{B}}$ per U atom at low temperature, but this rapidly declined at higher temperatures. This study showed relatively poorly defined interfaces, with diffusion between the layers.

\section{An overview of the growth of uranium-based films} \label{s:growthoverview}

\subsection{Deposition Techniques} \label{s:deposition}

Thin films, in a research sense, typically range from the Angstrom (\AA) to the micron scale, and involve the controlled deposition of the material of interest onto a prepared surface of a chosen substrate. There is a range of chemical or physical processes that one can employ, and many other reviews and textbooks have dealt with this subject comprehensively \cite{Smith1995, Reichelt1990, Vossen1991}. Here we will focus on just the subset of those techniques that have been used for U-based deposition. It is worth noting that deposition of U has some specific considerations, which depend on materials restrictions of a particular nation, and are centred around the basic radioactivity of the starting material of depleted U.

The choice of deposition method depends on the final application, whether a fundamental study of basic physics, or an applied nuclear materials investigation, this will influence the choice of material, metal, oxide, intermetallic etc. and the required physical and chemical structures. This Section will try and provide a strategic roadmap for new and existing research groups who wish to utilise uranium deposition, by comparing and contrasting the most successful examples in the literature.

{\bf Physical vapour deposition} (PVD) is by far the most frequently adopted technique in this field and can be generalised as the vaporisation of a starting material that is then condensed onto a substrate \cite{Reichelt1990}. PVD methods are flexible, they can be used for metal, compound, and multilayer deposition and one can control stoichiometry, phase and crystalline quality, to some degree. The drawbacks are the need for large apparatus, high or ultra-high vacuum, and bulk solid starting materials, and that the deposited material is often highly energetic, so some additional thermalisation energy at the substrate position is often required.

\begin{figure}[htb]
\centering
\includegraphics[width=0.6\linewidth]{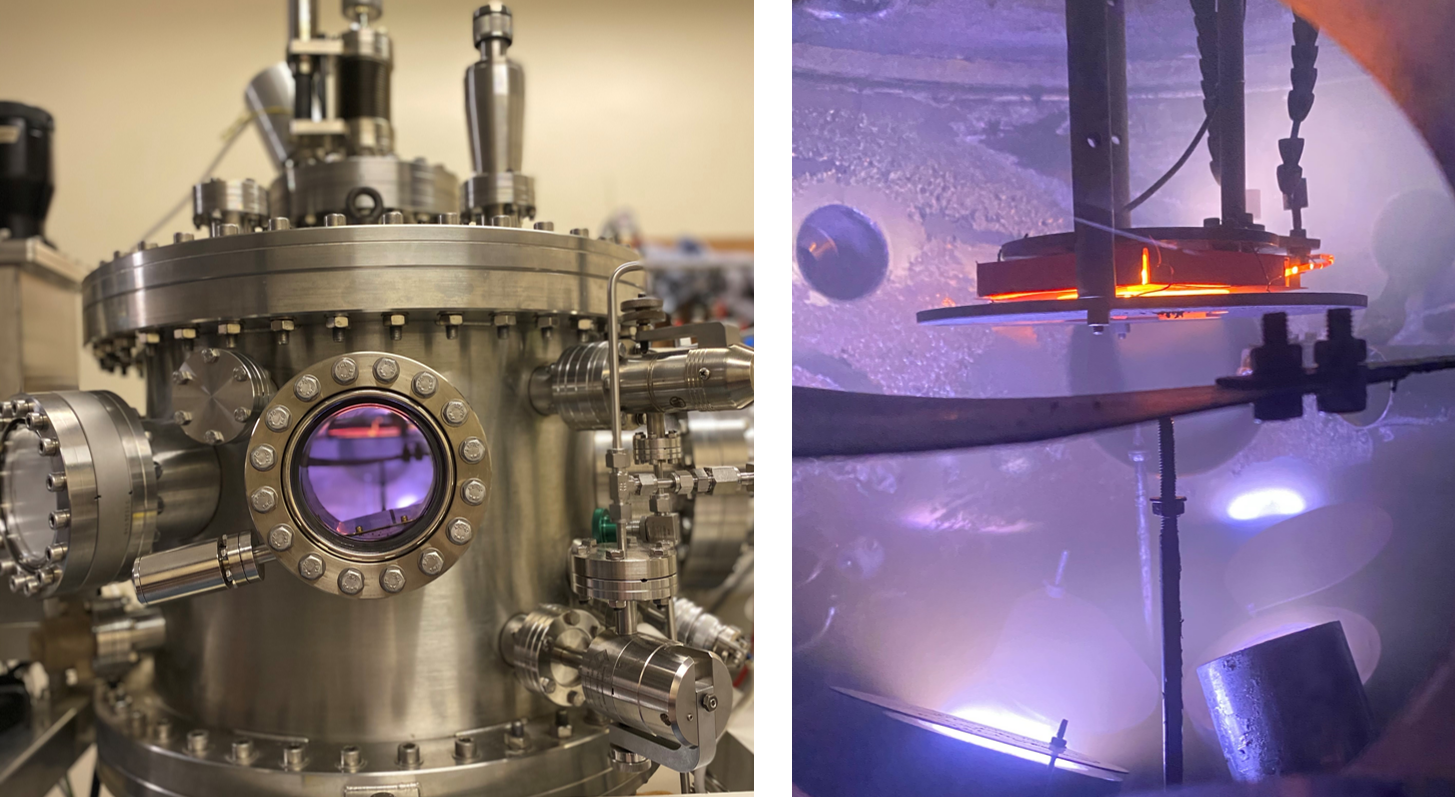}
\caption{External (left) and internal (right) photographs of the actinide DC magnetron sputtering system at Bristol. \label{fig2_1}}
\end{figure}

Of the many PVD options, {\bf sputtering} is the most prevalent and there are many examples of research groups employing this approach \cite{Gouder2001, Ward2008, Chen2010, Strehle2012, Havela2020}. {\bf DC magnetron sputtering} is commonly a UHV ($<10^{-9}$ mbar base vacuum) technique, which consists of discs/ingots of starting material (in these cases, either U metal or UO$_{2}$ ceramic, typically many centimetres in diameter) that are bombarded by a plasma of ionised noble gas, such as argon (pressures typically $<10^{-2}$ mbar). Note that ceramic targets require the use of pulsed-DC or RF sputtering techniques. Typical deposition rates vary from 0.1 to 2 \AA/s. This has been used for U metal deposition \cite{Ward2008}, bilayer spintronics \cite{Gilroy2021}, multilayers \cite{Springell2008, Springell2008a} and intermetallics \cite{Harding2023}.

A sophisticated modification to this technique is the use of {\bf triode sputtering}, which employs a tungsten filament for electron emission to stabilise the plasma at the source \cite{Gouder2001, Havela2020}. This has the advantage of using much smaller starting quantities, so more exotic starting materials are accessible, it is a more efficient use of material; the deposition `racetrack' produced by DC magnetron sputtering can yield efficiencies in the range of only 5\%. However, the lateral homogeneity at the substrate position is not as good. Both of these methods can be adapted for {\bf reactive sputtering} by feeding a small partial pressure of reactive gas into the chamber (pressures typically $<10^{-4}$ mbar), and this has been used to successfully grow oxides \cite{Bao2013, Elbakhshwan2017, Rennie2018a}, nitrides \cite{Black2001, Bright2018}, oxynitrides \cite{Eckle2004} and hydrides of U \cite{Havela2018, Havela2020}.

For the simplest polycrystalline films, one has control over magnetron power, sputter gas pressure, target to substrate distance and substrate temperature. For reactively grown compounds, we have the gas partial pressure as an added lever, and for binary and even tertiary systems, the relative powers of the co-depositing magnetrons become the crucial control mechanism for the formation of specific phases. Even then, it may be difficult to achieve phase pure materials. However, here one can utilise epitaxial matching to `lock-in' desired phases, which prove elusive, even in bulk systems: the line compounds of the U-Si system for example \cite{Harding2023}.

There are important differences with well-known literature examples of epitaxial thin film systems, which have extremely close lattice matches and result in high-quality crystals, often of single domains with mosaics below 0.05$^{\circ}$ (where the mosaic width is the rocking curve full width at half maximum [FWHM]) \cite{Fewster2015}.  Most U-based substrate matches are far from ideal, but are required to stabilise epitaxy of the many compounds, phases and orientations described in this review, hence they can have more complicated crystallographic domains and have mosaics from 0.1 - 2$^{\circ}$ \cite{Ward2008, Bao2013, Bright2018}. Also, the multilayers are not of the same quality as some of the famous rare-earth or tunnel junction heterostructures  and superstructures \cite{Goff2020, Barthelemy1990}, although these standards could be possible for UO$_2$/ThO$_2$ on CaF$_2$, for example.

{\bf Pulsed laser deposition} (PLD) is also a vacuum-based technique, but where the vaporisation is typically performed by a high-power pulsed excimer laser. A plasma `plume' then carries energetic material towards a substrate, which can then be thermalised to aid the crystalline growth of the film. It is also possible to deposit reactively, to make oxides, nitrides etc. The most notable PLD work at Los Alamos National Laboratory, used a KrF excimer laser ($\lambda=248$ nm, repetition rate 1-5 Hz) in varying partial pressures of oxygen, employing substrate heating and rotation, to stabilise UO$_2$, U$_3$O$_8$ and UO$_3$ oxides of uranium \cite{Enriquez2020, Sharma2022}. Although there is little work in the literature, using PLD for U-based deposition, it has many of the same attributes as sputtering. It could be used for metal deposition, and for multilayer and heterostructure synthesis. It typically requires bulk starting materials; the deposition rates are similar and crystalline quality is comparable. However, binary systems, such as silicides and carbides might be more complicated, whereby sputtering or an evaporation technique could be more suitable. Also, the energetics of the deposition process must be controlled sufficiently to prevent the insertion of defects into the growing film.

{\bf Molecular beam epitaxy} (MBE) is a UHV-based technique that uses Knudsen effusion cells or direct e-beam heating (for the more refractory materials such as U) onto small quantities of starting material to provide gradual sublimation. The energetics of this process are lower, and the atoms have longer mean free paths. This results in controlled deposition rates of fractions of an \AA$ $ per second and near layer-by-layer growth. MBE has an advantage in the ability to monitor the growing surface in real time using electron diffraction (RHEED – Reflection High Energy Electron Diffraction) without the requirement for differential pumping. Gas sources can be added to grow oxides, etc, often with cracker stages to produce confined beams of highly-reactive atomic oxygen, or ozone. This is an expensive technique that is focussed on the synthesis of high-quality epitaxial films and is more commonly found in the manufacture of semiconductor devices and magnetic memory; GMR and magnetic tunnel junctions, for example \cite{Barthelemy1990}, although MBE has also been used extensively to grow epitaxial rare-earth layers and superlattices \cite{Goff2020}. This technique can be used to deposit metals, oxides and more complex ternary/quaternary intermetallic materials. Some of the early work at Darmstadt, investigated the $\mbox{UPd}_2\mbox{Al}_3$and UNi$_2$Al$_3$ systems \cite{Jourdan1999, Jourdan2004} as epitaxial thin films. MBE requires a great deal of investment and is not good for high throughput studies, i$.$e$.$ for fast exploration of phase diagrams or a wide range of compositions, but is useful for particular studies where epitaxial quality is crucial.

Aside from PVD there are also chemical methods of deposition, which generally avoid the requirement for bulk solid U-metal or U compounds, and although the scope of these has been limited in terms of the range of U-based materials, they feature prominently in the literature. {\bf Chemical vapour deposition} (CVD) is a vacuum deposition technique that encompasses an enormous range of materials; it involves the reaction or decomposition of volatile precursors onto a substrate wafer. In terms of U-based materials, groups in Cologne \cite{Raauf2021} and UC Berkeley \cite{Straub2019} have successfully used CVD and Magnetic field-assisted CVD to make thin films of UO$_2$, employing the decomposition of U(IV) amidate and reduction of uranium {\it hexakis-tert-}butoxide, respectively. These methods are yet to yield epitaxial films but could be useful for investigating UO$_2$ grain morphologies in polycrystalline samples. The deposition rates are higher than most PVD techniques and these techniques could be used for efficient growth of $>\mu$m thick layers. Although studies have so far been focused on uranium oxides, it may also be possible to modify the methods to deposit metals, nitrides and carbides.

The {\bf sol-gel} process has also been used to prepare UO$_2$ films \cite{Meek2005}. This is most prevalent in metal oxide fabrication, such as TiO$_2$, where a colloidal solution or ‘sol’ is deposited onto a substrate, then becomes a two-phase wet gel, and liquid is removed slowly to allow for densification and eventual film synthesis. In the case of UO$_2$, uranyl acetate in methanol and acetic acid were heated together to form the precursor sol, which was then dropped onto substrates that were spun to coat evenly.  A final heating stage was used to drive off remaining liquid and form a dense film. This method has some drawbacks in terms of crystal structure control and synthesis of epitaxial films. However, it is possible to synthesise $>\mu$m thick layers, and to dope the uranium oxide with typical semiconductor dopant concentrations, which can be very difficult to achieve in most PVD methods.

{\bf Polymer assisted deposition} (PAD) is another chemical solution method, which is common for metal oxide thin film growth \cite{Burrell2007}. Precursors are made using metal ion-coordinated polymers. In this way, the polymer properties, such as viscosity, can be modified to control the metal ion distribution to form homogeneous films. U-based PAD has been successfully used by groups at Los Alamos National Laboratory \cite{Burrell2007, Scott2014} for more than a decade. They have reported epitaxial synthesis of a number of U-oxides, UN$_2$ and UC$_2$ \cite{Scott2014}, as well as the growth of PuO$_2$ epitaxial films \cite{Joyce2010, Wilkerson2020}. In the case of uranium oxides, for example, an aqueous solution of UO$_2$(NO$_3$)$_2$ is added to a polymer, which is then spin coated onto carefully chosen and prepared substrates with the desired lattice matches. These are then annealed in the presence of oxygen at 1000 $^{\circ}$C to form epitaxial films. This has similar advantages to other chemical processes, where film thicknesses are routinely larger than those made with PVD, but clear progress has been made with the PAD technique for uranium, such that epitaxial films of similar quality to those produced by the best PVD methods are possible.

There are a number of options when embarking on a new research programme in this field. In terms of overall strategy, one needs to consider some important questions. Firstly, what sort of material(s); pure metal, compound, heterostructure etc. as this will be the first limiting step in terms of synthesis choice. What amount of starting material is required/is accessible? Does the starting material need to be a compound first, or is it better to make the composition during the deposition process? Often, even if the correct composition is present in the starting material, it is necessary to adjust one or more of the components during growth. For example, a UO$_2$ target material will require additional oxygen to reach stoichiometry. How many samples are required? Some techniques are more suited to high throughput than others.

For metallic systems, especially those containing U, oxidation of the surface (which can be the entire depth of the film in some cases) is a major problem. This means that a capping layer is necessary – typical materials are aluminium, chromium, tungsten, etc., however, the authors recommend niobium, as this layer develops a thin passivating oxide layer of approximately 20\,\AA. Capping layers can sometimes be necessary for oxides also, as these will become hyper-stoichiometric over time and this can affect physical properties. In some cases, buffer layers are used to provide a chemical barrier and to mediate the lattice mismatch between substrate and film. The requirement of extra layers then has consequences for the choice of synthesis method.

Typical research projects/programmes involve preparation of starting material and substrate surfaces; this could be chemical cleaning, sonication, Ar plasma cleaning, or a combination of these steps. Compositions of desired materials are usually tested first on substrate standards, such as glass, or silicon, and where polycrystalline samples are enough then this can provide the basis for the remainder of the synthesis. However, where epitaxy is required, then the strategy is more refined, as careful substrate matching is required, considerations of temperature, thermal expansions of the different materials and possible interfacial mixing. It is not always easy to predict which substrate to use.

\subsection{Characterisation Techniques} \label{s:characterisation}

For most chemical-based synthesis routes the characterisation takes place ex-situ, once the films have been made. However, for all of the PVD techniques described, some in-situ characterisation is routinely used during the deposition process. For more detailed investigations of the physical structure of the films it is more likely that an ex-situ measurement will be employed, and this will depend on the length-scale of interest and whether lateral or longitudinal information is important. Here, we present the most common techniques in more detail with particular recommendations for U-based materials.

The deposition rate is in general the primary characterisation parameter for thin films as this determines the layer thicknesses, and there are many instruments and techniques that are found in the literature. The method of choice will depend on whether in- or ex-situ measurements are needed, the thickness regime, and the required precision; of course, it is often the case that a combination of methods is preferred. For thickness determination within deposition chambers/vessels, quartz crystal microbalances (QCM) are commonly employed, which exploit the Sauerbrey equation \cite{Huang2022}, relating the frequency of oscillation of a piezoelectric crystal (quartz for example) with the mass deposited, i$.$e$.$ as the thickness increases and more mass is deposited, the frequency decreases. The resolution is typically 1 Hz for resonant frequencies in the MHz range, which means that this technique has approximately, monolayer sensitivity. One major advantage is that this can be used during the growth process. However, this means that it has to be mounted in the deposition system itself, which can be complicated, and for a precise measurement it must be at the same position as the substrate, which can be spatially restrictive. Less common methods of thickness determination during growth include laser interferometry \cite{Wu1993}, RHEED, utilising the oscillating intensity of the specular diffraction spot \cite{Rijnders1997, Podkaminer2016}, and ellipsometry \cite{Harke1997}.

\begin{figure}[htb]
\centering
\includegraphics[width=0.8\linewidth]{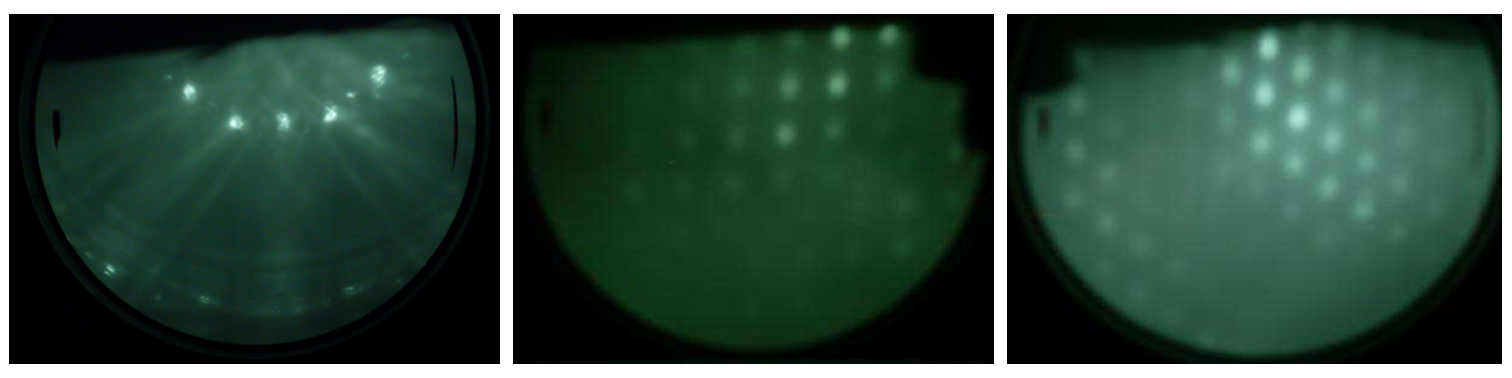}
\caption{The left hand image shows a RHEED pattern from a (001) LAO substrate, [110] azimuth. Kikuchi streaks are visible, and diffraction maxima on arcs where the 2D reciprocal lattice rods intersect the Ewald sphere, indicating a smooth surface. The right hand images are from a 1900\,\AA$ $ UO$_{2}$ film grown on LAO. The surface is rougher and 3D spot patterns are observed, formed by transmission through 3D islands. The [100] and [110] UO$_{2}$ azimuths are shown, respectively. RHEED images have been adapted from Ref. \cite{Ward2010} \label{fig2_2_1}}
\end{figure}

Both low energy electron diffraction (LEED) \cite{Hove2012} and reflection high energy electron diffraction (RHEED) \cite{Braun2007, Ichimiya_2004} are used to identify the crystal structure. It is possible to distinguish between polycrystalline, highly textured, and single crystal systems, and in the most advanced cases, even monitor strain as a function of growth. For MBE, electron diffraction can be acquired during growth, however, for PLD and sputtering, typically the synthesis process has to be paused to view the diffraction image, unless a double differentially pumped electron beam path is employed \cite{Rijnders1997, Podkaminer2016}. Fig.~\ref{fig2_2_1} shows typical RHEED images from a single crystal lanthanum aluminate ($\mbox{LaAl}\mbox{O}_{3}$ or LAO) substrate and UO$_2$ epitaxial film \cite{Ward2010}.

{\bf X-ray reflectivity} (XRR) is used to probe the electron density profile of the film \cite{Daillant2009}, which gives information about the sample morphology; the thickness, the roughness at each surface/interface \cite{Nevot1980, Holy1993}, and the value of the electron density itself, which can be used to infer the composition of the film. The geometry is typically in a specular or longitudinal mode, where the incident and reflected angles are equal (incorporating any offset angle due to sample surface misalignment). Therefore, the wave-vector momentum transfer (usually written Q or q$_z$) is along the surface normal, with no sensitivity to lateral features. There are many freely available resources for modelling this reflectivity spectrum \cite{Vignaud2019, Nelson2006, Bjorck2007}, which use Parratt’s recursive method \cite{Parratt1954}, and then employing a range of fitting algorithms in order to explore the parameter space and find the best global minimum.

\begin{figure}[htb]
\centering
\includegraphics[width=0.6\linewidth]{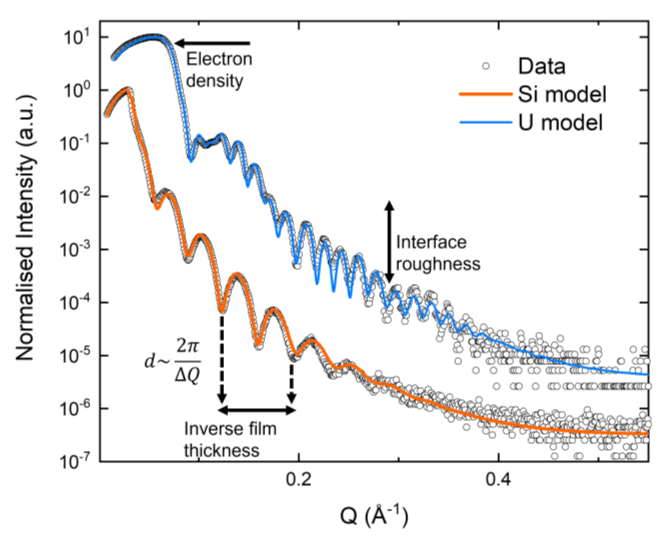}
\caption{Example reflectivity data for typical U and Si thin layers, taken from Ref. \cite{Hardingthesis}. The experimental data are represented by the open black circles and the fitted models of the reflected intensity are shown as blue and orange solid lines, respectively. Total thickness oscillations, Kiessig fringes, are visible for both samples. The U film also displays a longer wavelength periodicity, more visible at low Q, due to a thin oxide layer. The electron density dependence of the critical angles, is also obvious here as the U film $\theta_{C}$ is at a much higher Q than the silicon. \label{fig2_2_2}}
\end{figure}

Fig.~\ref{fig2_2_2} shows two reflectivity curves for thin Si and U films, with data shown as the open black circles and the fits as solid orange and blue lines, respectively \cite{Hardingthesis}. The oscillations are known as Kiessig fringes, which depend on the film thickness \cite{Daillant2009}; an approximate relationship between film thickness and fringe separation is highlighted in the figure. Models of the reflected intensity are often sensitive to changes in the layer thickness on the order of a fraction of an \AA$ $. However, one should note that for films greater than $1000\,$\AA$ $ thick, these fringes become very small and are eventually too difficult to resolve. For a perfectly smooth film the intensity decays as a function of $1/\mathrm{Q}^4$, surface and interface roughness cause this intensity to decay more rapidly \cite{Daillant2009}. The roughness is usually approximated as the root mean square of the thickness variation of a layer and appears as a Gaussian spread of electron density at an interface \cite{Nevot1980}. This means that interfacial roughness and interdiffusion manifest in the same way for specular XRR.

The roughness also affects the amplitude of the fringes, as does the electron density differences between the layers. The overall electron density of the topmost layers in a sample affects the position of the critical angle, $\theta_{C}$ (shown in Fig.~\ref{fig2_2_2}), where total external reflection ends, and the x-rays first start to penetrate the film. There is often a reduction in intensity associated with the smallest angles, below $\theta_{C}$, and this is known as a footprint effect (also visible in Fig.~\ref{fig2_2_2}, where the incident beam size becomes so large that a sizeable fraction of the photons are no longer incident on the surface of the sample.

XRR can also be used for more complex systems and in more complex geometries in order to extract even more information. For instance, bilayers, trilayers etc. will exhibit multiple periodic intensity oscillations that result from coherent reflections from each of the interfaces \cite{Daillant2009}; see the U metal film in Fig.~\ref{fig2_2_2}, where a thinner oxide layer appears as a much longer wavelength oscillation, more prominent at low Q values. Heterostructures and multilayers with repeat units will result in the appearance of sequential peaks of intensity, known as Bragg peaks, whose position will depend on the bilayer thickness and intensity on the number of repeat units \cite{Zabel1994}. The technique itself can be extended further by allowing the incident and reflected angles to vary at various positions in Q (along the specular ridge) to measure intensity as a function of Q$_x$. Modelling this intensity is more difficult and one has to employ the distorted wave born approximation, which maps height-height correlations to generate lateral coherence lengths, and a jaggedness factor, which describe the distribution of height variation and the smoothness of this variation as a function of lateral dimension in the sample \cite{Sinha1988, Daillant2009}.

In summary, XRR is an extremely powerful and versatile technique, which non-destructively investigates the physical structure of a thin film sample. Due to the often sizeable parameter space, and extreme variation in intensity, fitted models may look convincing, exhibiting excellent figures of merit, but they can often be misleading, so it is wise to synthesise a series of samples, where only one or two growth parameters are varied systematically.

{\bf X-ray diffraction} (XRD) has been used to study the crystalline nature of materials for over a hundred years and has some very special considerations when used to study thin films and heterostructures \cite{Fullerton1992, Schuller1992}. Thin films inherently have a small sample volume, and the majority of the photons will pass directly through the sample without scattering. However, the atomic form factor and therefore the scattering amplitude vary as a function of atomic number (Z), which means that the observed intensity varies as a function of Z$^2$ . This is an important advantage when considering U-based thin films, because U is such a strong scatterer that even films of just a few tens of \AA$ $ have measurable intensities on a standard laboratory x-ray source (Cu K$_\alpha\sim 1.54\,$\AA$ $, for example).

Typical measurements to determine phase and structure, are in a similar specular/longitudinal geometry to that described for XRR, however they use 2$\theta$ angles in ranges from 15 - 140$^{\circ}$. The positions of the peaks are the first indication of the crystal structure of the materials in the film, although the spectra can often be dominated by intensity from the substrate. At the two crystallographic extremes; glass gives an amorphous signature, which results in a long, damped periodic intensity background, which is relatively weak overall, but persists at all angles, whereas single crystal substrates only exhibit extremely strong intensity peaks at specific angles, relating to the $d$-spacing along the unique growth axis out of the plane.

Polycrystalline films are grown when there is no obvious lattice match between substrate and film, or when no thermalisation has been used to aid epitaxy. Usually, all of the reflections that one would expect for a powder are visible, and it is even possible to use the Scherrer equation \cite{Cullity2014} to determine grain size, which analyses the peak widths in the same way as for bulk samples; this works well if the grains are smaller than the film thickness, otherwise it is just a measure of the film thickness itself and does not indicate a lateral grain dimension (could be plate-like in shape for example). Normally, for bulk materials, it is standard practice to measure in a longitudinal geometry, and that is true in the first instance for thin-film measurements, however, if a researcher wants to improve their measured signal to be more surface sensitive, this can be achieved by fixing a small incident angle (few degrees in 2$\theta$) to fully illuminate the sample, and then scan the detector angle to achieve the desired 2$\theta$ range.

\begin{figure}[htb]
\centering
\includegraphics[width=0.6\linewidth]{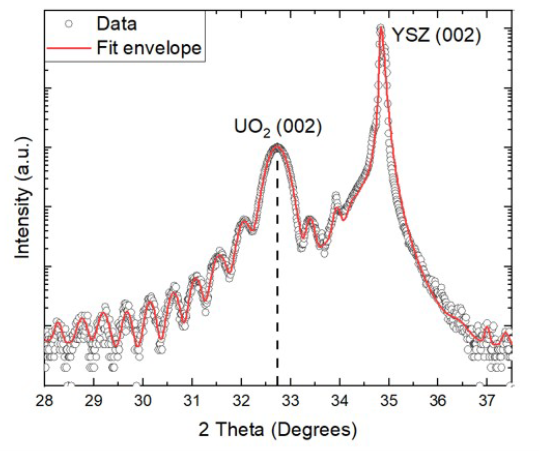}
\caption{Typical XRD spectrum for a $500\,$\AA$ $ UO$_2$ epitaxial thin film, deposited on yttria stabilised zirconia (YSZ), taken from \cite{Renniethesis}. The data are represented by the open black circles and a fitted model of the data is shown as a solid red line. The principal (002) reflection from the [001]-oriented YSZ substrate is very intense and has a mixture of Gaussian and Lorentzian contributions, and is well-modelled by a pseudo-Voigt function. The main (002) reflection of the UO$_2$ film is predominantly Gaussian, which is typical for thin films and there are Laue fringes, relating to the film thickness, visible either side. \label{fig2_2_3}}
\end{figure}

In many cases, the polycrystalline film will have a preferred orientation, or texture \cite{Cullity2014}, which manifests as a preferential intensity for particular reflections, which deviates strongly from the intensities expected from a theoretically ideal powder pattern \cite{Cullity2014}. For high symmetry structures, this is typically with the closest packed plane flat, facing upwards along the surface normal, so the [110] and [111] orientations for \textit{bcc} and \textit{fcc} crystals, respectively, for example. Due to the inherent energetics of most deposition processes and thermalisation, impurities and other structural defects, most polycrystalline samples will exhibit some form(s) of microstrain or residual stress. Microstrain depends on grain orientation, as local lattice spacing variations may vary from grain to grain and can be analysed using the Williamson-Hall analysis \cite{Williamson1952, Borbely2022, Cullity2014}, relating peak position, width and grain size to the microstrain. Residual stress results in an average change in lattice spacing, using $\sin^{2}\psi$ analysis \cite{Cullity2014, Murotani2000, Luo2022}, where lattice parameters are calculated from the $d$-spacing of a selected family of planes, measured as a function of sample inclination.

\begin{figure}[htb]
\centering
\includegraphics[width=0.8\linewidth]{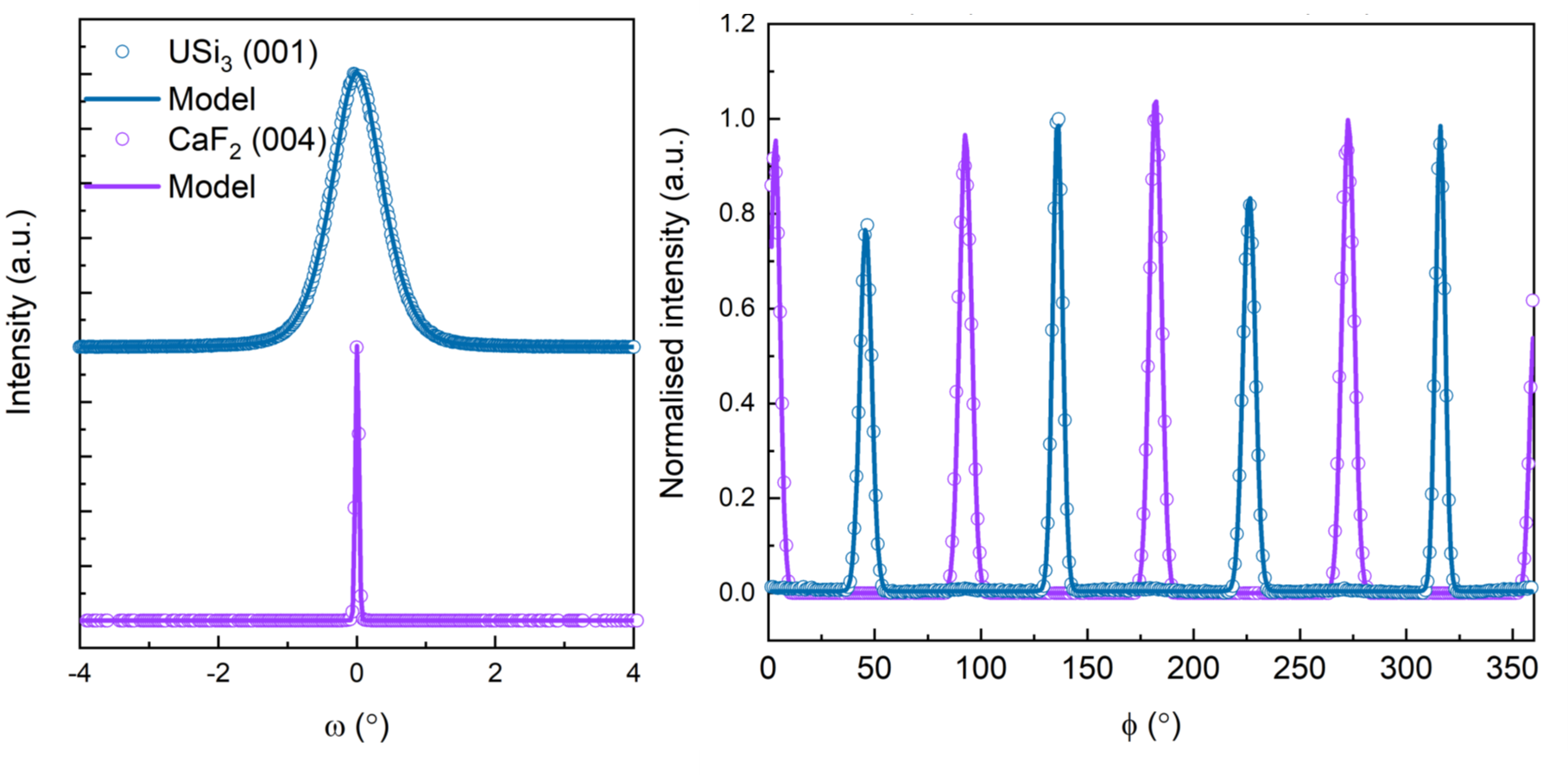}
\caption{The left hand panel shows typical rocking curve data and models for the calcium fluoride substrate and the epitaxial USi$_{3}$ film, in purple and blue, respectively.  The right hand panel shows an example $\phi$-scan, which is an azimuthal rotation about the specular direction at off-specular \{115\} and \{113\} reflections of CaF$_{2}$ and USi$_{3}$, respectively. Taken from \cite{Hardingthesis} \label{fig2_2_4}}
\end{figure}

Epitaxial films have a unique axis along a crystallographic surface normal and they are deposited onto single crystal substrates, which also exhibit a unique axis, see Fig.~\ref{fig2_2_3}. For some sample systems these axes might not be coincident, and an angular offset can be measured between the rocking curve peaks of the film and substrate. Usually, there will already be a predicted lattice match and the first high-angle scan is a survey scan with wide open diffracted beam slits to allow maximum intensity with low resolution. In this way, it is usually possible to observe all of the reflections aligned close to the surface normal of the film. Remember that the crystallographic directions will be different to the flat surface, since the polished substrate surface will not be perfectly aligned along a crystal plane. At this point, the slits can be narrowed to improve resolution and a more detailed measurement of the position of film and substrate peaks allow the determination of respective lattice parameters. Fig.~\ref{fig2_2_3} also highlights another inherent feature in a high-angle diffraction spectrum, which are unique to thin films, the phenomenon of Laue fringes. These are due to a beating frequency in the diffracted signal due to added interference of x-rays reflected at interfacial boundaries (not too dissimilar from the Kiessig fringes in XRR). The main peak for thin layers is also a lot more Gaussian in shape as the number of monolayers decreases. In fact, in this regime, it becomes possible to model the whole diffraction spectrum using discrete numbers of lattice planes to generate the observed peak widths \cite{Schuller1992}.

Once the orientation has been determined, a rocking-curve measurement \cite{Fewster2015, Cullity2014}, which varies incident and reflected beam angles at a fixed $2\theta$, is used to align at the maximum of the peak. The full width at half maximum (FWHM) of this rocking curve is also a standard measure of crystal quality (for films and bulk crystals) and is known as the mosaicity. Literature values for single crystals of U-based studies can be from 0.1 - 2$^{\circ}$, see the left hand panel of Fig.~\ref{fig2_2_4}. For many epitaxial thin film systems an unusual phenomenon is observed in the rocking curves \cite{Fewster2015}. They consist of two peak shapes, a sharp component, together with a wider contribution. There are a number of theories as to why this is present, and a recent paper by Wildes \cite{Wildes2020} gives a good discussion of such profiles. It is usual to take the wider component as a reflection of the bulk mosaicity of the deposited film.

To confirm that a film is indeed epitaxial and then to relate the rotational orientation of the substrate to the film, it is necessary to measure off-specular reflections and probe their azimuthal dependence, by rotation around the surface normal, a so-called phi scan \cite{Fewster2015}.  Fig.~\ref{fig2_2_4} shows a phi-scan for a [001]-oriented USi$_3$ film grown on a [001]-oriented CaF$_2$ single crystal substrate \cite{Hardingthesis}. The first point to note is that if this were simply textured then there would be no discrete azimuthal dependence. Second, the number of reflections in the 360$^{\circ}$ range indicates how many domains are present, as there may be more than one in-plane direction that results in a lattice match; in fact, this is very common for cubic materials. Finally, the azimuthal angular offset between substrate and film determines the in-plane lattice match relationship.

One could go further and measure truly in-plane diffraction, where the whole system is set at a slight tilt angle and the detector is moved out of the scattering plane, until the Q-vector is pointing almost along the surface of the sample, this is known as grazing incidence x-ray diffraction (GIXRD) \cite{Marra1979, Neuschitzer2012}. In this regime, the surface is severely truncated, and the diffraction spots become elongated and more rod-like, since the Fourier transform of a Dirac $\delta$ function (analogous to lattice for a single layer) is a constant (hence, rod of scattering). These are known as crystal truncation rods and can be modelled to give detailed in-plane information \cite{Marra1979, Neuschitzer2012}.

To summarise, there are many ways that XRD can be used to give a good understanding of the crystal structure(s), lattice parameters, lattice matches, stresses and strains etc. However, most x-ray spot sizes at the sample position will be several mm$^2$, which means that these techniques are not suited for local features, individual grain information, stresses, strains etc., and a microscopy method may be better.

\textbf{Scanning electron microscopy (SEM)} uses a focused beam of electrons to raster across the sample, which produces secondary and backscattered electrons, which can be used for imaging or diffraction imaging, achieving magnifications of approx. $\times$250,000 \cite{Reimer1998}. For epitaxial films, which are often smooth (rms roughness of $<$10’s\,\AA), the SEM image can look featureless, although it is possible to magnify the edge of a sample to observe the film/substrate interface and estimate the film thickness. This only works for thick films $>$1000\,\AA$ $  and the resolution is no better than 100\,\AA. However, there are other operational modes of an SEM, which can be used to give higher resolution images, compositional and crystallographic information. If the SEM system has an additional focussed ion beam column, then it is possible to cut thin cross-sectional foils through the surface and across the film/substrate interface and image them in transmission mode \cite{Tomus_2013}.

Compositional information can be gathered using energy-dispersive x-ray spectroscopy (EDX), where the electron beam stimulates the emission of characteristic x-rays, which relate to specific electron shell transitions in given elements \cite{Reimer1998}. This is not highly accurate quantitatively, but it can give a good approximation of the elemental composition. Moreover, the rastered electron beam gives lateral compositional data, so that a map of elemental composition can be constructed, which is especially useful for heterogeneous systems.

\begin{figure}[htb]
\centering
\includegraphics[width=0.4\linewidth]{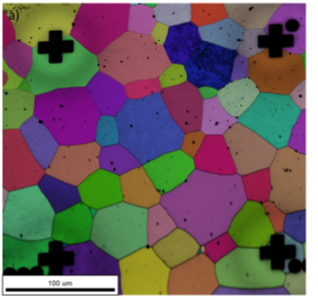}
\caption{EBSD data collected for a 1000\,\AA$ $ UO$_2$ film deposited on an annealed polycrystalline YSZ substrate, showing an average grain diameter of about 40\,$\mu$m. Each colour represents a particular grain orientation and these exhibit a random distribution, expected from a polycrystalline material. Taken from \cite{Wasikthesis}. \label{fig2_2_5}}
\end{figure}

Finally, \textbf{electron back-scatter diffraction} (EBSD) can be used to investigate the structure, phase and crystal orientation \cite{Dingley_2018}. For bulk materials, the surface has to be extremely smooth, but for thin film synthesis the resulting surface is often far better, in terms of rms roughness, than any mechanically prepared material, as long as the lateral features are larger than the lateral resolution ($\sim$100\,\AA). For single crystals, EBSD would just give a single orientation and a map of the surface would be uninteresting. However, for polycrystalline samples it is possible to generate a map of the grain structure and individual orientations, see Fig.~\ref{fig2_2_5} for an example UO$_2$ polycrystalline film, deposited on polycrystalline YSZ \cite{Wasikthesis}.

It should be noted that SEM is largely non-destructive for low resolution imaging and ESBD. However, the sample can undergo significant damage if images are taken at high resolution, or it may need to be coated if non-conducting. Also, the probing depth of the EDX process is on the order of microns, which means that signals are often dominated by the substrate materials.

\begin{figure}[htb]
\centering
\includegraphics[width=0.7\linewidth]{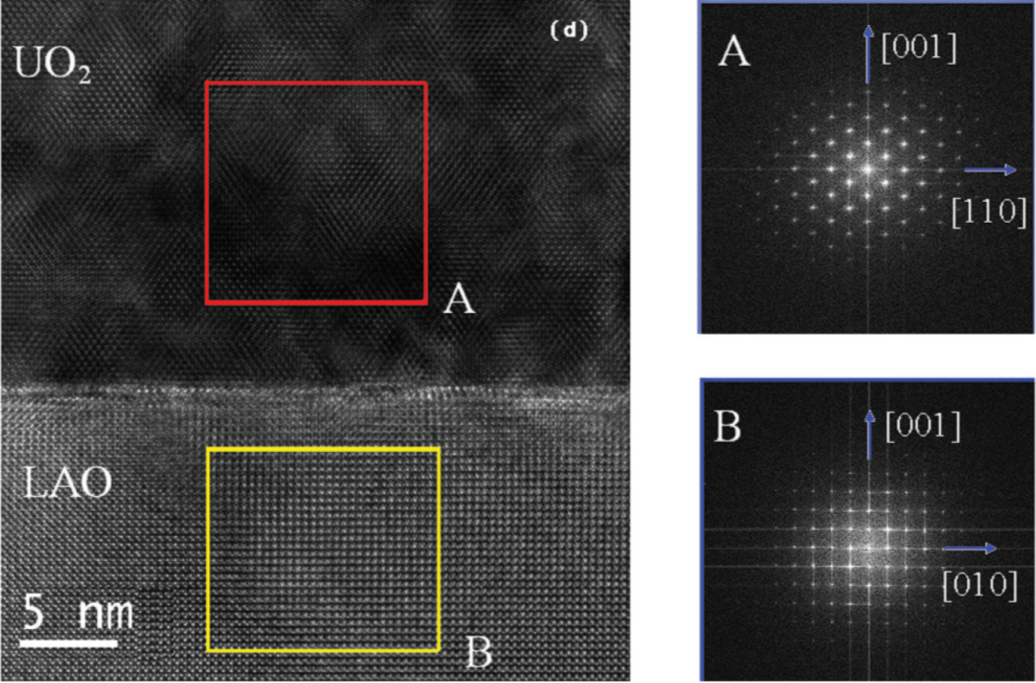}
\caption{Left hand panel is a high resolution TEM image of a UO$_{2}$/LAO cross-section taken from Bao \textit{et al.} \cite{Bao2013} and the right hand panel shows diffraction images for, A, the [001]-oriented UO$_{2}$ film and B, the [001]-oriented LAO.\label{fig2_2_6}}
\end{figure}

{\bf High-resolution Transmission electron microscopy} (HRTEM) images a thin cross-section of a sample, using a magnetically focussed, highly collimated, high energy ($\sim200\,\mbox{keV}$) electron beam in a transmission geometry \cite{Carter2016}. This is typically a UHV set up, although specialist systems are able to deliver gas environments \cite{Hansen2015}. Typical magnifications can reach $\times$1,000,000 so can image in the $10\,\mbox{\AA}$ regime, and can operate in a variety of different acquisition modes where the image contrast can be simply due to thickness variation, atomic number or crystallographic orientation. It is also possible to generate diffraction images and analyse the crystal structure at a very local level. Fig.~\ref{fig2_2_6} shows an example HRTEM high resolution image and diffraction image from an epitaxial UO$_2$ film deposited on lanthanum aluminate (LAO) as an example \cite{Bao2013}.

\textbf{Electron energy loss spectroscopy (EELS)} \cite{Hofer2016}, which uses an electron spectrometer to measure the energy lost by electrons due to inelastic scattering to probe core shell states, much in the same way as EDX, can be used to make high resolution elemental composition maps. Note that HRTEM usually relies on initial preparation via SEM and focussed ion beam milling. A cross-sectional lift out is prepared by careful etching and final platinum adhesion to a movable needle to remove the sample, before attaching it to a standard TEM grid. Specific consideration has to be made for U-based materials, as the electron density of U poses a particularly difficult challenge for the transmission of an electron beam. This means that foils of $<$1000\,\AA\,are preferable, and this requires added preparation time and care, compared with more standard materials.

\textbf{Ion beam analysis (IBA)} \cite{Jeynes2016} is a less common, but complementary field of materials characterisation that, particularly for the case of thin films, uses \textbf{Rutherford backscattering spectroscopy (RBS)} to map the depth-dependent profile of sub-micron layers with element specificity. This technique typically uses the energy profile of backscattered helium ions to build a model of the component species in a thin layer, where the depth resolution is limited by the energy resolution of the detector. This can be especially effective for heavy ions, such as uranium \cite{Kim-Ngan2010, Usov2014, Chiang2015}, where it is possible to operate in isotope detection limits of the ppm. The most advanced high resolution systems can operate with a depth resolution of less than 20\,\AA\, \cite{Robson2021}, which although not at the same scale as XRR, does provide simultaneous element-specific information.

\begin{figure}[htb]
\centering
\includegraphics[width=0.8\linewidth]{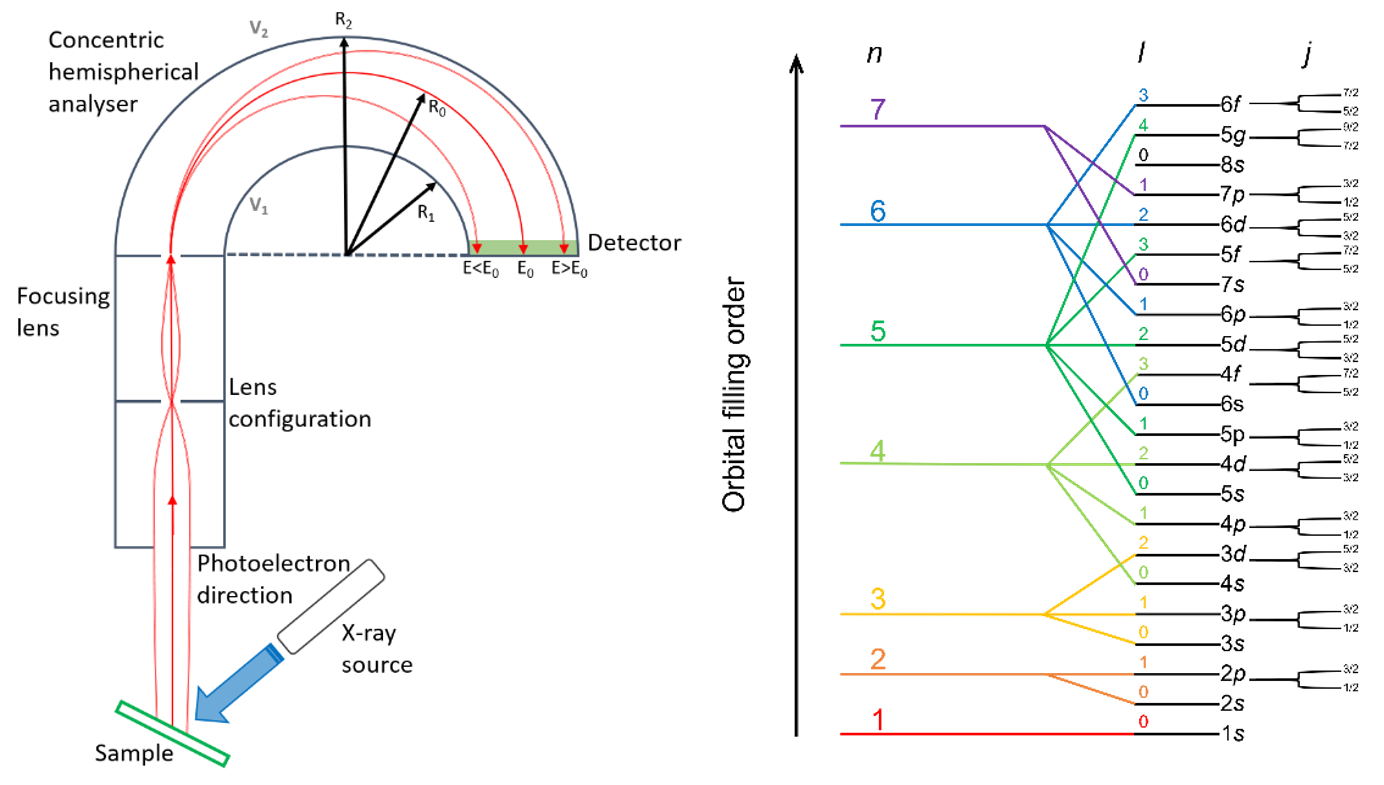}
\caption{Left hand panel shows a schematic of a typical XPS system, indicating the photelectron path within the concentric hemispherical analyser. The right hand panel gives a depiction of the electron configuration filling order for sub-shells for a uranium atom. Here, the principle quantum number, orbital angular momentum and total angular momentum are labelled, \textit{n}, \textit{l} and \textit{j}, respectively. Taken from \cite{Hardingthesis} \label{fig2_2_7}}
\end{figure}

{\bf X-ray photoemission spectroscopy (XPS)} is based on the photoelectric effect \cite{Hofmann2012, Krishna_2022}, which is the emission of electrons from a given material, due to an incident light source. Most lab-based facilities use an Al or Mg K-edge x-ray source with energies of 1486.6 and 1253.6\,eV, respectively, which can be monochromated to improve the final energy resolution of the resulting spectra \cite{Hofmann2012}. The electrons are emitted from core shells within the different atomic species in the film and their energies are then determined by an electrostatic hemispherical analyser. This means that the system has to be in UHV conditions, and in fact the main chambers are often in the 10$^{-11}$ mbar range. Fig.~\ref{fig2_2_7} shows a schematic of a typical XPS \cite{Hardingthesis}.

At a basic level, the spectra of the core levels of the constituent elements give a fingerprint of the composition, similar in information to EDX, however, the probing depth here is less than 100\,\AA, which means that the signal will result from interactions with the film in all but the thinnest layers. An analysis of the integrated areas of respective core-level peaks can also give quantitative compositional information \cite{Brundle2020}. Fig.~ \ref{fig2_2_7} shows the expected core levels for a U atom \cite{Hardingthesis}.

One of the most powerful uses of XPS is in characterising the binding state of the constituent elements, where specific chemical shifts in energy appear due to the existence of particular valence states \cite{Hofmann2012}. In some compounds the core levels of the metallic ions have features that are even more sensitive to the local coordination chemistry. For example, UO$_2$ has `so-called' shake up satellites that are sensitive to the oxygen stoichiometry, and with careful measurement it is possible to measure excess oxygen to better than about 3\%, i.e. x to $\pm0.06$ in UO$_{2\pm\,x}$ \cite{Allen1982, Santos2004}.

XPS is clearly a very powerful non-destructive technique for chemical analysis, but there are possible modifications from the standard that make it even more useful. Since many of the deposition systems are in UHV conditions, it is also possible to combine these facilities with XPS. Focussed x-ray sources, or focussed analysers, can achieve lateral resolutions of 10’s $\mu$m, so that it is possible to laterally map the chemical states of samples. Argon plasma sources can be used to gently etch through the sample, which gives depth profiling information, which could be crucial for understanding complex interfaces. Finally, the use of a much lower-energy ultraviolet source (UPS) means that the valence states are accessible with good resolution, which in conjunction with the XPS spectra, creates a more complete chemical picture \cite{Hofmann2012}.

\clearpage

\section{Uranium Metal Phases and Alloys} \label{s:metals}

\subsection{Introduction} \label{s:metalsintro}

\begin{figure}[t]
\centering
\includegraphics[width=\linewidth]{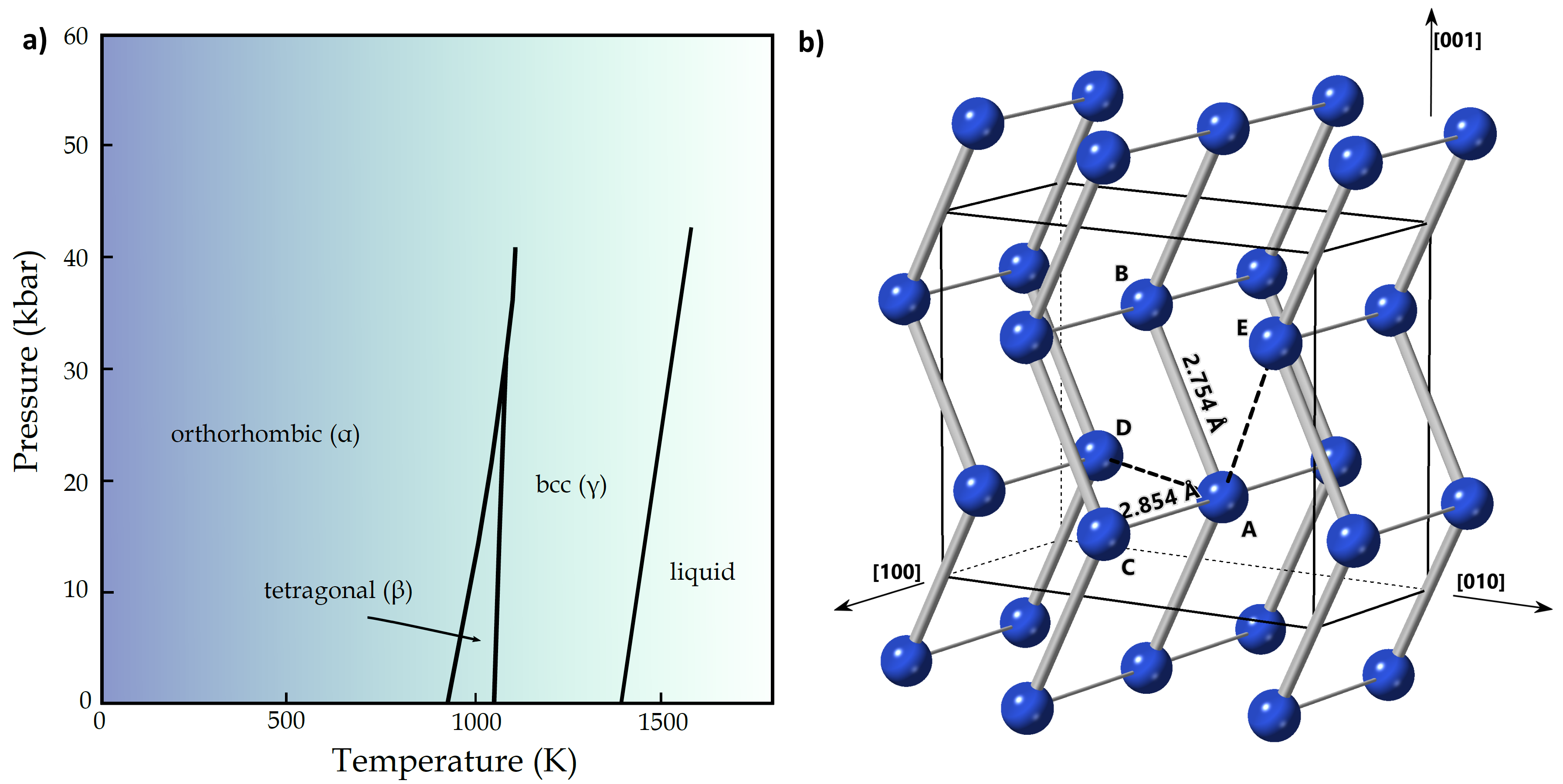}
\caption{Phase diagram of uranium as a function of pressure and temperature, with structure of $\alpha\mbox{-U}$ shown expanded outside the conventional unit cell to highlight the corrugated atomic planes along the [010] direction. For every atom there are four neighbouring atoms in the corrugated (010) sheets, two at $d_1 = 2.754\,\mbox{\AA{} (AB)}$ and two at $d_2 = 2.854\,\mbox{\AA{} (AC)}$, which are shown by thick and thin grey bonds, respectively. A further eight neighbouring atoms at distances $>3.2\,\mbox{\AA{}}$ are found in the adjacent corrugated planes, examples of the two types (AD) and (AE) are shown as black dashed lines. Adapted from \cite{Lander1994}} \label{fig2_2a}
\end{figure}

Before discussing the progress in created thin films of U metal and related alloys, we first discuss some of the fundamentals of bulk U metal, including a brief discussion of its phase diagram and some important electronic properties that have been studied over many decades. Much of this discussion will be focused on the basic physics, to frame the importance and power of studies with thin films, but it is important not to forget the use of metallic U in the earliest days of nuclear energy production, as well as of course weapons.

The phase diagram of U as a function of pressure and temperature is shown in Fig$.$ \ref{fig2_2a}. Similar to many metals U adopts a high-temperature body-centred cubic, \textit{bcc}, $\gamma$-phase stable from the melting point of $1132$ $^{\circ}$C down to $772$ $^{\circ}$C. At pressures below approximately $3.5\;\mbox{GPa}$ this temperature marks the transition to a tetragonal $\beta$ phase, stable between 772 $^{\circ}$C and 662 $^{\circ}$C,  with a complex atomic structure and 30 atoms in the unit cell. Efforts were made to solve the $\beta\mbox{-U}$ structure as early as the 1950's \cite{Donohue1974}, but it was only in 1988 with the advent of modern neutron powder diffraction, together with Rietveld analysis, that the the structure was eventually solved \cite{Lawson1988b}.

Significantly before this is in 1937 Jacob and Warren \cite{Jacob1937} solved the atomic structure of the room-temperature stable, orthorhombic $\alpha$-phase which is adopted below 662 $^{\circ}$C at atmospheric pressure or, above approximately $3.5\;\mbox{GPa}$,  directly from the \textit{bcc} phase at around $800$ $^{\circ}$C. The $\alpha$-phase allotrope, defined within the \textit{Cmcm} space group and shown in Fig. \ref{fig2_2a} is a highly open, consisting of series of corrugated atomic chains nested along the $[010]$ direction. Such a structure produces a series of highly anisotropic interatomic distances reflecting the complex role of the 5$f$ electrons in stabilizing the structure \cite{Brooks1995,Soderlind1995,Mettout1993}. At ambient pressure this low symmetry, highly open orthorhombic character is unique among the elements, however it has been found as a common stable structure for the high-pressure forms of the light rare-earths and is also closely related to the high-pressure forms of the heavier actinides \cite{mcmahon_jpcm_2021}. For a more in-depth description of the structures of the three bulk allotropes the reader is directed to Ref.~\cite{Donohue1974} and for a detailed account of extensive structural studies conducted on bulk $\alpha\mbox{-U}$ in the 1980's and early 1990's the reader is directed to the review by Lander \textit{et al.} \cite{Lander1994}.

Given the complexity of the U phase diagram and the unusual structures contained within it, it is perhaps unsurprising that large single crystals of U metal are difficult to prepare. Indeed, until the production of single crystals of $\alpha$-U from a molten salt process in the late 1990's \cite{McPheeters1997}, only the few crystals made by E. Fisher in the 1950's (known to contain measurable impurities of Si and Fe) were available for use \cite{Lander1994}. Obtaining single crystals of the two high temperature allotropes has proved even more elusive. Long lived metastable single crystals of the $\beta$ phase were first synthesised by A. N Holden in 1951 by quenching a $1.36$ at.\%U-Cr alloy from the $\beta$-phase in water \cite{Holden1952}. However these crystals were not truly single phase and contained a likely Cr rich precipitate. Later, the same authors produced phase pure single crystals by reducing the Cr content to 0.5 at.\% however such crystals transform to the $\alpha$-phase within a few hours at room temperature \cite{Holden1952}. To date, bulk single crystals of the \textit{bcc} $\gamma$ phase have never successfully been produced despite significant efforts.

Despite the challenges in crystal growth, subsequent decades saw great numbers of detailed and varied studies into the structural, electronic and thermodynamic properties of the bulk U allotropes. For the room temperature $\alpha$-phase, many of the experiments focused on pursuing the various possible long range ordered states that could exist at low temperature. The major findings of most studies these studies (pre-1994) are encompassed by the review article in Ref.~\cite{Lander1994}. Perhaps the greatest experimental effort was expended on probing the unexpected and mysterious phase transition near 43 K, first observed in temperature dependant measurements of the elastics constants in 1961 \cite{Lander1994}. Key to understanding general properties of $\alpha\mbox{-U}$, as well as the CDW state it hosts, was a pioneering experiment in 1979 \cite{Crummett1979} to measure the phonon dispersion in a bulk single crystal, and a subsequent experiment by Smith {\it et al.} in 1980 \cite{Smith1980} that extended the phonon studies to low temperature, and showed significant phonon softening near $q_{\mathrm{CDW}}$ highlighting the transition as soft-phonon driven. Robust \textit{ab-initio} calculations that accurately modelled the observed dispersion were not developed until 2008 by J. Bouchet \cite{Bouchet2008}.

At even lower temperatures a superconducting state was established, $T_{\mathrm{c}} = 0.7\;\mbox{K}$ \cite{Lashley2001} but there is still much ongoing debate as to the exact nature of the superconductivity, bulk or filamentary \cite{OBrien2002,Graf2009}, and the importance of sample purity considerations. The coexistence of these two ground states is unique amongst the elements, and reminiscent of more exotic highly correlated electron systems such as the high $T_\mathrm{c}$ cuprate family of superconductors. However, the exact relationship between the ordered states is still unclear. It has been determined that pressure initially suppresses the CDW and enhances the superconductivity, increasing $T_{\mathrm{c}}$ to over 2\,K, however further increase in pressure to $10\;\mbox{GPa}$ suppressed both phenomena \cite{Lander1994,Raymond2011}. Additionally, magnetism has long been discussed and sought in connection with U metal, early neutron work on thermal expansion even claimed to have found extra peaks \cite{Barrett1963}. Although later it was established that these came from multiple scattering processes and to date no magnetism has been found in bulk $\alpha$-U \cite{Lander1994}.

Considering the relative difficulty of obtaining bulk single crystals for the various allotropes and the obvious applications for epitaxial strain to modify the different types of long range order present, it becomes clear that the synthesis of U thin films, especially epitaxial films, is of great interest. The following Sections will lay out the general synthesis considerations for the key metallic systems before detailing a number of important case studies where metallic U thin films have been employed to provide scientific insight into this fascinating element and its many allotropes that would not have been possible by relying on bulk synthesis routes alone. In terms of superconductivity, there is ongoing work on elemental, and alloy thin films, but the situation at the time of writing is currently unclear in thin films. 

\subsection{Production of metallic uranium films} \label{s:metalsgrowth}

The number of laboratories world wide that are sufficiently equipped to deposit metallic U films is understandably small. Aside from the issues of sourcing appropriate purity starting material, the base vacuum within any potential deposition system must be sufficiently low to avoid the immediate formation of $\mbox{UO}_2$ or other compounds. As a result the ratio of metal film producing laboratories to oxide or other U compound producing laboratories is also small. As discussed in Section \ref{s:earlyefforts} the first U deposition for the explicit purpose of forming metallic films was conducted in the 1990's by T. Gouder in an effort to induce localisation \cite{Gouder1993} after which there was also early work on heavy fermion films \cite{Huth1994}, and attempts to stabilise the metastable hcp form of U in thin layers \cite{Molodtsov1998,Molodtsov2001}. Following these early examples, we will now discuss a number of the key U thin film systems that have been developed to date and are detailed in Table \ref{tab:metals}. The majority these systems were first fabricated using the PVD method DC magnetron sputtering, see Section \ref{s:deposition} and the optimum growth conditions discovered by an iterative process using many of the characterisation methods discussed in Section \ref{s:characterisation}.

\begin{figure}[t]
\centering
\includegraphics[width=0.6\linewidth]{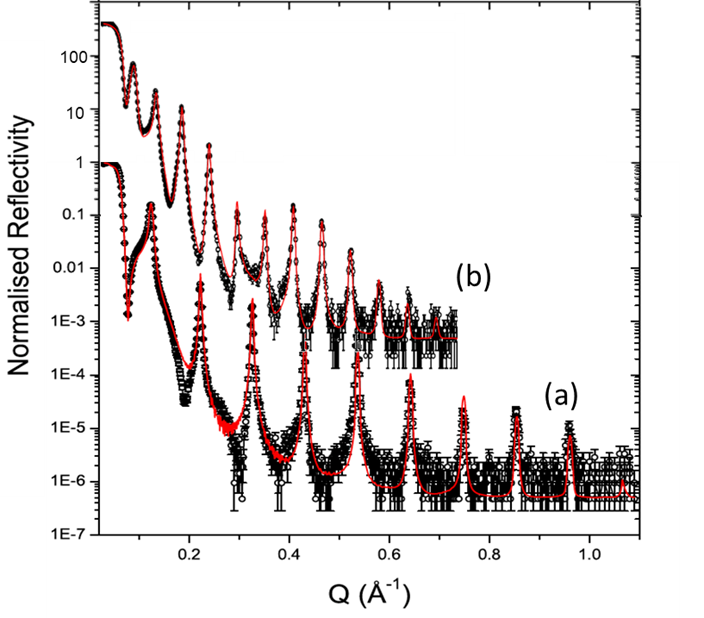}
\caption{X-ray reflectivities (Cu $\mathrm{K}_\alpha$ source) for  (a) [U(32 \AA{})/Fe(27 \AA)]$_{30}$, and (b) [U(89 \AA{})/Gd(20 \AA{})]$_{20}$. Data are shown as black open circles, and the red lines are best fit models. Adapted from Ref. \cite{Springell2008}.} \label{fig2_5}
\end{figure}

\subsubsection{Uranium containing multilayers} \label{s:ML}

The deposition of polycrystalline or amorphous U layers is substantially easier than the epitaxial systems described below. However, there has been significant and continued interest in such systems in the context of multilayers and bilayers to investigate the possibility of induced behaviour between U and other metals as a direct result of their proximity across the interface. The first work on uranium containing multilayers was conducted by Beesley \textit{et al.} \cite{Beesley2004,Beesley2004a}, which spurred significant further work detailed in Sections \ref{s:U_MLs} and \ref{s:Ubilayers}. The precise growth details for each system can be found in the references included in table \ref{tab:metals} however in general the U layers are grown by sputtering techniques with minimal substrate cleaning and little or no heating during growth. As induced effects typically exist over relatively short length scales, a key parameter in such films is the interfacial quality between the layers, and this can vary dramatically from system to system. Note that similar growth considerations are valid for both multilayer and bilayer systems; however, significantly more characterisation has been performed on the multilayer systems and thus they will form the focus of the following discussion. As described in Section \ref{s:characterisation}, XRR is an invaluable tool for characterising thin film systems. Typical XRR curves from U/Fe and U/Gd multilayers are shown in Fig.~\ref{fig2_5}.

\begin{figure}[t]
\centering
\includegraphics[width=\linewidth]{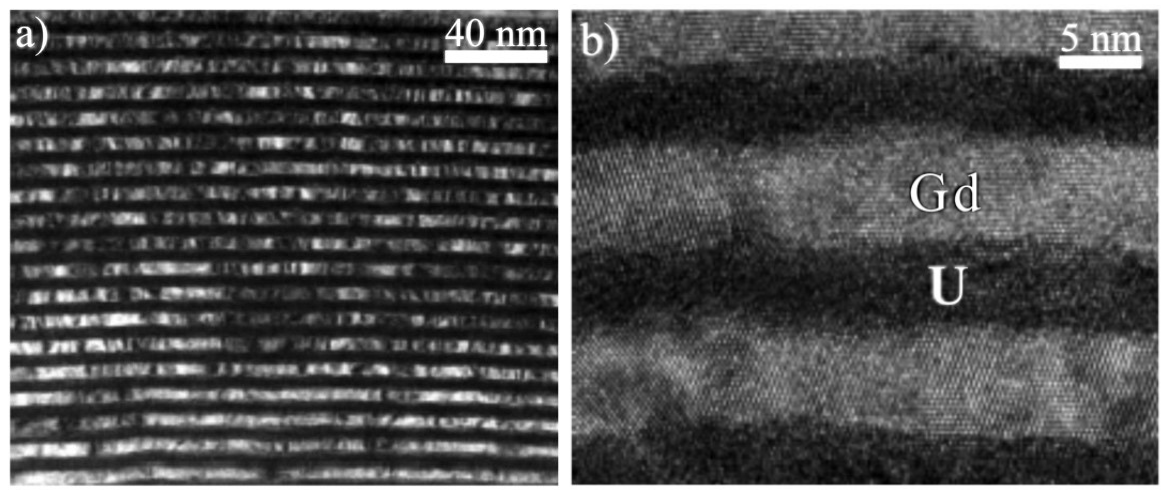}
\caption{TEM images of a [U(50 \AA{})/Gd(50 \AA{})]$_{20}$ multilayer. The mass contrast between the U and Gd can be easily distinguished; the lighter Gd scatter the electron beam less than the U atoms and thus appear brighter. The layers are labelled in the high-resolution image (b); (a) shows a lower resolution. Taken from Ref. \cite{Springell2010}.} \label{fig2_6}
\end{figure}

Analysis of these and related data demonstrate that the interfacial properties of the U/Fe, U/Co, and U/Ni multilayers are quite different from that of the U/Gd multilayers. In the case of the transition metal systems, the interfaces are not as chemically sharp, and there is an interdiffused region of thickness $\sim$ 15 \AA{} at each interface. In contrast, for the U/Gd multilayers, the interfaces are much sharper, and there appears no significant interdiffusion between Gd and U \cite{Springell2008}. This conclusion is supported by transmission electron microscopy studies - a typical image of one such film is shown in Fig. \ref{fig2_6} . Here the layers are well defined with relatively low roughness, and the Gd layers are strongly crystalline, whereas the U layers have small nano-crystallites or are amorphous. In many cases the brighter Gd crystallites extend across the layers suggesting sizes of as much as 50 \AA{} in the vertical growth direction.

Regarding the growth directions of the layers, in all cases the structure of the 3$d$ transition element follows the expected preferred orientation, i$.$e$.$  {\it bcc} for Fe, {\it hcp} for Co, and {\it fcc} for Ni. The U is in a poorly defined $\alpha$ structure. Surprisingly however, in the U/Gd multilayers, although the Gd forms in the expected {\it hcp} phase, the U forms also in a {\it hcp} symmetry, with a unique $c$ (hexagonal axis) growth axis. There is no ordering that could be found within the hexagonal planes, so these consist of a number of random domains of the hexagonal basal plane \cite{Springell2008}. The $c_{\mathrm{Gd}}$ is 5.84 \AA{}, which is close to the bulk value of 5.78 \AA{}, and the $c_{\mathrm{U}}$ = 5.60 \AA{}. If we assume that the atomic volume is the same as that of the of the $\alpha$-U form, then the $a_{\mathrm{U}}\sim 2.91$ \AA{}, giving a $c_{\mathrm{U}}/a_{\mathrm{U}}$ ratio of 1.92, which is much larger than that expected for a hard sphere model 1.633. The only element close to this is Cd with a ratio $c/a$ = 1.89. We return to the topic of {\it hcp}-U in Section \ref{s:hcpgrowth}.

\subsubsection{Achieving single crystal films of $\alpha\mbox{-U}$ } \label{s:sxalpha}

The key breakthroughs in the synthesis of epitaxial metallic U layers came with the realisation of the importance of unreactive epitaxial buffer layers deposited prior to the U layer, and the addition of substrate heating to allow sufficient mobility of deposited atoms. Refractory metal seed/buffer layers are routinely implemented in the growth of rare-earth systems to prevent interactions with the substrate and/or to bridge a large mismatch in lattice parameters between the substrate and overlayer in order to achieve successful adhesion of a crystalline thin film. As many substrates contains oxygen, which reacts readily with U metal, the addition of these (nominally) non-reactive buffers opens up a new region of phase space. It is fortunate that the refractory metals that prove good chemical buffers also demonstrate excellent epitaxial matches with U. As we will see below, however, the details of these epitaxial matches are not obvious in many cases. To date high quality epitaxial layers of, orthorhombic and hexagonal U metal have been achieved \cite{Ward2008,Springell2014,Springell2008c}, as well as pseudo-{\it bcc} U-Mo alloys \cite{Adamska2014a,Chaney2021}. As with the multilayer systems described above each system will be briefly explored here and further details for each system can be found in the references provided in Table \ref{tab:metals}.

Utilising this approach of high temperature growth and epitaxial refractory metal buffers the epitaxial growth of $\alpha\mbox{-U}$ was first achieved in 2008 via deposition onto thin, single crystal buffer layers of either Nb(110) or W(110) at 600 $^\circ$C \cite{Ward2008}. The optimal growth temperature was later refined to 450 $^\circ$C \cite{Springell2014}. In these cases, the substrates were single crystals of sapphire, Al$_2$O$_3$, epi-polished parallel to the (11.0) plane. This excellent, if non-intuitive, epitaxial match had already been identified \cite{Ward2003}. The resulting films were capped with a thin layer of one of the two refractory metals to protect the U layer from atmospheric degradation. It was found that Nb serves as a better capping layer as the passivating oxide, Nb$_2$O$_5$, is limited to approximately  20-30 \AA$ $ and can be stable for many years. Despite the use of similar $bcc$ buffer layers ($a_\textrm{Nb}=3.300$ {\AA} and $a_\textrm{W}=3.165$ {\AA}), there are significant differences in the orientations, domain structures and strains induced in the U overlayers for these two buffer materials.

\begin{figure}[t]
     \centering
     \subfloat[][]{\includegraphics[width=0.51\textwidth]{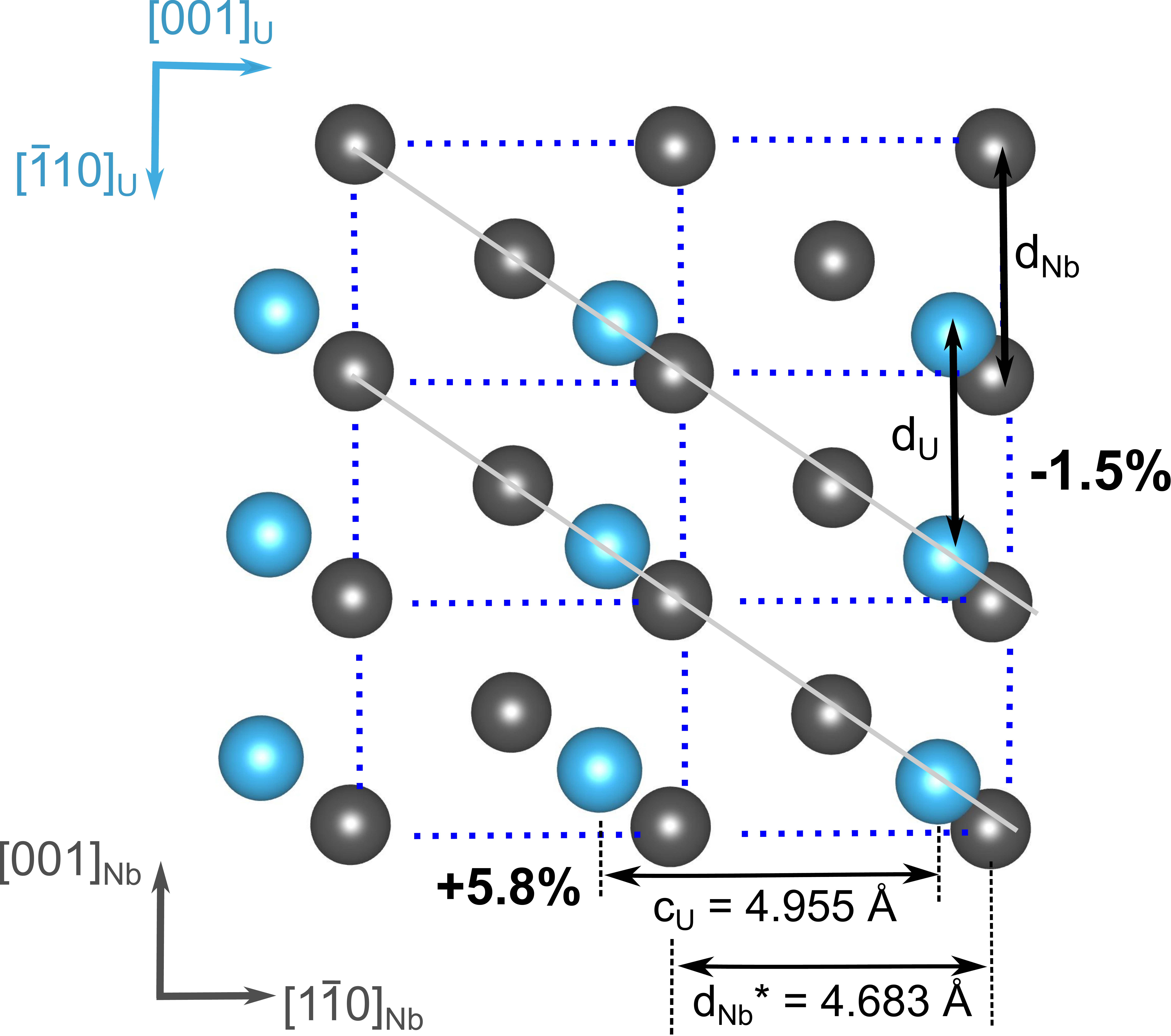}\label{fig2_3a}}
     \subfloat[][]{\includegraphics[width=0.49\textwidth]{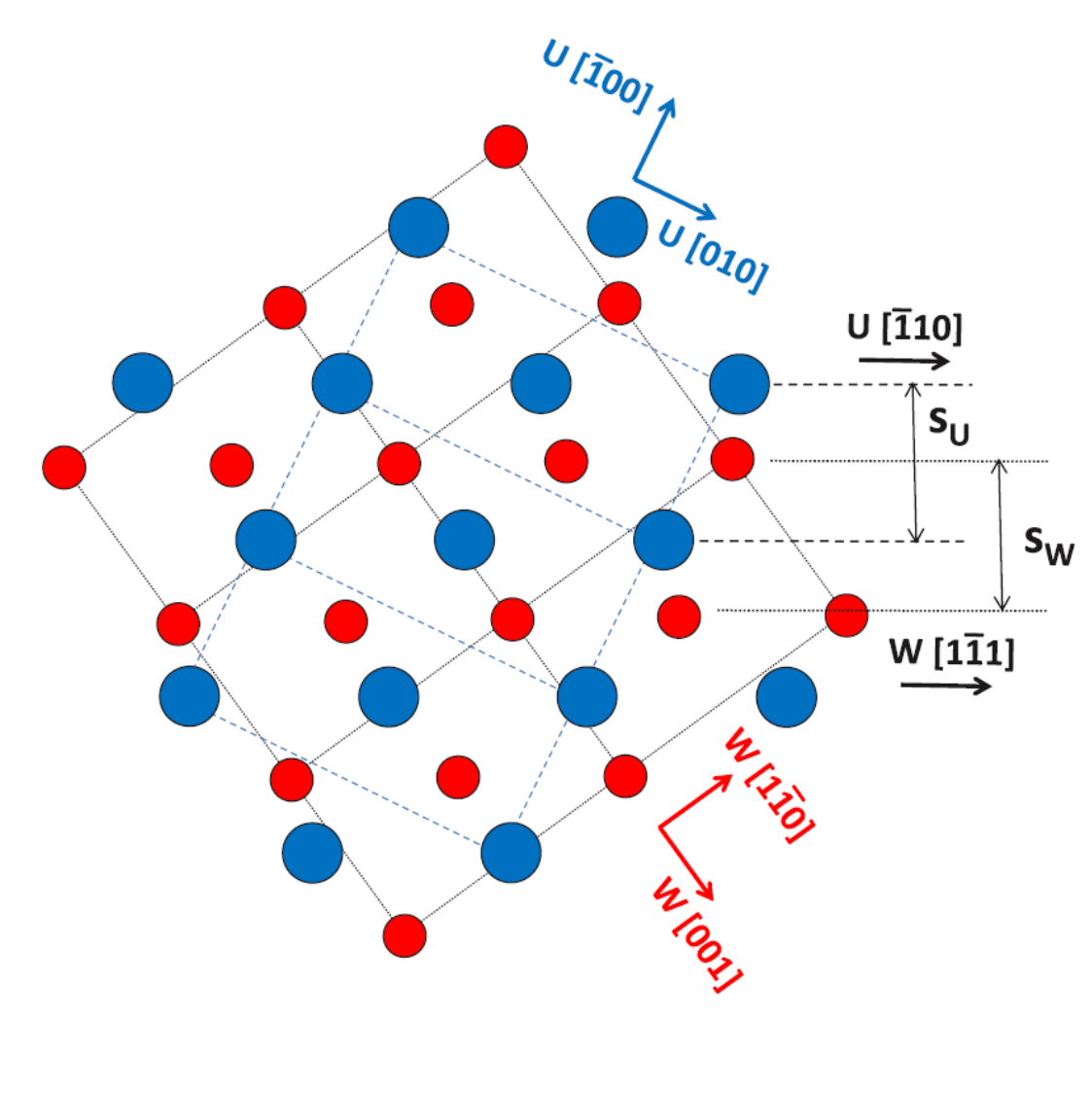}\label{fig2_3b}}
     \caption{Two-dimensional projection illustrating the growth of the $\alpha$-U(110) epitaxial film  on (a) Nb(110) (a=3.300 \AA{}) with Nb atoms shown in dark grey and uranium atoms in light blue. The close packing direction of uranium is shown by light grey, solid lines, dashed lines are guides for the eyes and (b) W(110) (a = 3.165 \AA{}) where U atoms are shown in blue and tungsten in red. The single-crystal sapphire ($\alpha$-Al$_2$O$_3$) substrate atoms are shown in either projection. Adapted from Ref.~\cite{Ward2008}}
     \label{U_epitaxy}
\end{figure}

Firstly, for the Nb(110)/$\alpha\mbox{-U}(110)$ epitaxial match the overlayer grows with the orientation relationship illustrated in Fig.\ref{fig2_3a}. The Nb buffer has a growth axis [110], and one of the in-plane [1-11] axes is aligned parallel to the sapphire [00.1] \cite{Ward2003}. The deposited U atoms self organise in order to align the close-packed rows of each layer, indicated by grey diagonal lines in Fig.~\ref{fig2_3a}. The epitaxy is driven by the close match between the distances $d_\textrm{Nb}= a_\textrm{Nb} = 3.311$ {\AA} and $d_\textrm{U}=\frac{1}{2}(a_\textrm{U}^2+b_\textrm{U}^2)^\frac{1}{2}=3.264$ \AA{} thus circumventing conventional wisdom that two systems with almost 6\% maximum lattice mismatch should never form an epitaxial system. As the system cools back to room temperature from the elevated growth temperature, the in-plane $c$-axis is locked into a state of tensile strain due to the positive thermal expansion coefficient of $\alpha$-U along the $[001]$-axis ($+30\times10^{-6}$ K$^{-1}$). It is expected that the $a$-axis, which is closest to the surface normal, should then contract slightly in order to maintain the unit cell volume. The low mosaic spread of both the Nb and U layers ($\Delta\omega=0.15^{\circ}$) in Ref.~\cite{Ward2008} indicated high quality epitaxial matches between the substrate, buffer and U overlayer.

Secondly, the W/U epitaxial system results in layers of $\alpha$-U(001) with multiple domains, and an $a$-axis that is held in a state of tensile strain \cite{Ward2008,Springell2014}. The complexity arises partly because the epitaxy of Al$_2$O$_3$/W(110) results in the formation of two domains that are related via a 70.5$^\circ$ in-plane rotation about the W[110] axis. The epitaxial match for one of eight predicted U domains \cite{Ward2008} is shown in Fig. \ref{fig2_3b}, such that the distances $d_U=d_{110}=2.556\,\mbox{\AA{}}$  and $d_W=2d_{211}=2.584\,\mbox{\AA{}}$ are aligned producing rows of U and W atoms in register. The eight possible domains arise since the condition $\mathrm{U}[110]\parallel\mathrm{W}(1\bar{1}1)$ can be satisfied in eight equivalent ways. However, the associated experimental data reported by Ward \textit{et al.} \cite{Ward2008} was inconclusive and showed only four domains, further reduced to two strong and two weak domains by the underlying sapphire orientation. It is noteworthy that these domains form a pseudo-hexagonal symmetry with angles between domains of either 52 or 63$^{\circ}$. These unusual angles arise because of the orthorhombic symmetry of U. However, this pseudo-hexagonal symmetry could have played a role in the earlier experiments reporting {\it hcp}-U \cite{Molodtsov1998,Molodtsov2001,Berbil-Bautista2004,Chen2019} as X-ray characterisation of the films was not performed.

\subsubsection{The hunt for hexagonal close packed uranium} \label{s:hcpgrowth}

\begin{figure}[t]
\centering
\includegraphics[width=\linewidth]{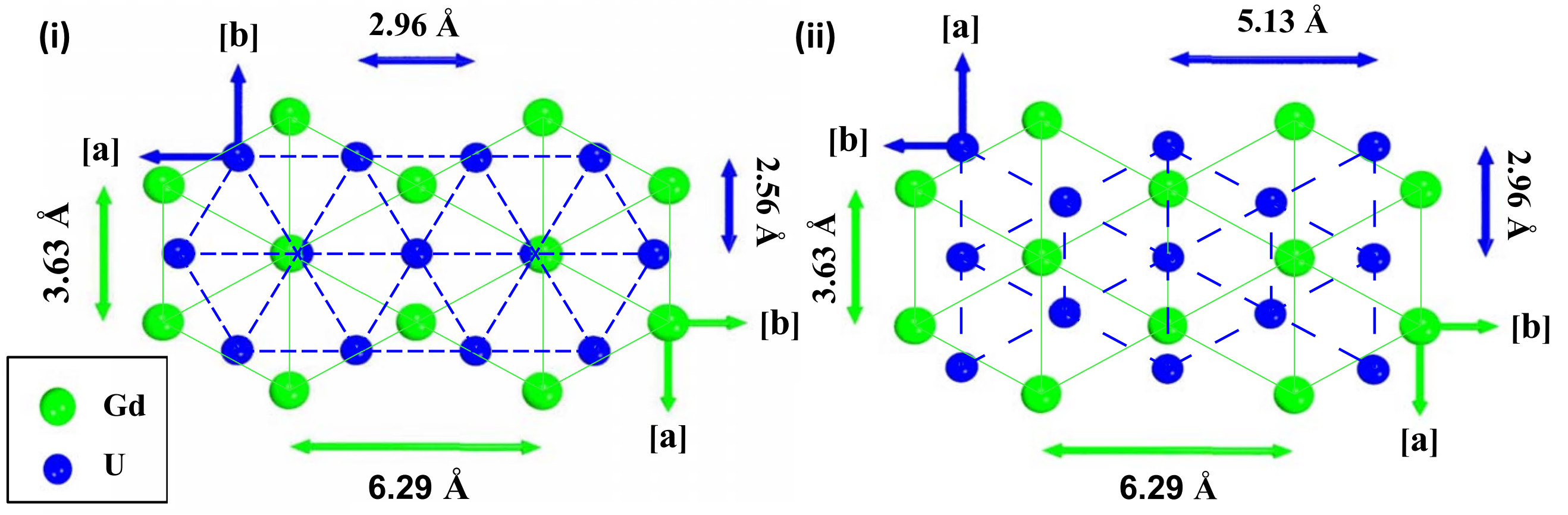}
\caption{Possible epitaxial matches of
of hcp-U on Gd. (i) Domain 1, which has
$[a]_{\mathrm{U}}$ parallel to $[a]_{\mathrm{Gd}}$ and a repeat
motif every nine lattice spacings
along $[a]_{\mathrm{Gd}}$. (ii) Domain 2 is oriented at 30 to domain 1 and has
$[b]_{\mathrm{U}}$ parallel to $[a]_{\mathrm{Gd}}$ and a repeat
motif every seven lattice spacings
along $[a]_{\mathrm{Gd}}$. Reprinted from \cite{Springell2008c}.} \label{hex_match}
\end{figure}

As explained above, and shown in Fig. \ref{fig2_2a} the phase diagram of U does not contain the hexagonal close packed ({\it hcp}) form, despite it being a common structure in elemental metals. However, as was first pointed out by Axe \cite{Axe1994}, the single soft phonon mode that relates the {\it bcc} and orthorhombic $\alpha$-forms passes first through a hypothetical {\it hcp} structure. As such one could imagine that the {\it hcp} structure could be stabilised in thin film form if an appropriate epitaxial match and growth conditions could be identified. On cooling these materials might well develop magnetic ordering, super-conductivity and/or CDWs; indeed, some of these phenomena have already been predicted \cite{Schonecker2012}.

There are reports that U may have been stabilized in such a \textit{hcp} phase \cite{Molodtsov1998, Molodtsov2001}. The first paper in 1998 \cite{Molodtsov1998} describes the fabrication of the films and their characterization with low-energy electron diffraction (LEED). After evaporation of the U metal onto a W(110) substrate, the films were annealed at $1127\,^{\circ}$C, and the subsequent LEED pattern suggested a close-packed structure of with an interatomic spacing of 3.2(1) \AA{}. They then state that: ``The LEED data, however, do not allow us to decide definitively whether the close-packed pattern relates to a cubic structure like $fcc\mbox{-Th/Ce}$, or to a hexagonal arrangement, as for most rare-earth metals.'' Because the films were fabricated in a closed system, and strongly oxidise if exposed to air, they were not able to perform an X-ray analysis on these films. This would, in any case, be difficult, as the films were only $\sim$ 80 \AA{} thick, which is close to the limit for analysis with a laboratory-based X-ray diffractometer. Subsequent resonant photoemission experiment were performed on these samples to show that the material had considerable 5$f$ weight at the Fermi level \cite{Molodtsov2001}, and scanning tunnelling spectroscopy \cite{Berbil-Bautista2004}, again showing the predominant feature of the 5$f$ states near the Fermi energy. In Ref. \cite{Berbil-Bautista2004} the authors report lattice parameters of $\bm{a}=3.5(3)\,\mbox{\AA{}}$  and $\bm{c}= 5.4(2)\,\mbox{\AA{}}$ , giving a $c/a \sim 1.61(8)$ , which is not far from the close-packed ideal value of 1.633.  More recently a 70 \AA{} film was also grown by Chen \textit{et al.} \cite{Chen2019} who reported (from LEED) a close-packed structure with a U-U spacing of 3.15(10) \AA{}. The $\bm{c}\mbox{-axis}$ lattice parameter was not stated, but using angular-resolved photoemission they reported evidence for the 5$f$ states hybridising with the conduction-electron states, as well as $f–f$ hybridisation. Again, a difficulty in this work is the absence of comprehensive X-ray structural characterisation.

\begin{figure}[t]
\centering
\includegraphics[width=\linewidth]{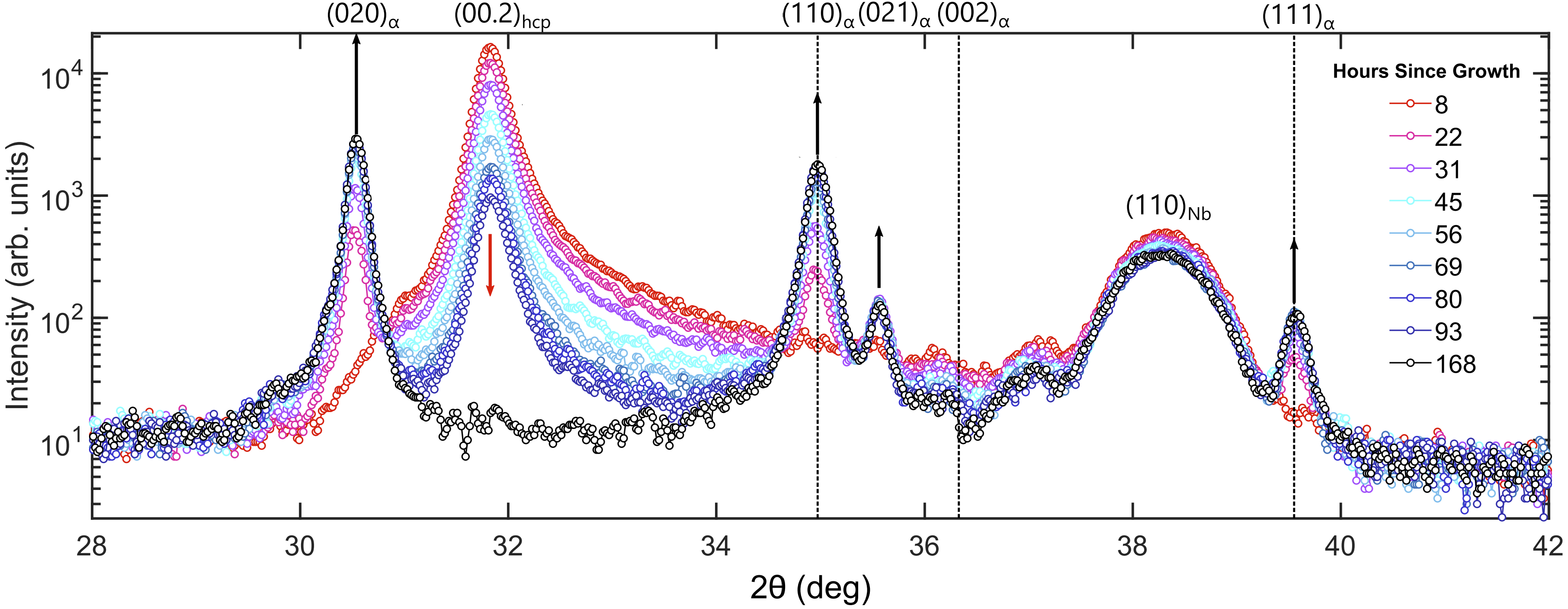}
\caption{Specular scans of a $1400\;\mbox{\AA}$ hexagonal-close-packed U film grown onto a polycrystalline $\mbox{Cu}{111}$ surface over the first seven days following growth. Arrows indicate the direction of growth/decay of the Bragg peaks. The dashed line represents the position of bulk $(002)_{\alpha}$. These data have not been corrected for absorption of the x-ray beam. Reprinted from \cite{Nicholls2022}.} \label{Nicholls2022}
\end{figure}

Regarding capped systems - that can be removed from UHV conditions for X-ray characterisation - as discussed in Section \ref{s:ML}, it was reported that the U layers (which were usually $<100\,$ \AA{} in thickness) in U/Gd multilayers formed in a hexagonal structure with $c=5.60(1)\,$ \AA{}. If an atomic volume similar to the $\alpha$-U phase is assumed, then this gives $a=2.91(1)\,$ \AA{} and a ratio $c/a$=1.92, a very large value. As reported in \cite{Springell2008c}, an attempt was made extend this result and form an epitaxial {\it hcp} layer by depositing a $500\,$ \AA{} U film onto a $500\,$ \AA{} epitaxial buffer of Gd grown on Nb. The $c$-axis was found to be $5.625(5)\,$ \AA{}, and using the off-specular family of (10.4) reflections $a$ was to found to be $2.96(2)\,$ \AA{}, giving an atomic volume close to the $\alpha$ phase and $c/a= 1.90(2)$, consistent with the value found in the multilayers. Some diagrams of possible epitaxy are shown in Fig.~\ref{hex_match}, but these were not experimentally confirmed and attempts to increase the U layer thickness resulted in exfoliation. Furthermore, in the successful, $500\,\mbox{\AA{}}$ thick system, the \textit{hcp}-U layer displayed poor mosaicity of 1.5$^{\circ}$, compared to $< 0.2^{\circ}$ for the $\alpha$-U films on Nb. Interestingly, theory \cite{Springell2008c}, has predicted a value for $c/a = 1.84$ with a similar atomic volume as $\alpha$-U. A more recent theory \cite{Schonecker2012}, has found $a = 3.01\,$ \AA{}, and a $c/a = 1.82$. When these authors fixed their $a = 2.96\,$ \AA{} to agree with the experiment \cite{Springell2008c}, they calculated $c/a=1.864$. Thus, the large $c/a$ ratio for the {\it hcp} phase seems to be a strong feature of both experiment and theory.

Recently an extensive study of preparing {\it hcp-}U has been reported by Nicholls {\it et al.} \cite{Nicholls2022} to try and understand whether this form of U can be made stable and thick enough to investigate other material properties e. g. electronic transport, or lattice dynamics. In addition to using the substrates discussed above, an effort was also made on Cu(111) and Ir(111) faces, as these {\it fcc} materials present a hexagonal face in this orientation that has lattice parameters close to those discussed above for {\it hcp} U. Fig.~\ref{Nicholls2022} shows that the initial phase deposited was indeed  {\it hcp}, but that a short time after growth ($\sim20$ minutes for a film of $1400\,\mbox{\AA{}}$) there is a rapid decomposition of the  {\it hcp} phase, which transitions to the stable orthorhombic structure with the the main grain orientations (020) and (110) \cite{Nicholls2022} . Thinner samples of {\it hcp} films are stable for somewhat longer times, but in all cases observe phase decomposition of the \textit{hcp} phase was observed as a function of time. It was also observed in the same study that the specific decay path {\it hcp} to orthorhombic U varies depending on film thickness.

\subsubsection{Achieving single crystal films of $\gamma\mbox{-phase}$ alloys} \label{s:sxgamma}

\begin{figure}[t]
\centering
\includegraphics[width=0.7\linewidth]{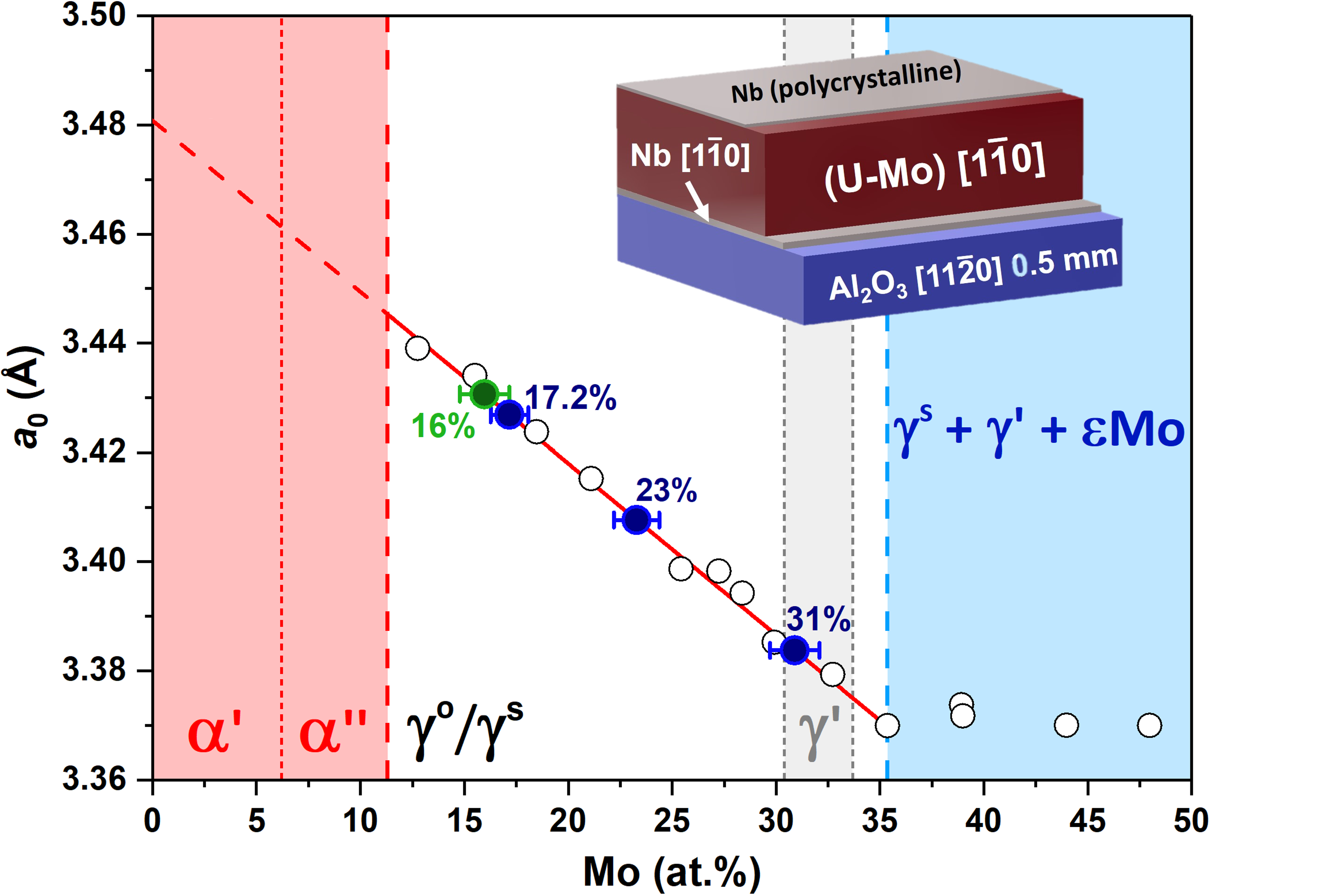}
\caption{Nominal {\it bcc} lattice parameter (measured at room temperature) as a function of Mo content as determined by lattice parameter and calculated using the empirical fit shown in red, data shown as open circles \cite{Dwight1960}. $\alpha$, $\gamma$ and phase separation fields are highlighted in red, white and blue, respectively. The first two contain all metastable phases with the indicated parent structure. The latter contains metastable $\gamma$ along with the $\gamma '$ phase, the tetragonal compound U$_2$Mo. For detailed discussion of the various metastable structures the reader is referred to Ref. \cite{Chaneythesis}. 3000 \AA$ $ and 700 \AA$ $ films are shown as blue and green solid points, respectively. (Insert) Schematic of film structure. Adapted from from Ref. \cite{Chaney2021}  }
\label{fig4_2_1}
\end{figure}

It has been known since the mid 1950's that $\alpha\mbox{-U}$ makes for a poor nuclear fuel as, due to its orthorhombic structure, it displays both anisotropic thermal expansion and poor dimensional stability under irradiation \cite{Lander1994,Roy1985}. As such, the high temperature, $bcc$, $\gamma\mbox{-phase}$ was highlighted early on as an enticing prospect for increased U density for low-enriched fuel solutions \cite{Berghe2014}. Long lived $\gamma\mbox{-like}$ metastable samples have been obtained through the alloying of various transition metals combined with rapid cooling techniques since the 1960's, up to the present day \cite{Tangri1961, Yakel1969, Tkach2012}. However, a direct result of such fabrication methods is the inability to produce single crystal samples thus severely limiting the experimental probes that can be brought to bear when attempted to understand the properties of this allotrope. Adamska \textit{et al.} \cite{Adamska2014} were the first to demonstrate that epitaxial matching could provide a mechanism to avoid rapid cooling and instead lock in the metastable $\gamma\mbox{-like}$ phase using epitaxial strain. This procedure makes use of the $\mathrm{Nb}/\mathrm{Al}_2\mathrm{O}_3$ match detailed in section \ref{s:sxalpha} to provide a $[110]$ surface onto which to co-deposit U and Mo at $800^{\circ}\mathrm{C}$, inside the stable $\gamma$ region, and the similarity in $\bm{a}_{\mathrm{Nb}} = 3.33\,$ \AA{} and $\bm{a}_{\gamma\mbox{\tiny-UMo}} \approx 3.41\,$ \AA{} allows a simple ``cube-on-cube'' epitaxial match that produces sufficient strain to preserve the $\gamma^{\mathrm{s}}$ phase upon cooling to room temperature. This work was built on further by Chaney \textit{et al.} \cite{Chaney2021} and refined to span the large majority of the $\gamma\mbox{-like}$ region in the metastable UMo phase diagram as shown in Fig.~\ref{fig4_2_1}. Mo was chosen as proof of concept as it provides the strongest stabilising effect of all the transition metals and is also the most likely future fuel candidate, however this growth approach can reasonably be extended to any $\gamma\mbox{-like}$ U-transition metal alloy system with high probability of success.

\begin{landscape}
\begin{table}[]
\centering
\caption{U allotrope, alloy, and multilayer thin films that have been produced, with references of their first mention in publication. *This structure was not confirmed with structural characterisation.}
\label{tab:metals}
\begin{tabular}{|lllll|} \hline
\multicolumn{1}{|l|}{Material} & \multicolumn{1}{l|}{Form}           & \multicolumn{1}{l|}{Substrate}                               & \multicolumn{1}{l|}{Deposition method}       & Reference       \\ \hline
\rowcolor[HTML]{EFEFEF}
Allotropes                     &                                     &                                                              &                                              &                 \\
\multicolumn{1}{|l|}{hcp* U}   & \multicolumn{1}{l|}{\hkl(001)?}     & \multicolumn{1}{l|}{\hkl(110) W}                                   & \multicolumn{1}{l|}{vapour deposition}       & \cite{Molodtsov1998}   \\
\multicolumn{1}{|l|}{$\alpha$ U}  & \multicolumn{1}{l|}{\hkl(110)}   & \multicolumn{1}{l|}{\hkl(110) Al$_2$O$_3$ with \hkl(110) Nb buffer}            & \multicolumn{1}{l|}{DC mag. sputtering} & \cite{Ward2008}        \\
\multicolumn{1}{|l|}{$\alpha$ U}  & \multicolumn{1}{l|}{\hkl(001)}   & \multicolumn{1}{l|}{\hkl(110) Al$_2$O$_3$ with \hkl(110) W buffer}             & \multicolumn{1}{l|}{DC mag. sputtering} & \cite{Ward2008}        \\
\multicolumn{1}{|l|}{$\alpha$ U}  & \multicolumn{1}{l|}{\hkl(021)/\hkl(110)  }     & \multicolumn{1}{l|}{\hkl(110) Al$_2$O$_3$ with \hkl(110) Nb and \hkl(001) Gd buffer} & \multicolumn{1}{l|}{DC mag. sputtering} & \cite{Ward2008}        \\
\multicolumn{1}{|l|}{hcp U}   & \multicolumn{1}{l|}{\hkl(001)}       & \multicolumn{1}{l|}{\hkl(110) Al$_2$O$_3$ with \hkl(110) Nb and \hkl(001) Gd buffer} & \multicolumn{1}{l|}{DC mag. sputtering} & \cite{Springell2008}   \\
\multicolumn{1}{|l|}{$\alpha$ U}  & \multicolumn{1}{l|}{\hkl(001)}   & \multicolumn{1}{l|}{\hkl(110) Al$_2$O$_3$ with \hkl(110) Nb and \hkl(110) W buffer} & \multicolumn{1}{l|}{DC mag. sputtering} & \cite{Springell2014}        \\
\rowcolor[HTML]{EFEFEF}
Alloys                         &                                &                                                              &                                              &                 \\
\multicolumn{1}{|l|}{U-Mo}     & \multicolumn{1}{l|}{poly}      & \multicolumn{1}{l|}{glass}                                   & \multicolumn{1}{l|}{DC mag. sputtering} & \cite{Adamska2014a}     \\
\multicolumn{1}{|l|}{U-Mo}     & \multicolumn{1}{l|}{\hkl(110)} & \multicolumn{1}{l|}{\hkl(110) Al$_2$O$_3$ with \hkl(110) Nb buffer}            & \multicolumn{1}{l|}{DC mag. sputtering} & \cite{Adamska2014a}     \\
\multicolumn{1}{|l|}{U-Zr}     & \multicolumn{1}{l|}{poly}      & \multicolumn{1}{l|}{glass}                                   & \multicolumn{1}{l|}{DC mag. sputtering} & \cite{Adamska2014}    \\
\multicolumn{1}{|l|}{U-Zr}     & \multicolumn{1}{l|}{\hkl(110)} & \multicolumn{1}{l|}{\hkl(110) Al$_2$O$_3$ with \hkl(110) Nb buffer}            & \multicolumn{1}{l|}{DC mag. sputtering} & \cite{Adamska2014}    \\
\rowcolor[HTML]{EFEFEF}
Multilayers \& Bilayers                    &                                &                                                              &                                              &                 \\
\multicolumn{1}{|l|}{UAs/Co}   & \multicolumn{1}{l|}{amorphous} & \multicolumn{1}{l|}{glass}                                   & \multicolumn{1}{l|}{mag. co-sputtering} & \cite{Fumagalli1993} \\
\multicolumn{1}{|l|}{U/Fe}     & \multicolumn{1}{l|}{poly}      & \multicolumn{1}{l|}{polyimide (Kapton) and glass}            & \multicolumn{1}{l|}{DC mag. sputtering} & \cite{Beesley2004}     \\
\multicolumn{1}{|l|}{U/Fe}     & \multicolumn{1}{l|}{poly}      & \multicolumn{1}{l|}{\hkl(110) Al$_2$O$_3$ with \hkl(110) Nb buffer}            & \multicolumn{1}{l|}{DC mag. sputtering} & \cite{Springell2008}   \\
\multicolumn{1}{|l|}{U/Gd}     & \multicolumn{1}{l|}{poly}      & \multicolumn{1}{l|}{\hkl(110) Al$_2$O$_3$ with \hkl(110) Nb buffer}            & \multicolumn{1}{l|}{DC mag. sputtering} & \cite{Springell2008}   \\
\multicolumn{1}{|l|}{U/Co}     & \multicolumn{1}{l|}{poly}      & \multicolumn{1}{l|}{\hkl(110) Al$_2$O$_3$ with \hkl(110) Nb buffer}            & \multicolumn{1}{l|}{DC mag. sputtering} & \cite{Springell2008}   \\
\multicolumn{1}{|l|}{U/Fe}     & \multicolumn{1}{l|}{poly}      & \multicolumn{1}{l|}{glass}                                   & \multicolumn{1}{l|}{DC mag. sputtering} & \cite{Gilroy2021}      \\
\multicolumn{1}{|l|}{U/Ni}     & \multicolumn{1}{l|}{poly}      & \multicolumn{1}{l|}{glass}                                   & \multicolumn{1}{l|}{DC mag. sputtering} & \cite{Gilroy2021}     \\ \hline
\end{tabular}
\end{table}
\end{landscape}

\subsection{Science with uranium metal thin films} \label{s:metalsscience}

This Section discusses: the early attempts to make multilayers containing one layer of pure U metal; the important breakthrough to achieve epitaxial $\alpha$-U metal; the discovery that interfacial strain changes the properties of the $\alpha$-U film; experiments attempting to stabilize {\it hcp}-U films; and the first efforts to make U/ferromagnet heterostructures for spintronics experiments. We end with a short description of experiments on thin films (not epitaxial) of plutonium and important photoemission experiments showing the delocalisation of the Pu 5$f$ electrons as a function of film thickness.

\subsubsection{Induced magnetism in  uranium-ferromagnetic multilayers} \label{s:U_MLs}

The fundamental question in multilayer samples which combine ferromagnetic and non-magnetic materials is whether the latter has any magnetic moment induced on it due to the proximity of the former. Early work by Beesley \textit{et al.} on U/Fe  multilayers \cite{Beesley2004,Beesley2004a} employed M\"ossbauer spectroscopy and polarized-neutron reflectivity to examine the magnetic moments at the Fe atomic site, and determined the easy axis of magnetisation was in the plane of the layers. Later, the element specific synchrotron-based techniques of X-ray magnetic circular dichroism (XMCD) \cite{Wilhelm2007} and X-ray resonant reflectivity (XRRR) \cite{Brown2008} were developed allowing information on the magnitude and profile of the much smaller induced moment at the U atoms sites. Notably it was found that the U moments are indeed induced, but are only significant close to the interface for multilayer systems containing transition metal atoms. As mentioned in Section \ref{s:ML} the interface sharpness was relatively poor for multilayers containing 3$d$ transition metals with approximately 15 \AA{} of interdiffusion between the layers. Furthermore, M{\"o}ssbauer experiments \cite{Thomas2008} suggest that, at least in the case of Fe, the structure of the Fe in this region was amorphous, and it is in this region that the largest induced U moments are found. This sharp decrease in the U moment as a function of distance away from the interface has been supported by theory \cite{Komelj2005}. The magnetic moments of the transition metals were slightly smaller than the bulk values, and the easy direction of magnetisation was in the plane of the films \cite{Springell2008a}.

The work with the transition-metal ferromagnets was summarized in Ref. \cite{Springell2008b} in 2008. It was noted that that the maximum U moment in the U/Fe system was $\sim$ 0.12 $\mu_{\mathrm{B}}$ and reduced for both the U/Co and U/Ni systems. This behaviour is ascribed to the position of the 3$d$ band relative to the Fermi level ($E_{\mathrm{F}}$).  Given that the 5$f$ band is centered close to the Fermi level, $E_{\mathrm{F}}$, a maximum in $3d\mbox{-}5f$ band overlap, and thus maximised hybridization, occurs when the transition metal $3d$ band also lies close to $E_{\mathrm{F}}$. This criteria is maximised for Fe and decreases for the higher elements, Co and Ni.

\begin{figure}[t]
     \centering
     \subfloat[][]{\includegraphics[width=0.5\textwidth]{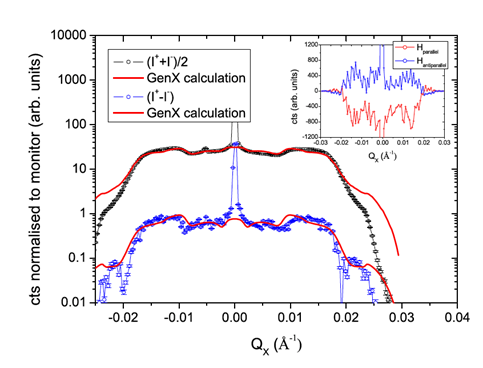}\label{fig2_7}}
     \subfloat[][]{\includegraphics[width=0.5\textwidth]{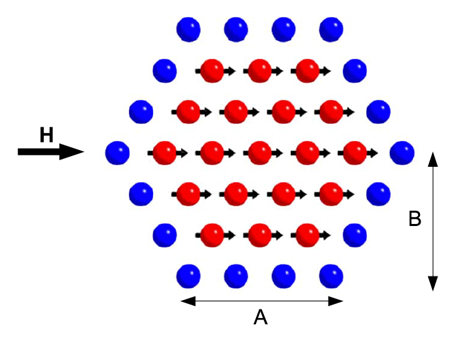}\label{fig2_8}}
     \caption{(a) Charge diffuse scattering (black) and the magnetic diffuse scattering (blue) at the Gd $M_5$ edge of a U/Gd multilayer of composition [U(50 \AA{})/Gd(50 \AA{})]$_{20}$. Data are shown as open symbols and lines are modeled fits. Inset shows the magnetic diffuse scattering from the two X-ray polarization states. (b) Schematic of columnar growth of Gd in the layers. The red atoms have the full moment of $\sim$ 7 $\mu_{\mathrm{B}}$ parallel to the applied field, whereas the atoms at the edge have their moments in some random direction that averages to zero over the whole layer. The reduction of the moment as a function of $D = (A + B)$ may be easily calculated. Both figures reprinted from Ref. \cite{Springell2010}.}
     \label{steady_state}
\end{figure}

In contrast, considering now the lanthanide/uranium multilayer system, U/Gd \cite{Springell2010}, there is no overlap in energy between that of the U 5$f$ band and that of the localised Gd 4$f$ electrons, since the latter are well removed from $E_{\mathrm{F}}$. Hence the induced U moment is much smaller, $\sim$ 0.02 $\mu_{\mathrm{B}}$ per atom, but it does oscillate through the U layer, as might be expected if this is driven by a RKKY-type interaction mediated via the conduction electrons. Notably the Gd magnetic moments are much reduced to $\sim$ 4 $\mu_{\mathrm{B}}$ from the elemental value of 7.5 $\mu_{\mathrm{B}}$ known for pure Gd.

This largeapparent reduction of the magnetic moment in U/Gd multilayers has been investigated using various techniques \cite{Springell2010}. One possible scenario to explain the reduced magnetisation is that the particular type of columnar growth observed by TEM, and discussed in Section \ref{s:ML}, leads to the pinning of Gd moments at the boundaries of the columns. To investigate this requires tools that probe the length scales laterally within the layers, rather than the vertical growth direction. By tuning polarized X-rays to the $M_5$ edge of Gd at 1187 eV it was possible to measure sufficiently far away from the specular ridge to learn something about length of both the chemical and magnetic correlations \cite{Springell2010}. When data are obtained with the X-ray circularly polarized in the two opposite directions the magnetic components change sign. Hence the charge diffuse scattering, which represents the chemical configuration, can then be obtained by adding the two contributions and dividing by two, whereas the magnetic contribution is obtained by subtracting the two contributions. The results are shown in Fig. \ref{fig2_7}, and it can be easily seen that the intensities fall abruptly at the same value of Q$_x$, the component of the moment transfer in the plane of the layers. This shows that the chemical and magnetic diffuse scattering define a similar length scale over which these correlations occur, which is not surprising given the proposed link to the columnar growth. A detailed analysis shows that this distance is $120 \pm 20$ \AA{}, whereas the interlayer distance is $\sim$ 50 \AA{}. This information then allows a simple model to be constructed based on the idea that the columnar growth resembles the model in Fig. \ref{fig2_8}, which explains the strong reduction of the Gd saturated moment, and confirms the columnar growth of Gd \cite{Springell2010}.

\subsubsection{Uranium metal bilayers for future spintronic applications}	\label{s:Ubilayers}

\begin{figure}[htb]
\centering
\includegraphics[width=0.5\textwidth]{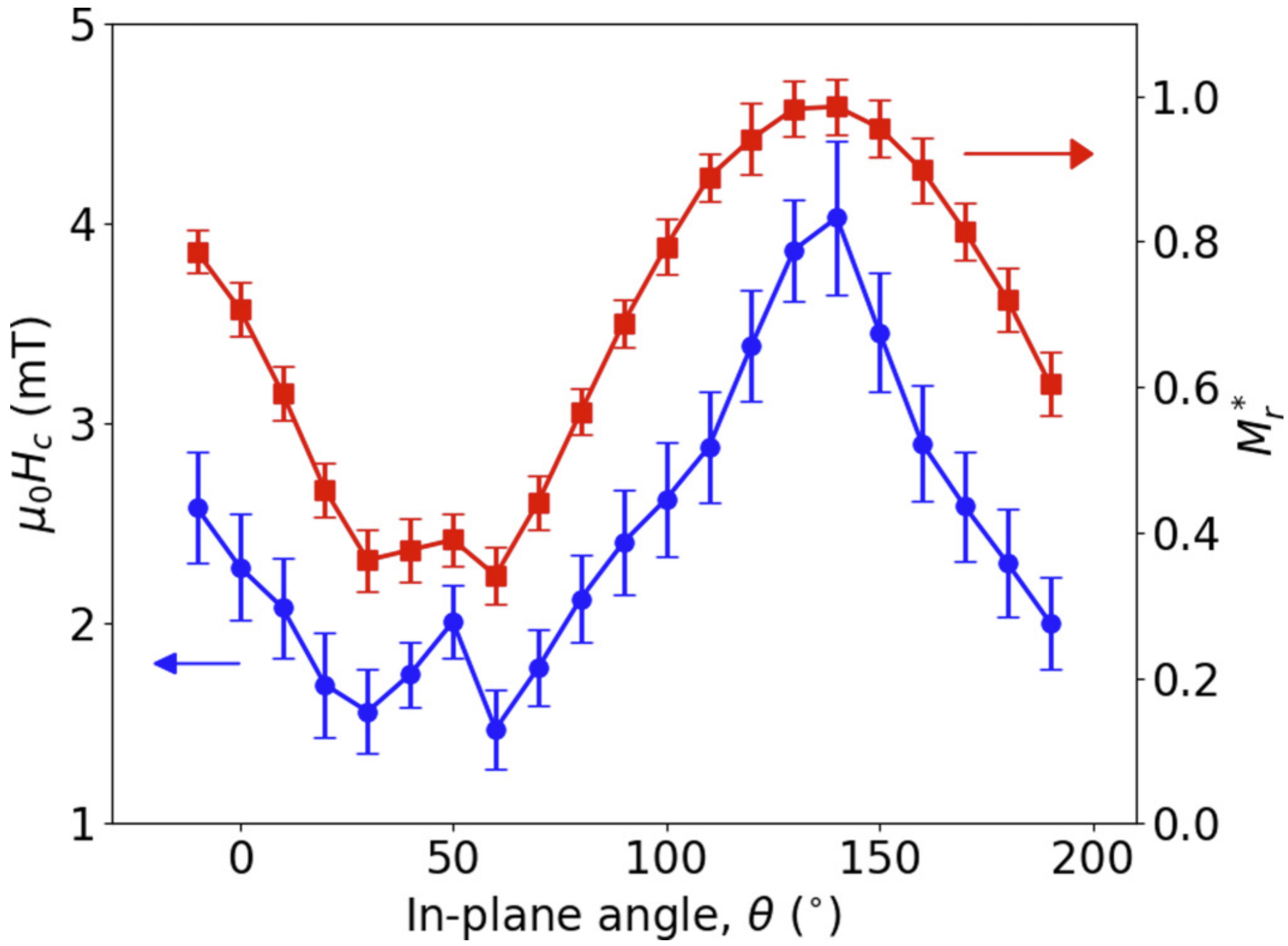}
\caption{In-plane angular dependence of the coercive field $\mu_{0}H$ and normalized remnant moment $M_{r}^{*}$ for $\mathrm{FM}=\mathrm{Fe}$ and $d_{\mbox
{U}} = 65\mbox{\AA}$. Lines are a guide to the eye. Taken from Ref.~\cite{Gilroy2021}.} \label{Gilroy4}
\end{figure}

In the field of spintronics, where information is carried by the spin magnetic moment of the electron rather than charge, one of the key parameters is the spin-orbit coupling parameter of the material, which can have a significant impact on both the generation of spin currents (via the spin Hall effect for example), and the spin lifetime of carriers in non-magnetic systems. Since the spin-orbit interaction increases in the periodic table as approximately $Z^4$, it is of interest to see whether U can be used in such devices, and be competitive with other materials. The first experiments with U \cite{Singh2015} were done on a bilayer consisting of a 125 \AA{} of permalloy $(\mbox{Ni}_{0.8}\mbox{Fe}_{0.2})$ ferromagnetic layer deposited on glass with a layer of 30 Å of uranium metal (non-magnetic layer) on top, capped with a 30 \AA{} film of Nb to prevent oxidation. The key parameter to determine is the spin Hall angle $\theta_{\mathrm{H}}$, where this is defined as the ratio of the injected spin current and resulting charge current. It is this quantity which should increase strongly with heavier elements. The method applied uses dynamical spin pumping \cite{Ando2011}. In spin pumping, the precessing magnetisation of an externally excited ferromagnet (FM) undergoing ferromagnetic resonance (FMR) is dynamically coupled to the charge carriers in an adjacent non-magnetic (NM) system, resulting in a net transfer of spin angular momentum across the ferromagnetic/non-magnetic interface. The measured value of $\theta_{\mathrm{H}}$ for U is positive and has a value of +0.004. This value is smaller than the value measured of +0.006 \cite{Pearson2010} for Pt, which is somewhat surprising. Since both the $5f$ and $6d$ shells of U are less than half filled, one would, to first approximation, expect a negative $\theta_{\mathrm{H}}$ for U, as is found for transition metals. In detail, however, the situation is more complex: the extrinsic and intrinsic contributions to the spin Hall angle need to be isolated, which depends on the presence of impurity scattering in the samples \cite{sagasta_prb_2016}. The crystal structure and resultant electronic band structure play an important role in the magnitude of the intrinsic contribution, and can be calculated in terms of the Berry phase curvature. Theoretical work by Wu \textit{et al.} examined the $\alpha$, $\gamma$, and {\it hcp} phases of U, and suggested that the {\it hcp} phase shows the largest spin Hall conductivity near to $E_{\mathrm{F}}$. In the same work they also showed that the nature of any 3$d$ transition metal ion impurities (which might be present at a disordered interface) can significantly affect the magnitude and even the sign of $\theta_{\mathrm{H}}$ \cite{Wu2020}. In more recent work the same authors show the peculiar result that spin accumulation in uranium films is highest at the side of the film opposite to the impurity position \cite{Wu2021}. Clearly, there is more to do on this subject to understand the value of U in spintronic applications.

In another study the anisotropic magnetic properties of thin U/Fe and U/Ni bilayers \cite{Gilroy2021} were investigated. Here the transition metal layers of $\sim$ 100 \AA{} were deposited on glass, and the U layer thicknesses ($d_U$) were varied from 0 to 80 \AA{}. The growth was carried out at room temperature, and the layers were polycrystalline, as expected. The U appeared to be mostly [001] textured as the preferred growth axis. Magnetisation measurements at room temperature were made of the coercive field as a function of angle in the plane, to investigate the impact of the heavy U on the magnetic anisotropy of the transition metal layer: since magnetocrystalline anisotropy is inherently linked to spin-orbit coupling, the proximity of enhanced spin-orbit effects in the U layer might be expected to have a significant impact. Some data are shown in Fig. \ref{Gilroy4}.

From these results the uniaxial anisotropy coefficient $K_{\mathrm{eff}}$ may be calculated and this is found to have a maximum in the U/Fe system at $d_U \sim 55$ \AA{}. Further examination suggests that quantum-well states in the U may be affecting the changes in $K_{\mathrm{eff}}$, suggesting some impact of the electronic structure of neighbouring U layer on the magnetic layer. Once again, the interface probably needs further investigation, especially since an induced moment on the U may be present, and as discussed in Section \ref{s:U_MLs}, an interdiffused  layer of at least 15 \AA{}, might be present.

\subsubsection{Manipulating the CDW States in $\alpha\mbox{-uranium}$ using epitaxial strain} \label{s:alphacdw}

\begin{figure}[tb]
\centering
\includegraphics[width=\textwidth]{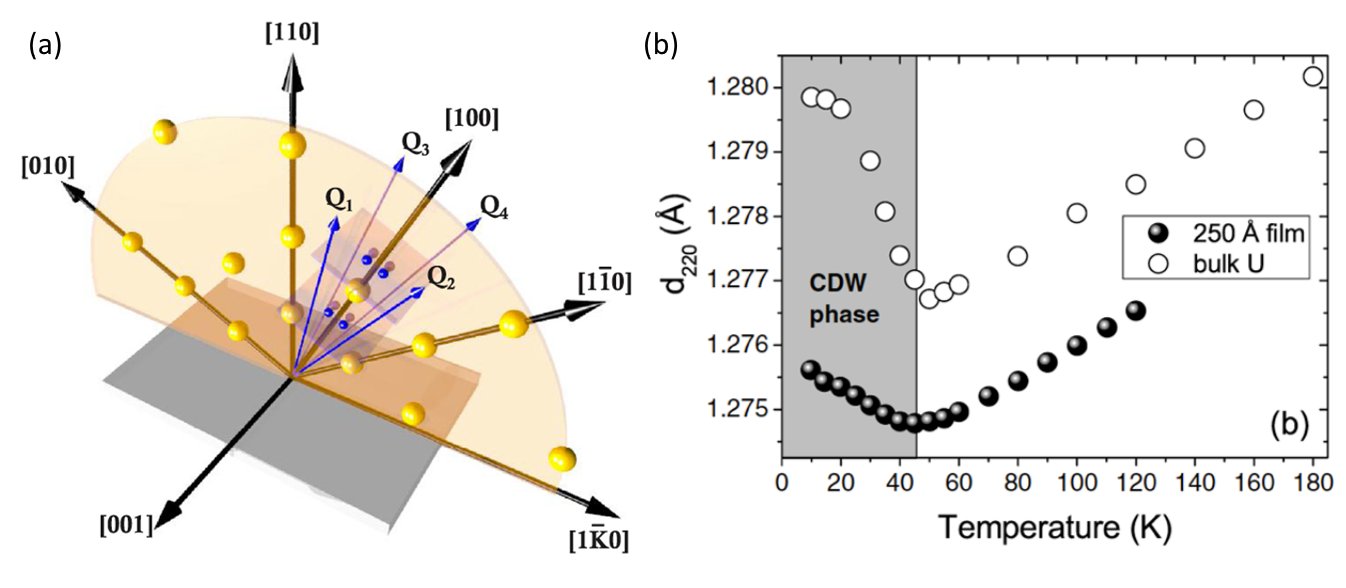}
\caption{(a) shows the U/Nb growth axis as [110] with the plane of the film in grey containing the [001] axis. In the bulk material \cite{Lander1994} the CDW is formed along the propagation vectors, {\bf Q$_1$}, {\bf Q$_2$}, {\bf Q$_3$}, and {\bf Q$_4$} so they all lie slightly away from the yellow plane containing [100] and [010]. (b) shows the $d$–space of the (220) (specular) charge peak as a function of temperature for both the bulk and from a 250 \AA{} epitaxial film. Taken from Ref. \cite{Springell2014}.} \label{fig2_9}
\end{figure}

 As explained in Section \ref{s:metalsintro}, significant research efforts were expended to understand the low temperature charge ordering in $\alpha\mbox{-U}$ through which it was established that there exists a series of CDW transitions starting at 43 K and ending at 22 K with a fully commensurate CDW state described by the wavevector $q=\frac{1}{2}\bm{a}^* + \frac{1}{6}\bm{b}^* +\frac{5}{27}\bm{c}^*$ \cite{Lander1994}. It was also determined that one of the key parameters determining CDW behaviour is the length of the $\bm{a}\mbox{-axis}$, as this is the direction along which the primary component of the CDW distortion lies \cite{Lander1994}. It is known from work on the phonons under pressure that on compression the soft-phonon mode in the [100] direction hardens (i.e. increases in energy) increasing the energy required for the CDW to form, suppressing the CDW state and enhancing $T_{\mathrm{c}}$ from $\sim0.5$ K at ambient pressure to a maximum of $\sim2$ K at 1.2 GPa \cite{Raymond2011}. In a na\"ive picture of competing CDW/SC ground states, tensile strain along the $a$-axis should therefore enhance the CDW stability and suppress superconductivity. Given this, the difference in both the magnitude and sign of the strain exerted on the $\alpha$-U $a$-axis in the epitaxial Nb/U and W/U systems described in Section\:\ref{s:sxalpha} opened the door to the potential to achieve something impossible in bulk, manipulating the CDW state via epitaxial strain.

\begin{figure}[t!]
\centering
\includegraphics[width=\linewidth]{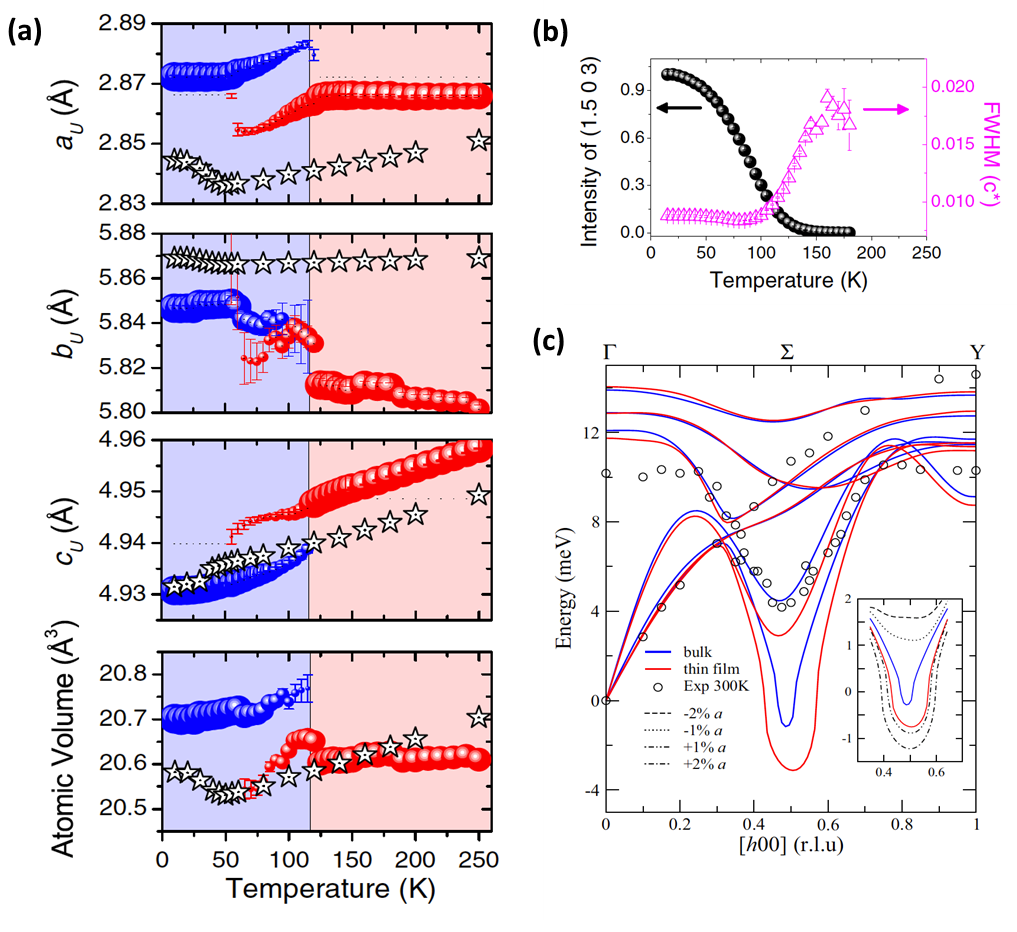}
\caption{(a) The lattice parameters of the strained uranium film as a function of temperature. The size of the symbols indicates approximately how much of the sample has that lattice parameter (or volume) at the given temperature. The bulk values are given by stars. (b) Temperature dependence of the intensity (filled circles) and full-width at half maximum (FWHM) (open triangles) for the (1.5 0 3). (c) Theoretical phonon dispersion for bulk vs thin film calculated at 0 K using  methods reported in ref. \cite{Bouchet2008}. Experimental points at 300 K \cite{Crummett1979} are shown as open circles for comparison. Insert shows the effect of compression (negative) or expansion (positive) of the $a$-axis on the unstable mode. All panels taken from Ref. \cite{Springell2014}.} \label{fig2_comb}
\end{figure}

The initial exploratory studies in this area were conducted in 2008 \cite{Springell2008c} and found that the two systems do indeed display vastly different CDW states. Starting with the Nb/U films system, the CDW reflections were found in the bulk positions \cite{Springell2008c} but with a dramatic change in the relative reflection intensities compared to bulk where all satellites are equal in intensity \cite{Lander1994}. Fig.~\ref{fig2_9}b gives the clue as to why the CDW is so close to that found in the bulk, but is in fact ``weaker'', as measured by the smaller change of the $d_{220}$ in the film compared to the bulk.  The $d_{220}$ spacing, which includes a component from the $\bm{a}$-axis, is smaller for the film than for the bulk, i.e. $\bm{a}$-axis is under compression in the film. This implies a weaker CDW interaction, hence less intensity in the CDW peaks, and less change of lattice parameter on cooling below $T_{\mathrm{CDW}}$ although the latter stays at about the same value.  Since the film was grown on a Nb buffer, which is superconducting at $\sim 9\,\mathrm{K}$, no effort was made to see whether the film was superconducting, since a simple resistive measurement would not detect the lower $T_{\mathrm{c}}$. Regarding the change in relative reflection intensities; for each layer of satellites, $h\pm\frac{1}{2}$, the CDW reflections are defined by the vectors Q$_1$ to Q$_4$, shown in Fig.~\ref{fig2_9}a, and it was found that for $h+$ the satellites could be divided into two clear pairs of comparable intensity, a strong pair {\bf Q}$_1$/{\bf Q}$_3$ and a weak pair {\bf Q}$_2$/{\bf Q}$_4$. The matching intensity within a pair is expected as the only difference is $\pm k$ which is a true mirror plane, however to explain the intensity variation between pairs the authors note that the stronger pair is closely aligned with the growth axis [110]. It can thus be inferred that a major effect of epitaxial strain is to promote or suppress the formation of CDW domains depending on their relative alignment with the growth axis. The variation of the CDW wavevector was also measured as a function of temperature and it was found that, unlike bulk, there were no lock-in transitions of the individual components, $\bm{q}_x$, $\bm{q}_y$, and $\bm{q}_z$ \cite{Springell2008c}.

The situation in the U/W system is quite different to that in the U/Nb system, and the complex domain structure discussed in Section \ref{s:sxalpha} needs to be suppressed in order to simplify reciprocal space for measurements. By depositing a thin (100 \AA{}) layer of niobium between the substrate and tungsten layer, the W(110) buffer is limited to a single domain and the total number of observable $\alpha$-U domains is reduced from eight to four. The domains are unequally populated and only two strong reflections (separated by an in-plane rotation of $\sim56^\circ$) tend to be observed \cite{Springell2014}.  As shown in Fig. \ref{fig2_comb}a, at 250 K the atomic volumes are comparable, but the $a$ and $c$ axes have expanded, and the $\bm{b}$ is contracted. The CDW occurs at $\sim$ 120 K (three times that of the U/Nb film and the bulk) and results in a large further expansion of the $\bm{a}$-axis; at the same time the $\bm{c}$-axis (which is not constrained, as this is the growth direction) contracts to allow the atomic volume to approach that of the bulk sample. Further to the dramatic increase in $T_{\mathrm{CDW}}$, the most remarkable result of the tensile epitaxial strain is to fundamentally change the CDW character, such that it becomes fully commensurate with the lattice, and changes from three- to one-dimensional losing both $\bm{q}_y$ and $\bm{q}_z$ components retaining only the $\bm{q}_x = 0.50$ component. Fig. \ref{fig2_comb}b shows the development of this peak, and its width, as a function of temperature.

This work showed conclusively that the expansion of the $\bm{a}$-axis is the key to the marked increase in $T_{\mathrm{CDW}}$. In the U/W films the $a$-axis is under tension, allowing its expansion. It is also confirms the theory developed first in Ref. \cite{Raymond2011} that the soft-mode phonon behaviour, and the associated strong electron-phonon interaction, are crucial to the development of the CDW. In the context of these different types of long range order, further strain tuning of epitaxial films could provide invaluable information regarding the interplay between these two low temperature ground states.

\subsubsection{Correlated disorder in pseudo-{\it bcc} uranium alloys} \label{s:corrdisor}

As introduced in Section \ref{s:sxgamma} there has been growing interest in \textit{bcc} uranium – transition metal alloys over the last 20 years with a re-assessment of their potential role as advanced nuclear fuels \cite{Ajantiwalay2020}. However, as traditional synthesis methods are incapable of producing single crystals, detailed structural analysis has been severely limited. The first seminal work on this problem was presented by Yakel in 1969 \cite{Yakel1969,Yakel1974} who obtained diffraction from single grains at the tip of a polycrystalline needles. He observed significant diffuse scattering in the $\gamma^{\mathrm{s}}$ phase, and attributed it to local structural modifications which preserved the global \textit{bcc} structure as a configurational average refuting previous work suggesting chemical order \cite{Tangri1961}. However, his proposed solution was both complex and largely ignored by the majority of the U-alloy community who, for the most part, have considered $\gamma^{\mathrm{s}}$ as fully stabilised \textit{bcc}. This is understandable as the characteristic diffuse intensity is at least eight orders of magnitude weaker than the Bragg reflections and therefore when working with polycrystalline samples and standard powder diffraction refinement techniques is impossible to detect \cite{Hengstler2010,Lopes2013,Brubaker2019}.

\begin{figure}[t]
\centering
\includegraphics[width=\linewidth]{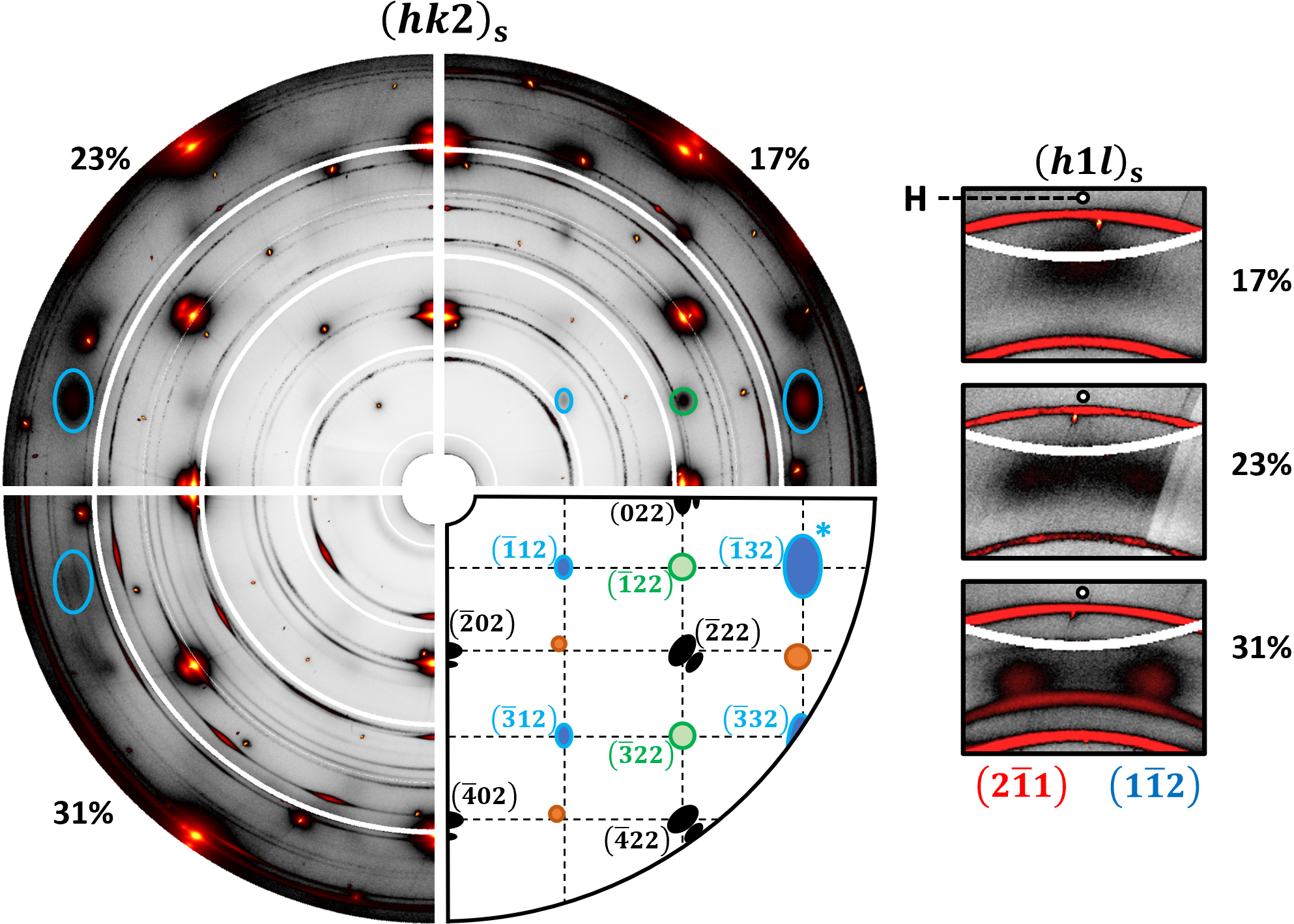}
\caption{Diffuse X-ray scattering patterns taken with $\lambda = 0.98$ \AA{}$ $  at room temperature from the three different Mo concentrations as marked outside quadrants. (Main) Reciprocal space reconstructions of the $(hk2)_{\mathrm{s}}$ plane produced using the superstructure unit cell with $b_{\mathrm{s}}\parallel b_{\mathrm{p}}$. Reflections are categorized and indexed in the bottom right quadrant, with indices in superstructure notation. Reflections are indicated in black (Mo parent Bragg peaks), blue ({\bf N} – domain 1), green ({\bf N} – domain 2) and orange ({\bf H}). The Nb buffer reflections are next to the Mo parent peaks. Powder rings arise from the polycrystalline Nb cap, whereas narrow intense peaks correspond to substrate Bragg reflections, none of which are included schematically. The ($\bar{1}32$) reflection, marked with an asterisk, is highlighted in all data sets to show mirror symmetry relations. (Insert) Reconstructions for the $(h1l)_{\mathrm{s}}$ plane highlighting the evolution of H type signal. The H point is marked with a circle. 31 at\% peaks are indexed with $\mathrm{U}_{2}\mathrm{Mo}$ with c parallel to $a_{\mathrm{p}}$ (red) and $c_{\mathrm{p}}$ (blue). Adapted from Ref.~\cite{Chaney2021}.}
\label{fig4_2_3}
\end{figure}

As explained in Section \ref{s:sxgamma} the single crystal synthesis problem was recently overcome using epitaxial matching as a stabilising force in substitute of rapid cooling \cite{Chaney2021, Adamska2014} allowing the $\gamma^{\mathrm{s}}$ phase to be studied in detail using modern synchrotron techniques for the first time. Using the diffuse x-ray scattering diffractometer at the ESRF \cite{Girard2019} Chaney \textit{et al.} confirmed the \textit{bcc} structure is not preserved locally and instead a three-dimensional structural modulation gives rise to a well defined diffuse X-ray pattern, shown in Fig.~\ref{fig4_2_3}. Further, the exceptional flux and resolution provided by modern synchrotrons allowed for another key breakthrough, the separation of the diffuse signal into two distinct types of different origin. The first group (\textbf{H}) were re-indexed as a coherent precursor structure of the intermetallic $\mbox{U}_2\mbox{Mo}$ allowing the remaining diffuse alloy reflections (\textbf{N}) to be explained by a substantially simplified model. The unique structural solution was given by a local orthorhombic structure, formed by atomic displacements along the $\langle001\rangle_s$ with no requirement for chemical ordering. The superstructure is C-face centered and defined with four atoms in the unit cell at $(0,\nicefrac{1}{4}\pm\delta,\nicefrac{1}{4})$ where $\left|\delta\right|=0$ corresponds to the \textit{bcc} structure. Not only is this a simpler solution, it also removes the unusual condition present in the previously suggested structure, whereby there are two atomic sets with distinctly different coordination and no chemical ordering.

\begin{figure}[t]
\centering
\includegraphics[width=\linewidth]{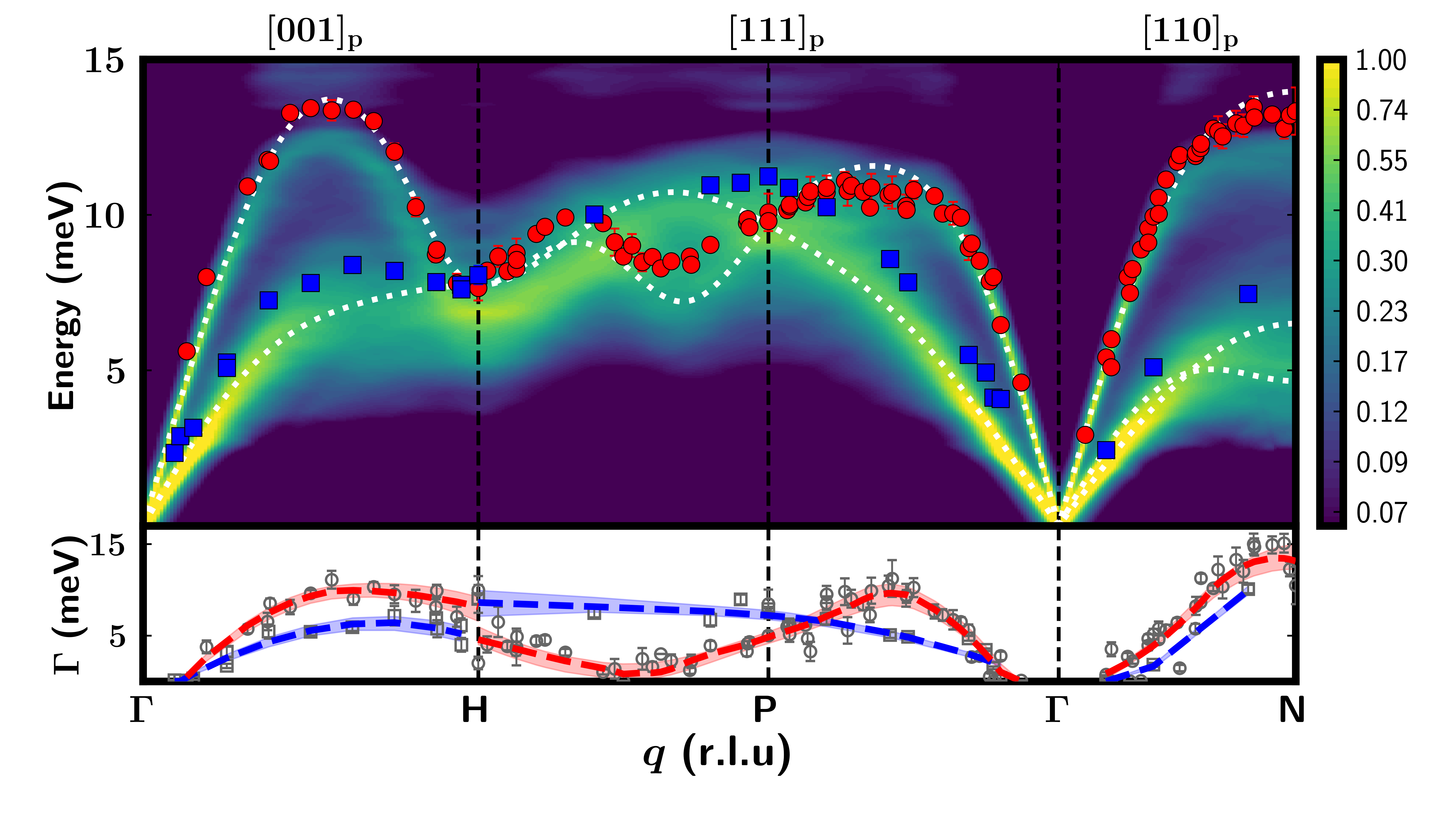}
\caption{Experimental and theoretical phonon dispersion curves for the U 23 at.~\% Mo alloy at room temperature. The transverse (longitudinal) acoustic modes are shown as blue squares (red circles). The VCA theory is shown as dashed (TA) and dotted (LA) white lines. The light shading shows the widths (from theory) after allowing for the alloying; note this is not on a linear scale. The soft mode $\mathrm{TA}_{1}[110]_{\mathrm{p}}$ predicted by theory at the {\bf N} point could not be observed, as the diffuse (elastic) scattering dominates the signal. The panel below the main figure shows the intrinsic widths (after deconvoluting with the instrumental resolution function and the theoretical width) of the phonon modes, which have both a directional, as well as an energy dependence. Adapted from Ref.~\cite{Chaney2021}.} \label{fig4_2_4}
\end{figure}

Given all atoms are displaced from their \textit{bcc} positions, the individual displacements, which transform from the parent to superstructure, can be viewed as a frozen phonon. The responsible modes being of general type $\mathrm{TA}_{1}[110]_{\mathrm{p}}$, with polarization along the fourfold axis $[001]_{\mathrm{p}}$ and are the exact modes predicted to destabilize the high-temperature \textit{bcc} phase \cite{Bouchet2008, Soderlind2012}. Twelvefold degeneracy generates six equivalent and equally occupied superstructure domains, however, given the diffuse nature of the reflections it is clear that the identity of any one choice of domain is only preserved over a relativity short distance, $\sim30$ \AA{} along $\bm{b}_{\mathrm{s}}/\bm{c}_{\mathrm{s}}$ and $\sim22$ \AA{} along $\bm{a}_{\mathrm{s}}$. As such, Chaney \textit{et al.}. \cite{Chaney2021} suggest that the local superstructure is better thought of, not as local order, but as correlated displacive disorder that lowers the local symmetry with three dimensional correlations between nearby atoms governed by rules represented as a frozen phonon. Twelvefold degeneracy maintains the higher average symmetry, while allowing  anisotropic neighbour distances reminiscent of $\alpha\mbox{-uranium}$ to be recovered. This situation can be understood intuitively as U has a desire, rare among the elements, to occupy highly open structures that produce extreme anisotropic local environments \cite{Soderlind1995, Mettout1993}. Thus, by attempting to stabilise U onto a highly symmetric \textit{bcc} lattice the mismatch in preferred symmetry creates an intrinsic conflict within the system. In the absence of sufficient thermal energy, this conflict creates a structural instability resolved by the formation of correlated disorder. The same experiments also showed this correlated disorder was intrinsically tunable. The authors use a qualitative metric combining correlated volume and magnitude of atomic displacement to capture the ``correlated disorder strength'' and show strong tunability with alloy composition, as evidenced by the variation of intensity and FWHM of the diffuse reflections in Fig.~\ref{fig4_2_3} as a function of minor alloy content.

Building on the discovery of correlated displacive disorder Chaney \textit{et al.} \cite{Chaney2021} explored the phonon dispersion for the 23 at.~\% Mo system using the grazing incidence inelastic X-ray scattering technique pioneered at the ID28 beamline \cite{ID28}, ESRF \cite{Rennie2018b,Serrano2011}. It is well understood that correlated disorder can couple to other periodic phenomena such as lattice dynamics \cite{Overy2016} and as mentioned in Section \ref{s:metalsintro} the uranium phonon dispersion has been subject to many notable theoretical efforts including high-temperature modelling \cite{Soderlind2012,Bouchet2017} and more recently calculations for the UMo alloy itself \cite{Castellano2020}. However, due to the lack of suitable single crystal samples no experimental data existed making this a significant gap in understanding. The first dispersion was published in 2019 by Brubaker \textit{et al.} \cite{Brubaker2019} who extracted a single enlarged grain from an annealed polycrystalline 20 at. \% sample by laser cutting. This work, along with the studies by Chaney \textit{et al.} \cite{Chaney2021}, agree well with each other as well as with theoretical predictions. The full dispersion as measured by Chaney \textit{et al.} \cite{Chaney2021} including room temperature \textit{ab initio} calculations is shown in Fig.~\ref{fig4_2_4}. Both experiments also observed extraordinary phonon lifetime suppression away from zone centre. As an alloy some intrinsic phonon broadening is expected, however this contribution was simulated and determined to be $<2\,\mbox{meV}$ for all $\bm{q}$, thus, the authors concluded that almost all phonon lifetime suppression in the system is attributable to disorder-phonon coupling.

\begin{figure}[t]
\centering
\includegraphics[width=0.5\linewidth]{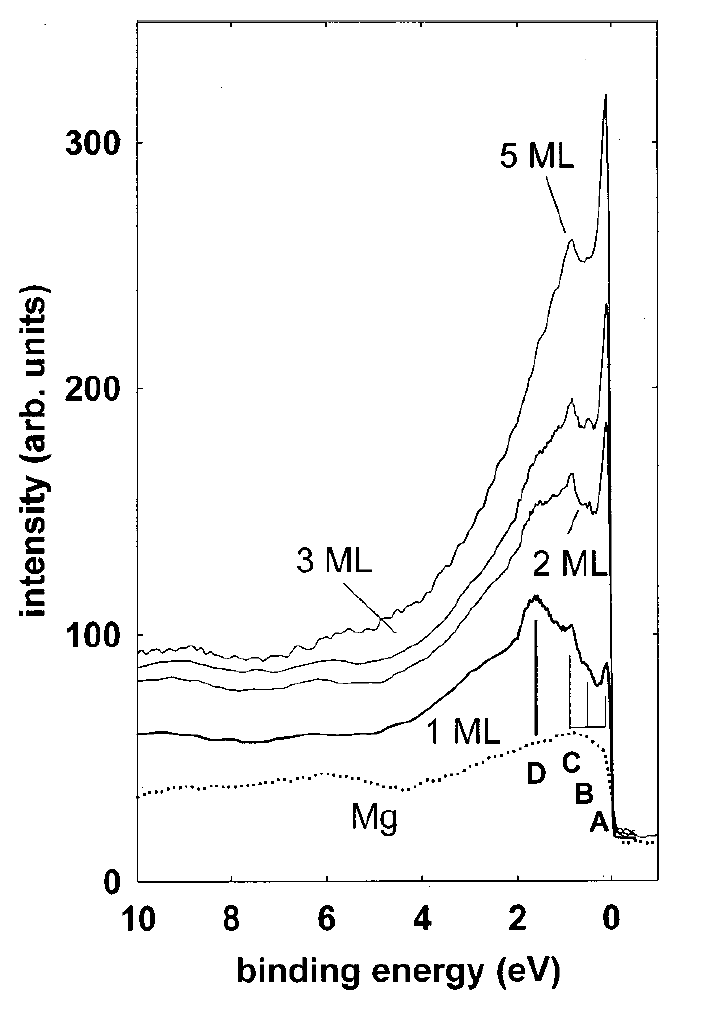}
\caption{Valence-band spectra of pure Mg substrate and increasing coverage of Pu. The thickness is indicated in the number of monolayers. Spectra were obtained with the photon energy $hv=40.8\;\mbox{eV}$. The vertical bars indicate the position of the $5f\mbox{-localized}$ peak and the triplet of narrow features. Reprinted from \cite{Havela2002}.} \label{Pu_thin_films}
\end{figure}

\subsubsection{Thin films of transuranium metals} \label{s:transuranics}

The number of laboratories that can perform experiments on transuranium films is, of course, extremely limited. To our knowledge the only laboratory to produce \textit{epitaxial} thin films is Los Alamos National Laboratory, where epitaxial transuranium oxide films have been grown and will be covered in Section \ref{s:UO2_PuO2_Photoemission}. However, the controversy over the electronic structure of the light actinides has been such that various efforts in photoemission and related spectroscopies have been undertaken over the past 20 – 30 years, many of which are discussed in a recent Plutonium Handbook\cite{Puhandbook}. For our purposes in this review we highlight one particular measurement that was able to monitor the localization of the 5$f$ electrons in Pu metal films by monitoring the photoemission spectra as a function of sample thickness \cite{Gouder2001,Havela2002}.

The key spectra are reprinted in Fig.~\ref{Pu_thin_films} and show an appreciable shift for the thinnest sample. Of course, in this work no X-ray characterization could be made of the samples, so the assumption is that the Pu metal was in the $\alpha$-Pu form and was polycrystalline. These spectra have been often cited in work on the electronic structure of Pu metal. The 3-peak (A, B \& C) structure are characteristic of 5$f$ localization, together with the peak D that represents the position of the 5$f$ states below $E_{\mathrm{F}}$.

\clearpage 

\section{Uranium Oxide Systems} \label{s:oxides}

\subsection{Introduction} \label{s:oxidesintro}

As the difficulties of manufacturing pure U metal fuel elements, as well as the safety issues raised by low-melting temperature of 1132 $^{\circ}$C, became more evident in the 1950's, uranium dioxide was introduced as an alternative nuclear fuel in the early 1960's. It remains today the fuel of choice of a majority of power reactors. A complete industry is devoted to the manufacture of such nuclear fuels and their treatment post-irradiation. The properties of UO$_2$ have thus been studied extensively in both a fundamental, as well as an applied, sense \cite{Hurley2022}. A major issue is related to the surface properties of UO$_2$, as it is at the surface that reactions will initially occur, and, with irradiated fuels, the potential danger arises from material escaping from the surfaces of such fuels. Films in the thin limit are essentially simply surfaces, so it is reasonable to expect that they can help with the understanding the behaviour of UO$_2$ surfaces. We shall show one example of research in this area. Another key aspect is related to interfaces involving UO$_2$, for example the interface between the fuel and the cladding, and here too, thin bilayers should offer new insights.

From a more fundamental point of view, UO$_2$ is a semiconductor with a band gap of $\sim$ 2 eV, which is almost twice that of silicon. It has garnered interest in applications (especially in satellites) requiring high reflectance in the wavelength range range of 40 - 100 \AA{}. Both UO$_2$ and UN films reflect significantly more \cite{Sandberg2004,Allred2002} than all known materials such as Au and Ir, and UO$_2$ is also cheaper. Additionally, there has been some effort to consider UO$_2$ for solar cells, as, if doped with oxygen, UO$_2$ is a p-type semiconductor \cite{Sandberg2004,Allred2002,Chen2010}.

\subsection{Surface studies of UO$_2$}	\label{s:oxidessurfacestudies}
An excellent review of the chemical reactions at the surface of UO$_2$ was given by Idriss \cite{Idriss2010} in 2010. Those studies (so far) have not used thin films, but are done on specially prepared surfaces of single crystals from bulk samples. Many of the properties deduced are relevant to thin-film research, and frequently can be extended with suitable films. On the chemical structure of the surface, earlier work was reported in the 1980s \cite{Ellis1980,Ellis1981}. An understanding of the chemical structure of the surfaces is important because with the fluorite CaF$_2$ structure of UO$_2$ the only primary surface that is non-polar is the (110). The other two surfaces, (100) and (111), are polar, so that with these latter two surfaces there will always be an atomic rearrangement at the surface to resolve the polar discontinuity with the vacuum \cite{Bottin2016}. We will need some of these results to understand later studies of the dissolution of UO$_2$.

Upon further oxidation the chemical formula can be generalised as UO$_{2+x}$, where $x$ = 0 at stoichiometry, and proceeds through many different phases, U$_4$O$_9$ ($x$ = 0.25), U$_3$O$_8$ ($x$ = 0.67), and finally arrives at UO$_3$ ($x$ = 1). As oxidation starts at the surface, various experiments have been performed by Stubbs and collaborators using the technique of x-ray surface truncation rods to understand the initial oxidation process \cite{Stubbs2015,Stubbs2017}. Other notable experiments on surfaces have been reported by Seibert \cite{Seibert2011}, and Spurgeon {\it et al.} \cite{Spurgeon2019} on UO$_{2+x}$. There has also been a series of theoretical investigations on the surface structure of UO$_2$ \cite{Tan2005,Maldonado2014,Wellington2016}, with the last two concentrating on the interaction of water with the surface.

Another area that has been studied is the magnetic structure and phase transition at the surface of UO$_2$. In the bulk this aspect has been studied since the 1960s \cite{Santini2009,Lander2020}, but the first observation of the magnetic structure from the surface was with resonant magnetic scattering \cite{Watson1996,Watson2000}. This work showed an unusual aspect of the phase transition at 30 K from the paramagnetic to antiferromagnetic state, and later experiments clearly defined this as a so-called “surface transition”, i$.$e$.$ a transition that is induced directly by the presence of the reduced symmetry at the surface \cite{Langridge2014}. Perhaps counter-intuitively, we shall show an example where a study of the magnetism, has been able to illustrate some important information about the interface of UO$_2$ with the substrate.

\subsection{Growth of oxides} \label{s:oxidesgrowth}
Given the abundant work reported on UO$_2$ surfaces in Ref. \cite{Idriss2010}, it is perhaps surprising that it took so long before epitaxial films of UO$_2$ were made systematically. As reported in Sec. \ref{s:earlyefforts}, epitaxial films were already made in the 1960's \cite{Steeb1961} and 1970's \cite{Navinsek1971,Nasu1972}, but relatively little was done with them.

The first recorded epitaxial UO$_2$ films were made at Los Alamos National Laboratory and reported in 2007 \cite{Burrell2007}. The substrates were LAO, which can be bought commercially. Since that time a whole series of papers have reported the growth of epitaxial UO$_2$ and also higher oxides with a wide variety of methods on many different substrates \cite{Elbakhshwan2017,Enriquez2020}. These are given in Table \ref{tab:oxides}.

Epitaxial UO$_2$ films were grown by DC-magnetron sputtering at Oxford in 2010 \cite{Ward2010} using substrates of LAO following the success of Burrell {\it et al.} \cite{Burrell2007}, but a paper on these samples was not published until 2013. In the meantime, a paper by Strehle {\it et al.} \cite{Strehle2012} appeared in 2012, also using DC-sputtering, and reporting epitaxial samples of UO$_2$ and U$_3$O$_8$. Strehle {\it et al.} \cite{Strehle2012} published much useful information on the growth of UO$_2$-based epitaxial films, and they also demonstrated that epitaxial films could be grown on yttrium-stabilized zirconia (YSZ), even though the mis-match was a large 6\%. This is important in UO$_2$-film research as YSZ substrates can be readily obtained in the three important orientations (100), (110) and (111), allowing epitaxial films of UO$_2$ to be grown in all three principal orientations. Strehle {\it et al.} also showed by chemical analysis with RBS and XPS that the (1–1.2) sapphire substrate material is unstable with respect to Al transport into the uranium-oxide overlayers.

\begin{landscape}
\begin{table}[]
\centering
\caption{Uranium oxide thin films that have been produced, with references of their first mention in publication. *No substrate given in the paper.}
\label{tab:oxides}
\begin{tabular}{|l|l|l|l|l|}
\hline
Material            & Form       & Substrate                          & Deposition method                & Reference       \\ \hline
UO$_{2+x}$          & \hkl(001)  & \hkl(001) MgO                      & vapor deposition                 & \cite{Steeb1961}       \\
UO$_{2+x}$          & \hkl(001)  & \hkl(001) NaCl                     & DC sputtering                    & \cite{Navinsek1971}    \\
UO$_{2+x}$          & \hkl(001)  & \hkl(001) NaCl, KCl, KBr, NaF, LiF & vapor deposition                 & \cite{Nasu1972}        \\
UO$_{2+x}$          & poly       & *                                  & pulsed laser ablation            & \cite{Gibson1999}      \\
UO$_2$, UO$_{2-x}$  & poly       & Si wafer                           & reactive DC sputtering           & \cite{Miserque2001}    \\
doped UO$_2$        & poly       & Al$_2$O$_3$, MgO                   & sol-gel                          & \cite{Meek2005}        \\
UO$_2$              & \hkl(011)? & \hkl(100) LaAlO$_3$                & polymer-assisted deposition      & \cite{Burrell2007}     \\
U$_3$O$_8$          & \hkl(100)  & \hkl(001) Al$_2$O$_3$              & polymer-assisted deposition      & \cite{Burrell2007}     \\
UO$_2$              & \hkl(001)  & \hkl(001) YSZ                      & reactive DC mag. sputtering & \cite{Strehle2012}     \\
UO$_2$              & \hkl(001)  & \hkl(012) Al$_2$O$_3$              & reactive DC mag. sputtering & \cite{Strehle2012}     \\
$\alpha$-U$_3$O$_8$    & \hkl(001)  & \hkl(012) Al$_2$O$_3$              & reactive DC mag. sputtering & \cite{Strehle2012}     \\
UO$_2$              & \hkl(001)  & \hkl(001) LaAlO$_3$                & reactive DC sputtering           & \cite{Bao2013}         \\
UO$_2$              & \hkl(001)  & \hkl(001) CaF$_2$                  & reactive DC mag. sputtering & \cite{Bao2013}         \\
$\alpha$-U$_3$O$_8$ & \hkl(001)  & \hkl(012) Al$_2$O$_3$              & polymer-assisted deposition      & \cite{Scott2014}       \\
$\beta$-U$_3$O$_8$  & \hkl(001)? & \hkl(001) Al$_2$O$_3$              & polymer-assisted deposition      & \cite{Scott2014}       \\
UO$_{2+x}$          & poly       & Ni                                 & electrodeposition                & \cite{Adamska2015}     \\
UO$_2$              & \hkl(110)  & \hkl(100) TiO$_2$                  & reactive DC mag. sputtering & \cite{Elbakhshwan2017} \\
UO$_2$              & \hkl(111)  & \hkl(001) ZnO                      & reactive DC mag. sputtering & \cite{Elbakhshwan2017} \\
UO$_2$              & \hkl(001)  & \hkl(001) SrTiO$_3$                & reactive DC mag. sputtering & \cite{Rennie2018b}      \\
UO$_2$              & \hkl(011)  & \hkl(011) YSZ                      & reactive DC mag. sputtering & \cite{Rennie2018a}     \\
UO$_2$              & \hkl(111)  & \hkl(111) YSZ                      & reactive DC mag. sputtering & \cite{Rennie2018a}     \\
UO$_2$              & \hkl(001)  & \hkl(001) LSAT                     & pulsed laser deposition          & \cite{Enriquez2020}    \\
$\alpha$-U$_3$O$_8$ & \hkl(001)  & \hkl(001) Al$_2$O$_3$              & pulsed laser deposition          & \cite{Enriquez2020}    \\
$\alpha$-U$_3$O$_8$ & \hkl(001)  & \hkl(001) LSAT                     & pulsed laser deposition          & \cite{Enriquez2020}    \\
$\alpha$-U$_3$O$_8$ & \hkl(001)  & \hkl(001) MgO                      & pulsed laser deposition          & \cite{Enriquez2020}    \\
$\alpha$-UO$_3$     & \hkl(001)  & \hkl(001) MgO                      & pulsed laser deposition          & \cite{Enriquez2020}  \\
\hline
\end{tabular}
\end{table}
\end{landscape}

\subsection{Science with UO$_2$ thin films} \label{s:oxidesscience}

\subsubsection{Photoemission experiments on UO$_2$ and PuO$_2$ epitaxial films} \label{s:UO2_PuO2_Photoemission}

We shall highlight here the ARPES results obtained from UO$_2$ and PuO$_2$ epitaxial films. Whereas polished single crystals of UO$_2$ exist and were already used for ARPES measurements, there are no sizable single crystals of PuO$_2$, so that effort is unique and important. The central question posed by the theory presented also in Ref. \cite{Scott2014} is to what extent are the actinide 5$f$ electrons hybridizing with the oxygen 2$p$ electrons? The hybrid density functional theory discussed predicts that in UO$_2$ the mixing between these two orbitals is present, but relatively small, whereas it should be much stronger in PuO$_2$. Below in Figs. \ref{fig5_1_1a} and \ref{fig5_1_1b}  we show the UO$_2$ and PuO$_2$ results from the ARPES measurements.

\begin{figure}[htb]
\centering
\includegraphics[width=0.55\linewidth]{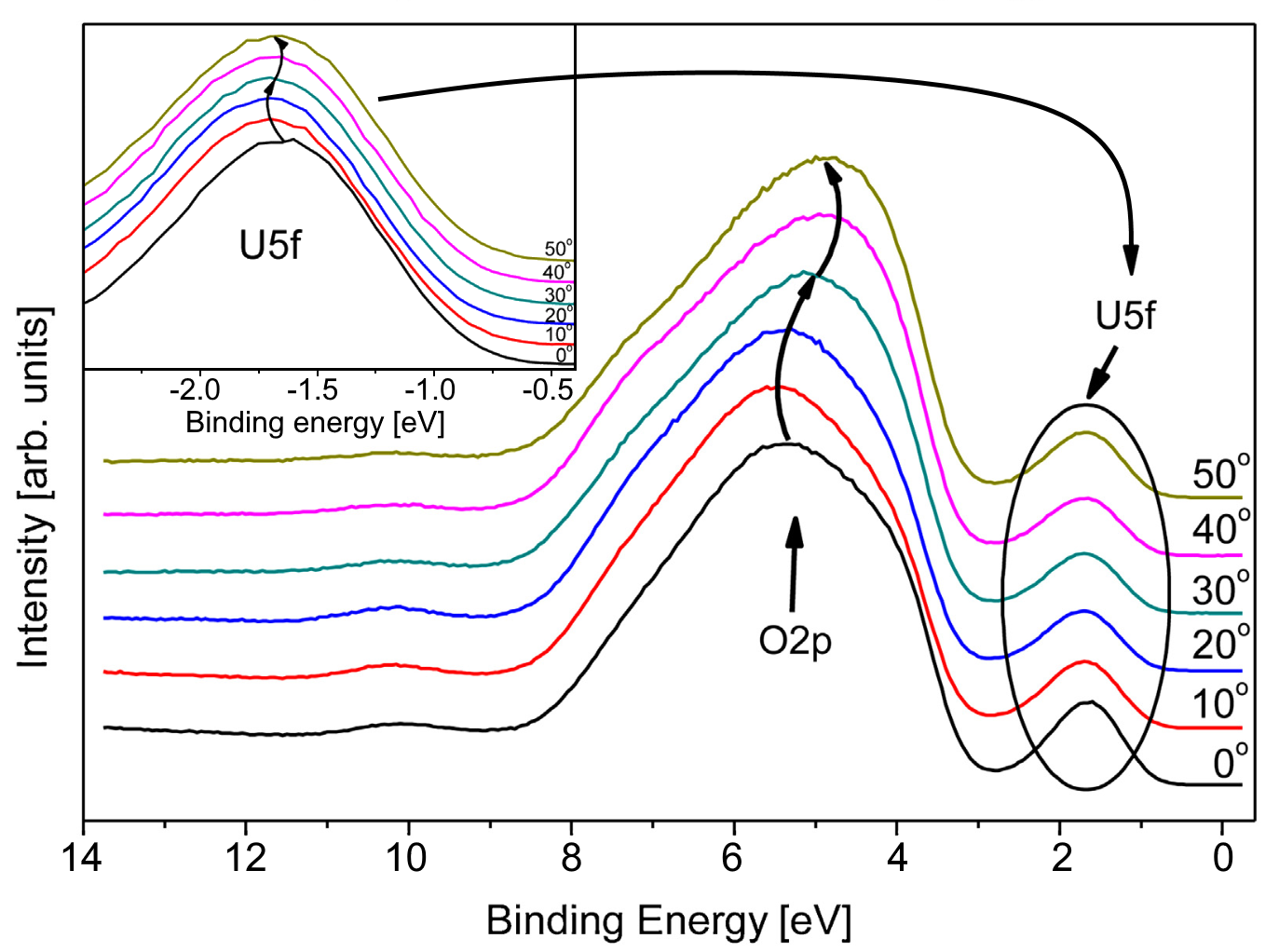}
\caption{The small dispersion of $\sim$ 100 meV across the zone in the 5$f$ band for UO$_2$. Adapted from Ref. \cite{Scott2014}} \label{fig5_1_1a}
\end{figure}

\begin{figure}[htb]
\centering
\includegraphics[width=0.85\linewidth]{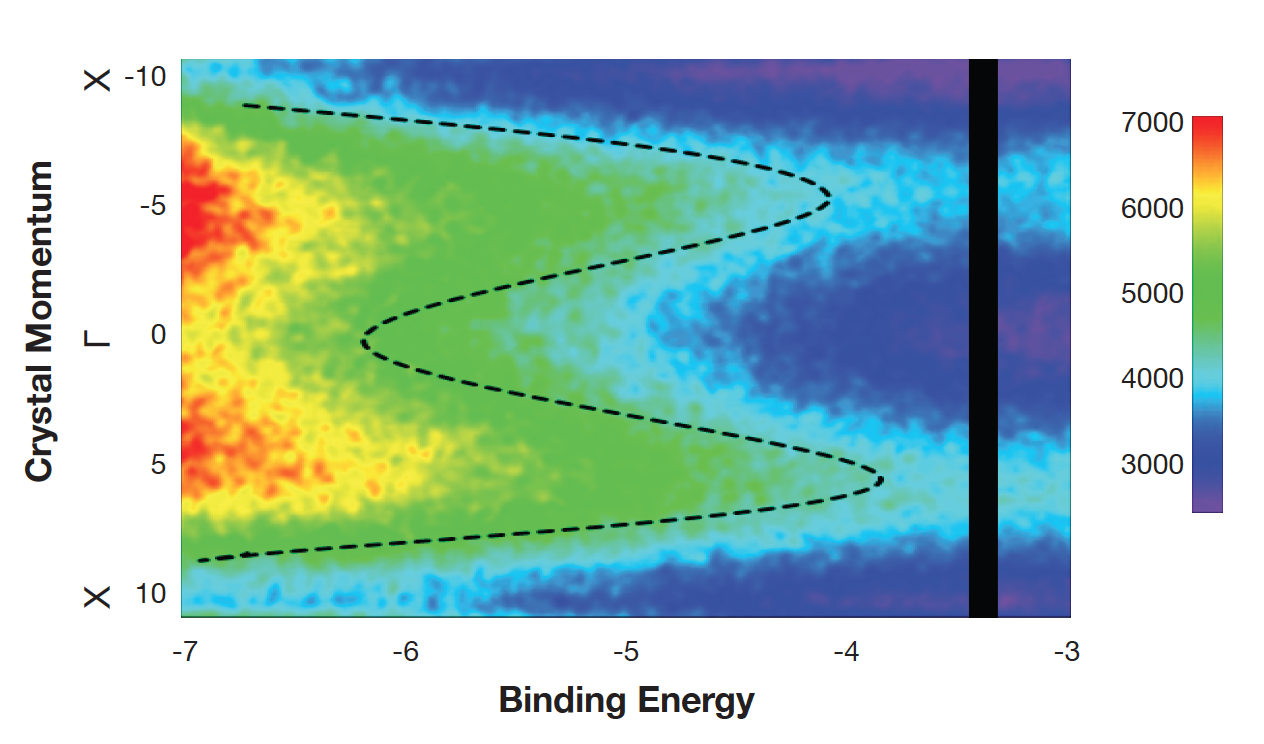}
\caption{ARPES work on PuO$_2$ showing that the dispersion in PuO$_2$ is $\sim$ 3 eV. The black bar indicates approximately the position and width of the similar U 5$f$ states found in UO$_2$. Taken from Ref. \cite{Joyce2010} }
\label{fig5_1_1b}
\end{figure}
Clearly, the much greater dispersion ($\sim$ 3 eV) in the 5$f$ level in PuO$_2$ \cite{Joyce2010} than found in UO$_2$ of $\sim$ 0.1 eV is a key finding confirming the theory used in this investigation of the actinide dioxides.

\subsubsection{Antiferromagnetism of UO$_2$ epitaxial films} \label{s:AF_UO2}

The work published on the films from Oxford/Bristol about the low-temperature antiferromagnetic (AF) structure of UO$_2$ films illustrates the challenges of working with epitaxial films, which must be grown on substrates. The films were produced at ITU, Karlsruhe, as well as at Bristol. All initial films used LAO substrates and the depositing temperature was 650 $^\circ$C. A careful examination \cite{Bao2013} of the UO$_2$ films showed tetragonal symmetry, even for films over 2000 \AA{} thick. The $a$ and $c$ axes are shown in Fig. \ref{fig5_1_2} below.
\begin{figure}[htb]
\centering
\includegraphics[width=0.55\linewidth]{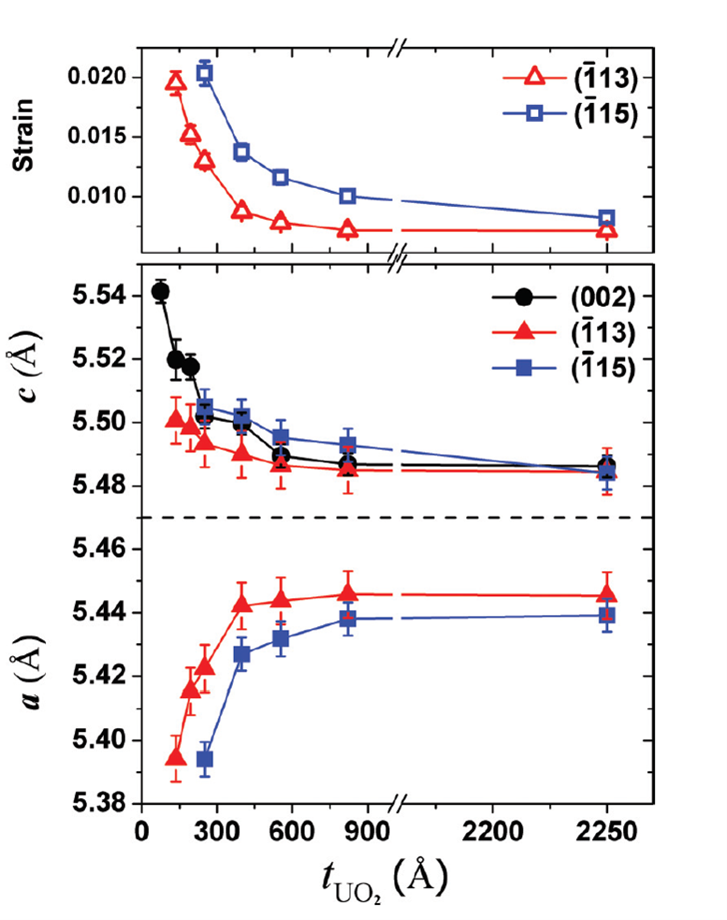}
\caption{Evolution of the out-of-plane, $c$, and the in-plane, $a$, lattice parameters determined from reciprocal space mapping measurements with UO$_2$ thickness (deposited on LAO substrates). The dashed line is the bulk value of $a$ and $c$. The upper panel shows the strain, determined as $\frac{2(c-a)}{(c+a)}$. Taken from Ref. \cite{Bao2013} }
\label{fig5_1_2}
\end{figure}

\begin{figure}[htb]
\centering
\includegraphics[width=0.55\linewidth]{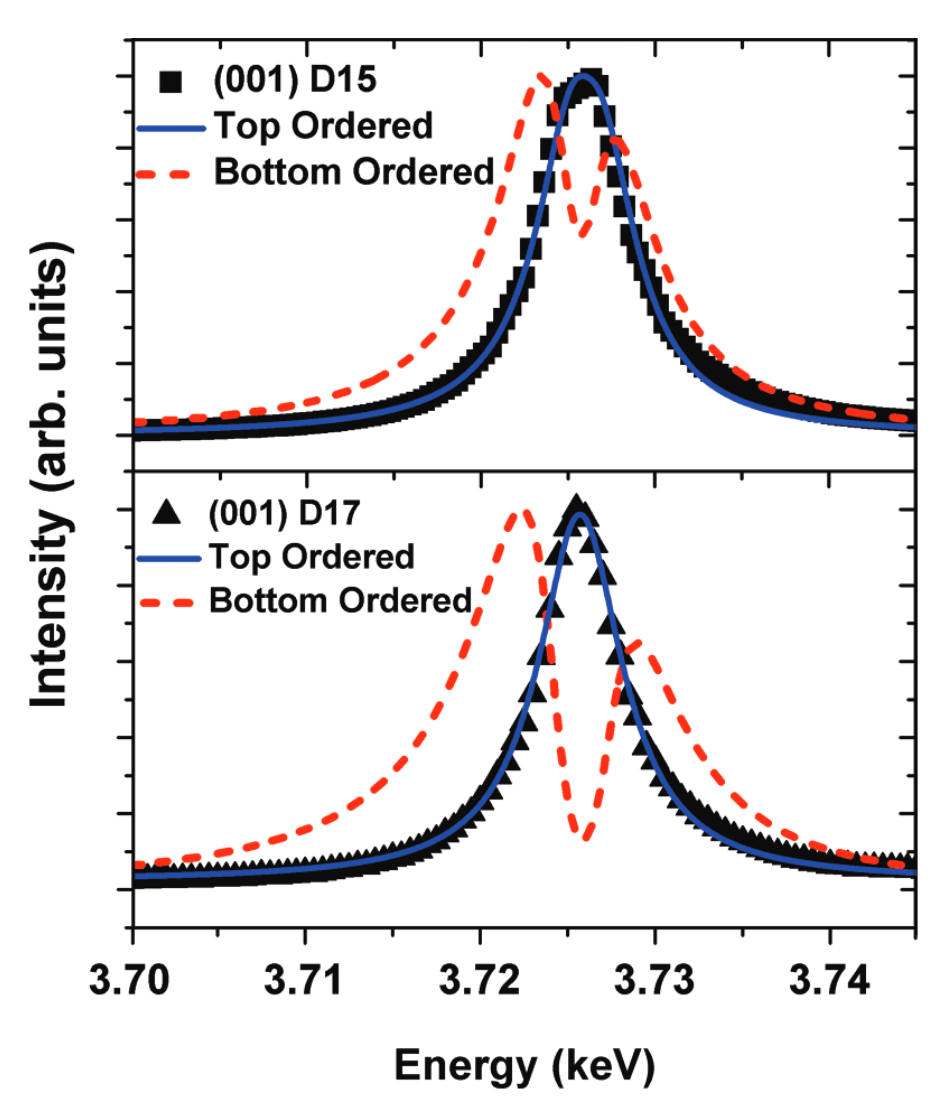}
\caption{Experimental data as solid points give the intensity of the AF peak as a function of incident energy across the $M_4$ resonant energy. The solid (dashed) lines are from a model, using the theory presented in Ref. \cite{Bernhoeft1998}, in which there is a dead magnetic layer below (above) the AF ordered region in the film. The thicknesses are 2250 and 4500 \AA{} for D15 and D17, respectively. Taken from Ref. \cite{Bao2013}.  }
\label{fig5_3}
\end{figure}

The UO$_2$ films had a rocking curve width that exceeded 1$^{\circ}$, (2$^{\circ}$ for the thinnest samples) although the substrate peaks were very sharp. Some of the thinner samples showed no signs of AF magnetic ordering. This certainly is related to the tetragonal symmetry found and demonstrated in Fig. \ref{fig5_1_2}. The AF structure of UO$_2$ is associated with a cubic lattice \cite{Santini2009}, and the distortion, which is due to the interaction with the substrate, breaks that condition. On cooling these samples to base temperature ($\sim$ 10 K) one should see the (001) AF peak of UO$_2$ along the specular direction, as this direction is [001]. For the thinnest samples ($t \leq 250$ \AA{}) no AF peak was observed. $T_{\mathrm{N}}$ of UO$_2$ is 30 K, and this is a robust value, i$.$e$.$ it takes an effort to change this, so this result was surprise. The energy dependence of the AF peaks observed from the thicker films was also measured, as shown in Fig.\ref{fig5_3}.

A model proposed by Bernhoeft {\it et al.} \cite{Bernhoeft1998} was used that can calculate the profile of the energy dependence of the peak as a function of depth that is ordered magnetically. The model shows that the peaks come from a volume that is tied to the surface of the UO$_2$ film, rather than an ordering tied to the interface with the substrate. It appears that there is a “dead magnetic layer” next to the substrate. After performing the experiments, the authors discovered that LAO has a ferroelastic transition at 560 $^\circ$C, which is below the depositing temperature of 650 $^\circ$C used for these films. Thus, on cooling, the transition occurs in the substrate effecting the interface. The thinner UO$_2$ films have a larger strain (see Fig. \ref{fig5_1_2}), which prevents the UO$_2$ from ordering magnetically, because the AF structure of UO$_2$ requires a cubic UO$_2$ chemical structure. Above the dead layer of $\sim$ 500 \AA{} the relaxation allows the local cubic symmetry, and ordering occurs.

A semi-proof of this hypothesis was obtained by depositing UO$_2$ on CaF$_2$, which has a lattice parameter and crystal structure almost identical to UO$_2$, and finding that the AF ordering was throughout the film \cite{Bao2013}. Of course, in the original choice of LAO for a substrate using the PAD method \cite{Burrell2007}, the films are deposited at room temperature, so this unexpected problem did not arise. This is a cautionary tale; one is not only producing an epitaxial film, but also an interface at which unexpected interactions may occur.

\subsubsection{Enhanced paramagnetism in strained epitaxial UO$_2$ films} \label{s:UO2paramagstrain}

\begin{figure}[htb]
\centering
\includegraphics[width=0.95\linewidth]{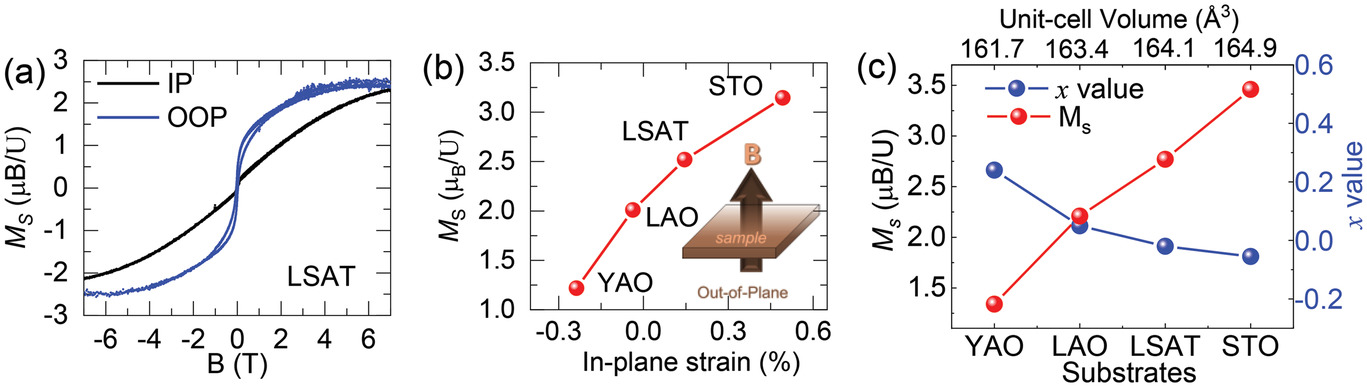}
\caption{a) Comparison of out-of-plane and in-plane $M(H)$ for the UO$_2$ film on LSAT measured at 10 K. The magnetic easy axis is oriented along the OOP direction in compressive-strained UO$_2$ film on STO. b) In-plane strain-dependent OOP saturated magnetization ($M_S$) for UO$_{2+x}$ films grown on YAO, LAO, LSAT, and STO substrates measured at 10 K. Inset shows the OOP $M(H)$ measurement configuration. c) The growth substrate and unit-cell volume of UO$_{2+x}$ films dependent $M_S$ and $x$ value. Taken from Ref. \cite{Sharma2022}.  }
\label{fig4_4_3}
\end{figure}

An interesting paper appeared in 2022 \cite{Sharma2022} with a report that in thin ($<$ 200 \AA{}) epitaxial UO$_2$ films prepared by PLD on various perovskite substrates the uranium ions exhibit enhanced paramagnetism. The authors used YAO (YAlO$_3$), LAO, LSAT {(La,Sr)(Al,Ta)O$_3$}, and STO, all of which have lattice spacings close to that of of UO$_2$, and where epitaxial UO$_2$ growth has a $\sqrt{2}$ arrangement on the substrate surface. In this configuration the lattice mismatch, compared to UO$_2$, covers a range of strain from -3.86\% (for YAO) to +0.91\% (for STO). They performed reciprocal space mapping to determine the individual in-plane and out-of-plane strains in the UO$_2$ films. From the unit cell volume, compared to UO$_2$, the authors propose that these films represent hypo- (i.e. $x$ $<$ 0) or hyper- ($x$ $>$ 0) stoichiometric samples of UO$_{2\pm{}x}$ in the fluorite structure, with $x$ in the range from -0.06, (UO$_{1.94}$) for STO to +0.23, (UO$_{2.23}$) for YAO substrates. This assignment is based on the lattice parameters of bulk UO$_{2\pm{}x}$. The films have a large induced paramagnetism, as summarized in Fig.~\ref{fig4_4_3}.

A key question here is whether there is true ferromagnetism in these samples, as the authors suggest. When compared to the bulk magnetic behaviour of UO$_2$ there are a range of important points to note. The sizes of the induced magnetic moments is exceptional, ranging from 1.2 to 3.2 $\mu_{\mathrm{B}}$. In pure stoichiometric UO$_2$ the antiferromagnetic moment is only 1.74 $\mu_{\mathrm{B}}$ and to break the AF coupling at least 14 T is needed, and then no large induced {\it paramagnetism} is observed up to 100 T \cite{Jaime2017}. Indeed the paramagnetic susceptibilities that they report are at least 100 times greater than found in pure bulk UO$_2$, and also found by susceptibility measurements \cite{Arrott1957} in non-stoichiometric samples.

It is well accepted in {\it bulk} UO$_{2+x}$ that the AF ordering is suppressed for $x$ $>$ $\sim$ 0.10 \cite{Arrott1957}, and yet the authors report large induced moments of $>$ 1 $\mu_{\mathrm{B}}$ for UO$_2$ on YAO, which they propose is equivalent to UO$_{2.23}$. An earlier study (see previous Section \ref{s:AF_UO2}) of a number of UO$_2$ films deposited (by sputtering) on LAO was reported by Bao {\it et al.} \cite{Bao2013} in 2013. No magnetization studies were done on these samples, so a direct comparison with Ref. \cite{Sharma2022} is not possible. On the basis of the ground-state known for UO$_2$ the maximum moment for the $\Gamma_5$ triplet \cite{Santini2009,Lander2021} ground state is $\sim$ 2 $\mu_{\mathrm{B}}$. The splitting to the higher levels is such that no moment greater than this should be observed. It remains to be understood if the crystal field present in non-stoichiometric UO$_2$ could be so heavily distorted by strain, and the relatively small movement away from stoichiometry.

Since these results are so unexpected, more studies will undoubtedly be needed to provide more insight. Key measurements include temperature-dependent measurements to clarify the existence or otherwise of a Curie point, and element specific measurements (e.g. X-ray circular magnetic dichroism at the uranium $M_{4,5}$ edges) to demonstrate that the magnetism arises from U ions \cite{Wilhelm2007}. We note that the easy direction for the moments is out of plane, which is unusual, as normally with thin films the easy axis is {\it in-plane} on account of the shape anisotropy. Interfacial anisotropy is known to drive the moments out of plane, famously in Co/Pt and Co/Pd superlattices \cite{yakushijia_apl_2010}, but an intriguing point here is that polarised-neutron reflectivity measurements have shown that these moments are distributed uniformly across the film thickness.

\subsubsection{Search for exchange bias using UO$_2$ thin films} \label{s:UO2exchangebias}

Exchange bias is a phenomenon in which the hysteresis loop of a ferromagnetic material is offset in the field by interface interaction with another system, usually an antiferromagnetic material. Discovered more than 60 years ago \cite{Meiklejohn1956} it has many applications in devices. The production of epitaxial UO$_2$ films opened the possibility of examining exchange bias with an anisotropic antiferromagnet. The first samples were made with magnetite (Fe$_3$O$_4$) as the ferromagnetic material \cite{Tereshina2014}. 300 \AA{} of UO$_2$ were deposited on LAO substrates, and then varying thicknesses (90 to 700 \AA{}) of Fe$_3$O$_4$ were deposited on top of the UO$_2$, with a 500 \AA{} cap of Mg deposited to prevent any further reaction occurring. Cross-sectional TEM scans showed partial coherence across the UO$_2$/Fe$_3$O$_4$ boundary, but there were a number of domains in the Fe$_3$O$_4$.

The hysteresis loops were then measured as a function of field at various temperatures down to 5 K. Exchange bias of 2.6 kOe was found in the thinnest ($\sim$ 100 \AA{}) Fe$_3$O$_4$ samples at 5 K, but rapidly diminished for thicker samples. An unusual feature was the presence of reasonable exchange bias up to $\sim$ 50 K, i.e. considerably above the $T_{\mathrm{N}}$ = 30 K of UO$_2$.

A second attempt was made with (polycrystalline) permalloy (Ni$_{80}$Fe$_{20}$) films replacing the Fe$_3$O$_4$ in the previous study. Also, the LAO substrates were replaced by CaF$_2$ because of the known problems with the former \cite{Bao2013}. In this system the exchange bias was considerably smaller than found in the films with Fe$_3$O$_4$, presumably because the UO$_2$/permalloy interface was not as coherent as the one made with Fe$_3$O$_4$ and the effect was not present above the $T_{\mathrm{N}}$ of UO$_2$. Rather surprisingly, when the magnetization was performed perpendicular to the plane of the film (which is the hard direction of magnetization), there was considerable hysteresis, and an order of magnitude difference in the exchange bias, measured as 220 Oe, still much smaller than found with Fe$_3$O$_4$. Moreover, the $t^{-1}$ dependence  (where $t$ is the permalloy thickness) of this exchange bias shows that the origin of the effect is interfacial in nature \cite{Tereshina2015}.

Although it would appear that neither of these measured effects is particularly dramatic, this project, as with the metal bilayers mentioned in Sec. \ref{s:Ubilayers}, depends to a large extent on the interfacial nature of the bilayers. Until those can be well characterized and improved, the observation of effects such as exchange bias must be taken with some caution. In addition, UO$_2$ is a complicated 3{\bf k} antiferromagnet \cite{Santini2009,Lander2020}, and the effect might be greater with an actinide collinear antiferromagnet.

\subsubsection{Dissolution studies of UO$_2$}	\label{s:UO2diss}

In the spirit of thinner UO$_2$ films approximating to the surface, explorations have been carried out of the interaction of very thin epitaxial films ($<$ 100 \AA{}) with water as a function of pH. It has been known for many years that the corrosion of UO$_2$ proceeds through a process converting the stable U$^{4+}$ ion, which is almost insoluble in water, into a uranyl U$^{6+}$ ion, which is then highly soluble in water \cite{Shoesmith2000}. The strategy in this context is that by treating the surface of thin films, it might be possible to observe, either with Bragg scattering or reflectivity, some change when the surface was treated with oxidising agents. Starting simply by wiping the UO$_2$ thin films with ionised water, which normally should have no effects as the U$^{4+}$ ion is very stable; it was found that when the synchrotron beam illuminated the film and water, there was strong radiolysis of the water, the production of both H$_2$O$_2$ and OH$^{-}$ radicals, and a concomitant conversion of U$^{4+}$ to U$^{6+}$, and dissolution of U from the sample. The intensity of the Bragg peak of the sample of thickness $\sim$ 40 \AA{} immediately reduced. With a bulk sample this reduction of the Bragg intensity would not be visible, and even with a sample of 1000 \AA{} the effect is small.

\begin{figure}[htb]
\centering
\includegraphics[width=0.95\linewidth]{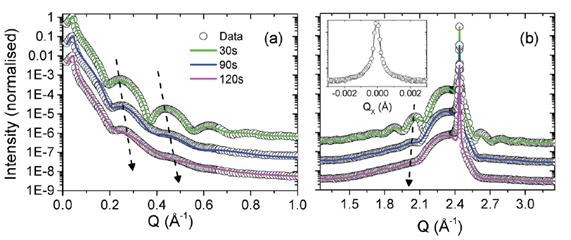}
\caption{Panel (a) shows X-ray reflectivity data and (b) shows high-angle diffraction data from a 40 \AA{} thin UO$_2$ (100) film (deposited on YSZ) at exposure times as marked. The experimental data are represented by open circles and the fitted calculations by the colored lines. The insert of panel (b) shows the rocking curve of the (002) Bragg reflection from the film. The black arrows indicate an increase in the fringe spacing as a function of exposure time, and hence a concomitant loss of material, i.e. reduced thickness. Taken from Ref. \cite{Springell2015} }
\label{fig5_3_1}
\end{figure}

\begin{figure}[htb]
\centering
\includegraphics[width=0.95\linewidth]{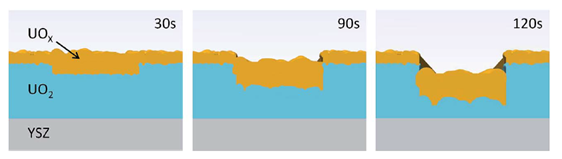}
\caption{Pictorial representation of the increase of roughness and UO$_x$ thickness, and amount of dissolution as the surface undergoes corrosion. Note that without a cap a UO$_2$ thin film always has a finite layer of material that can be described as UO$_x$, where $x >$ 2.00. This extends to $\sim$ 4-8 monolayers, i.e. 10 - 20 \AA{}. Taken from Ref. \cite{Springell2015}.}
\label{fig5_3_2}
\end{figure}

Using a combination of the reflectivity and Bragg scattering, Springell {\it et al.} \cite{Springell2015} were able to construct a picture of the dissolution as a function of exposure, as pictorially shown in Fig.~\ref{fig5_3_2}.

During the experiment the pH of the water was changed. As expected, the dissolution is faster in acidic water (pH $\sim$ 2) than in an alkaline (pH $\sim$ 11) solution. In the latter dissolution was almost halted. The experiments also tested whether using X-rays at the $L_3$ absorption edge at 17.116 keV increased the effect due to photocatalytic processes (i.e. the large number of excited electrons emitted at this energy), but no measurable effect was observed.

The authors also remarked on the clean image of the beam on the thin UO$_2$ film when using a scanning electron microscope. If the main product of the radiolysis was H$_2$O$_2$ alone, then one would expect some dilution of the edges of the ridge cut out by the beam in Fig. \ref{fig5_3_2}, as the lifetime of H$_2$O$_2$ is quite long, so the process should continue after the beam is shut off. The sharpness of the profile suggests that other products, possible OH free radicals, might be important, as they have a short lifetime and would disappear once the beam is turned off. This question has not yet been answered with further experiments.

The next question to be probed with this method was the directional dependence of the dissolution \cite{Rennie2018a}. These experiments showed quite conclusively that the most stable surface of UO$_2$ is the (111) planes, as has been theoretically predicted. As discussed earlier, the surface of UO$_2$ is only non-polar for the (110) plane, i.e. there are layers containing two oxygen atoms for every one uranium atom at the surface, so that the surface layer has no charge (non-polar). For the other two directions (100) and (111) there will be a re-arrangement at the natural surface layers so that charge neutrality occurs \cite{Bottin2016}. It will therefore depend on the stability of this latter process which plane is the more stable.

As can be seen from the figure, the (111) plane is the most stable. Interestingly, the dissolution appears to proceed and after a short while is passivated. The most unstable surface appears to be the (110).
\begin{figure}[htb]
\centering
\includegraphics[width=0.95\linewidth]{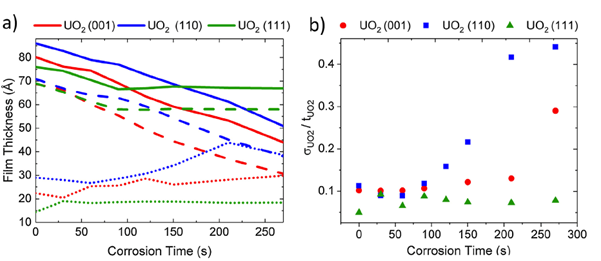}
\caption{(a) The layer thicknesses for the (001), (110) and (111) UO$_2$ thin films, shown in red, blue, and green, respectively, are compared as a function of corrosion time. The UO$_2$ and total film thickness are displayed by the dashed and solid line, respectively. The dotted line indicates the dissolution zone.
The error bars here are about $\pm$ 3 \AA{}.  (b) The roughness of the UO$_2$ layer normalized by the total UO$_2$ thickness for the 3 different directions. Errors are $\sim$0.05. Taken from Ref. \cite{Rennie2018a}. }
\label{fig5_3_3}
\end{figure}
These measurements can be compared to oxidation studies (not dissolution) performed by Stubbs and collaborators for both the (111) \cite{Stubbs2015} and (100) \cite{Stubbs2017,Spurgeon2019} surfaces, as well as a number of theoretical studies that approach this subject \cite{Rak2013,Maldonado2014,Wren2005}. All agree on the stability of the (111) surface. The dissolution methodology has also been used for other systems, notably UN and U$_2$N$_3$, which we shall cover later in this review.

\subsubsection{Studies of phonons with irradiated UO$_2$} \label{s:UO2phonons}

Thin films give another perspective when discussing properties of irradiated materials. In a reactor the fuel elements are in a high flux of fast neutrons that penetrate deep into all materials and cause damage throughout. Of course, differences will appear for different materials, and also as a function of the temperature. The materials in a reactor become extremely hot, i.e. radioactive, emitting both alpha particles, and high-energy gamma radiation from unstable nuclei in the fission process and from fission products themselves. Examining such materials requires high-tech “hot laboratories” that are exceedingly expensive to maintain and need specialised staff.

Some (but not all) of the work on radiation damage can be simulated by using high-energy irradiation sources, such as He$^{2+}$ and heavier ions, to cause the damage. Whereas such methods leave the samples essentially inactive, such sources of radiation (which are charged) do not (unlike neutral neutrons) penetrate deep into materials. Depending on the energy, they can perhaps penetrate 2-10 $\mu$m. Such damage is therefore ideally matched to thin films, where the whole sample can be damaged in a homogeneous manner. A good example of such a use for thin films is the work on the examination of the phonons in irradiated thin films of UO$_2$ by Rennie {\it et al.} (2018) \cite{Rennie2018b}.

It has been known for many years that the thermal conductivity of UO$_2$ is strongly reduced on irradiation in a reactor \cite{Ronchi2004a, Ronchi2004b}. As the thermal conductivity falls, the radial temperature gradient across the fuel pin becomes more substantial, leading to enhanced cracking and deformation. Consequently, the decay in thermal conductivity not only reduces the reactor efficiency but also contributes to the degradation in structural integrity of the fuel; together these effects ultimately act to limit the fuel lifetime. The thermal conductivity in UO$_2$ is contributed almost exclusively by the phonons, at least at temperatures below $\sim$ 1500 K, where the contributions from polarons are small \cite{Hutchings1987}. This is also the region of interest for reactor operations. The phonons in stoichiometric UO$_2$ were first measured in 1965 in a pioneering experiment by Dolling, Cowley, and Woods at Chalk River National Laboratories, Canada \cite{Dolling1965}. Since then, they have been measured many times, but recently Pang {\it et al.} \cite{Pang2013a,Pang2014a} have published a study at different temperatures where they show quantitatively how each phonon branch contributes to the thermal conductivity.

\begin{figure}[htb]
\centering
\includegraphics[width=0.95\linewidth]{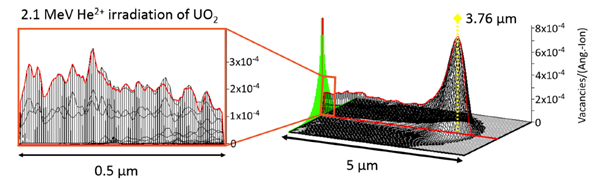}
\caption{The irradiation damage profiles calculated using the monolayer method in SRIM for the irradiation of a 5000 \AA{} (0.5 $\mu$m)  UO$_2$ layer and bulk UO$_2$ sample with 2.1 MeV He$^{2+}$ ions, using displacement energies of 20 and 40 eV for oxygen and uranium, respectively. The dashed yellow line represents the peak of the damage, located at 3.76 $\mu$m, i.e., in the substrate. Taken from Ref. \cite{Rennie2018b}. }
\label{fig5_4_1}
\end{figure}

The thin epitaxial films chosen for the experiment had a thickness of 3000 \AA{} and were deposited on SrTiO$_3$ substrates. (YSZ based materials exhibit a considerable amount of diffuse scattering, so these were avoided). Fig. \ref{fig5_4_1} illustrates the situation of the damage when the films were exposed to a 2.1 MeV He$^{2+}$ accelerated beam. The first 1 $\mu$m has a roughly homogeneous damage profile, whereas the so-called Bragg peak of the damage is located at just under 4 $\mu$m, i.e. well into the substrate in this situation.

One of the most difficult parameters to determine was the amount and type of radiation damage to produce in the films. If the damage is too extensive and the lattice itself is partially destroyed, then, clearly, we are unable to measure the phonon spectra as related to crystal directions; on the other hand, too little damage risks observing only small or no changes in the phonons. After the above irradiation, which calculations determined to be $\sim$ 0.15 dpa (displacements per atom), a sizeable change in the lattice parameter corresponding to an expansion of $\Delta a/a$ = +0.56 (2) \% was observed. Since the full widths at half maxima (FWHMs) were almost the same for the two films, in both the longitudinal and transverse directions, the damage was judged to be uniform across the 3000 \AA{} of the film, and the crystallinity remains almost intact. This can be compared with other experiments on bulk materials, \cite{Wiss2014} where the swelling of $\sim$ 0.7 \% corresponded to an irradiation of $\sim 5 \times 10^{17}$ $\alpha$ particles/g and the thermal conductivity was reduced after irradiation by $\sim$ 50\%.

\begin{figure}[htb]
\centering
\includegraphics[width=0.95\linewidth]{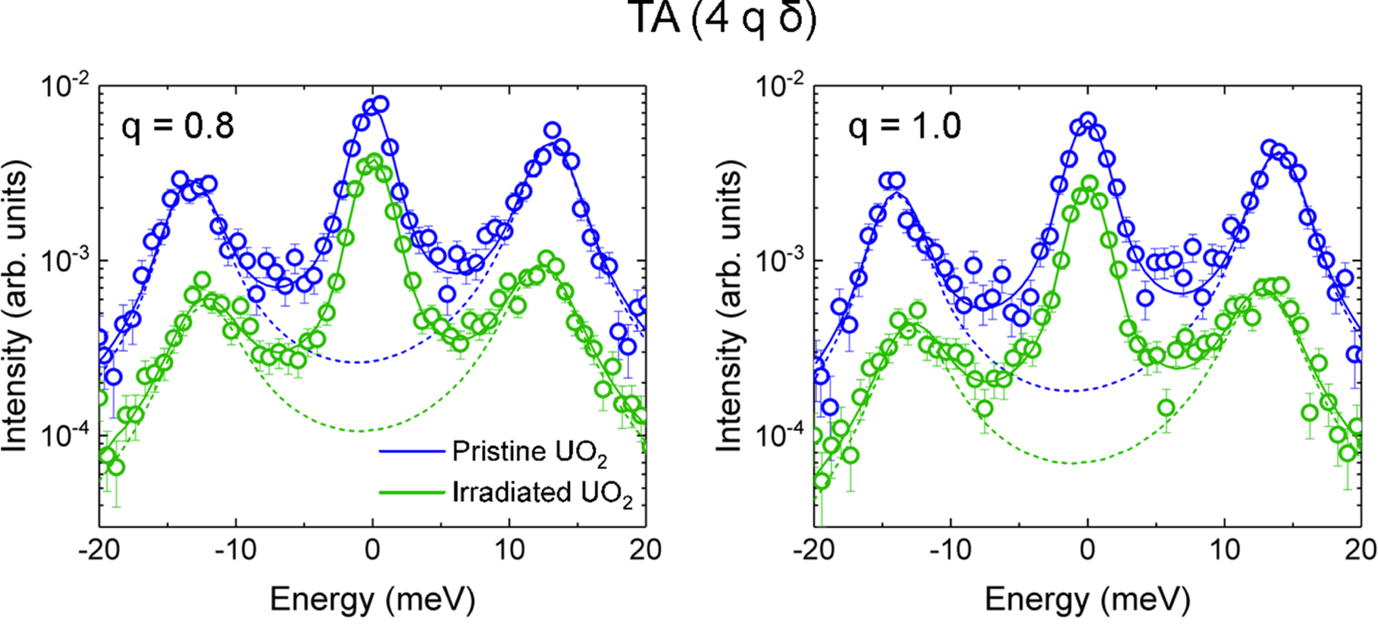}
\caption{Experimental data (at room temperature) for the $\mathrm{TA}[010]$ phonon from UO$_2$ thin films. The phonons are at $q = 0.8$ and 1.0 rlu. The pristine sample is shown in blue, and the irradiated sample in green. They have been displaced vertically for clarity. The dotted lines are the fits including a central elastic contribution (fixed to have the resolution function of 3 meV FWHM), and the Stokes and anti-Stokes phonons fitted as Lorentzians (weighted by the Bose factors) with a damped harmonic oscillator to give the width of the phonons. The L component of the phonon, here represented as $\delta$, is between 0.05 and 0.15 rlu.
Taken from Ref. \cite{Rennie2018b}.}
\label{fig5_4_2}
\end{figure}

The experiments at the ESRF ID28 facility \cite{ID28} used grazing incidence inelastic X-ray scattering to determine the phonon dispersion curves. Technically this is challenging as to achieve $\sim$ 3 meV resolution, which we need to be able to determine any broadening of the phonons, the incident energy used at the instrument is 17.794 keV, which uses the Si (999) reflections for the analyser. By chance, this energy is close to the U $L_3$ absorption edge of 17.166 keV. At this energy the 1/e penetration of the photon beam in UO$_2$ is $\sim$ 10 $\mu$m. The angle of incidence was in all cases $<$ 1 deg, but the film was slightly tilted to give a further penetration of the beam of $\sim$ 1500 \AA{}. In spite of this, the total mass of the UO$_2$ illuminated by the X-ray beam can be estimated at $\sim$ 100 ng. The experimental data for two points close to the zone-boundary of the $\mathrm{TA}[010]$ phonon (the {\bf X} point) are shown in Fig. \ref{fig5_4_2}. The phonons are well defined and their frequencies as a function of wave-vector {\bf q} do not change compared to the bulk experiments \cite{Dolling1965,Pang2013a}, and their widths can be measured. Notice that the central peak is {\it more} intense (compared to the phonons) for the irradiated (green) samples. This is because of the additional {\it elastic} diffuse scattering from the defects in the case of the irradiated sample.

\begin{figure}[htb]
\centering
\includegraphics[width=0.55\linewidth]{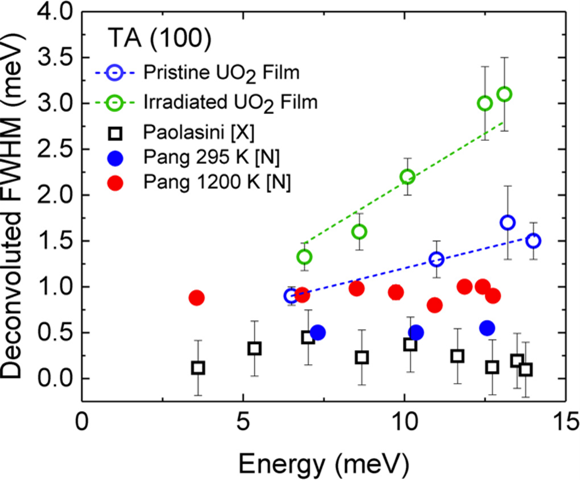}
\caption{Values of the FWHM deduced from analysis of the
phonons measured in the $\mathrm{TA}[100]$ direction after using the resolution function. The values tabulated in Ref. \cite{Pang2014a} by neutron inelastic scattering are shown as blue (295 K) and red (1200 K) solid circles. Our values using inelastic X-ray inelastic scattering are shown as blue (pristine) and green (irradiated) open circles. Values determined from a small bulk UO$_2$ crystal at room temperature determined on the same X-ray instrument are shown as open black squares \cite{Paolasini2021}.}
\label{fig5_4_3}
\end{figure}

The final results are shown in Fig. \ref{fig5_4_3}. Appreciable broadening of the TA phonons occurs for the irradiated sample, and it is observed to be a function of the phonon energy. Although the pristine films do appear to have slightly larger widths than found in bulk experiments, the change with irradiation is unmistakable. A similar effect is found for the longitudinal modes, and the net result in terms of thermal conductivity can be judged to be about a factor of two reduction on irradiation.

Using such thin films Weisensee {\it et al.} \cite{Weisensee2013} have irradiated them with 2 MeV Ar$^+$ ions (at room temperature) and by using a method of time-domain thermal reflectance have shown that the thermal conductivity is reduced by $\sim$ 50\%. This is in agreement with the above studies. Unfortunately, for all the power of the X-ray technique, there is still the subject of sensitivity to the oxygen atoms. Because of the large mass difference between uranium and oxygen, the acoustic modes tend to be dominated by U motions, and the optic branches dominated by oxygen motions. In this case, the low number of electrons around the oxygen nucleus means that the X-ray technique is not sensitive to the optic modes. A good example is the study by X-rays of the phonon dispersion curves in NpO$_2$ \cite{Maldonado2016}, which was done with a small bulk sample and not with the grazing incidence technique.

As Pang {\it et al.} showed \cite{Pang2013a}, the LO$_1$ optic mode is one of the main carries of heat in the UO$_2$ system, but this was not accessible in our experiments, so that a total estimate of the thermal conductivity of the irradiated sample could not be made.

\subsubsection{Studies of irradiated films} \label{s:UO2irrad}

Another series of experiments was conducted during work for a PhD degree at Cambridge University by A. J. Popel and collaborators. Thin films of UO$_2$ deposited on YSZ substrates in the three principal directions were irradiated at GANIL (Caen, France) with 110 MeV $^{238}$U$^{31+}$ for irradiation up to about $5 \times 10^{12}$ ions/cm$^2$. The first paper \cite{Popel2016} showed that this dose was not sufficient to destroy the crystallinity of the films. The most stable (least radiation damage) was with the (111) plane of UO$_2$. There is no mixing of U and Zr at the substrate interface. A second series of UO$_2$ thin films deposited on LSAT was irradiated, again at GANIL, with 92 MeV $^{129}$Xe$^{23+}$ ions to a fluence of $4.8 \times 10^{15}$ ions/cm$^2$ \cite{Popel2016a}. In this case considerable damage was done to the UO$_2$ films. The surface roughness was increased and there was evidence of the “cauliflower-like” structures, which are a feature of high-dose irradiations. In addition, aluminium was found to have diffused from the substrates into the UO$_2$ films, a phenomenon already observed by Strehle {\it et al.} \cite{Strehle2012}. In both these studies film thicknesses were $\sim$ 1000 – 1500 \AA{} and the radiations penetrated the whole film thickness.

A third paper \cite{Teterin2016} was published using the same films as described above but focusing on analyses with core X-ray photoemission spectroscopy (XPS) using the O 1$s$ line at $\sim$ 503 eV and the U 4$f$ spin-orbit split lines at $\sim$380 and $\sim$ 391 eV. The position of these transitions (and the accompanying satellites) allows an estimate to be made of the excess oxygen in the sample, i.e. the value of $x$ in UO$_{2+x}$. After irradiation these values ranged from 0.07 to 0.11 on LSAT films (Xe irradiation) and 0.17 to 0.23 on the YSZ substrates (with U irradiation). Thus, despite the larger fluence (three orders of magnitude) with Xe ions, and the clear damage to the lattice structure, the excess oxygen was actually greater for the U irradiation of films on YSZ. Whether the substrate has any role in this process is one of the questions raised by this work.

A fourth paper \cite{Popel2017} examined the Xe irradiated films for dissolution in water. The experiments found that the irradiated samples showed a decrease in the amount of dissolved uranium, as compared to the corresponding unirradiated samples. This somewhat counter-intuitive result was ascribed to irradiation-induced chemical mixing of the UO$_2$ films with the substrate elements, which resulted in stabilization of the UO$_2$ matrix and increased its aqueous durability. The last paper in this series \cite{Maslakov2018} returned to the analysis by XPS (and also the valence band UPS) and used a 1500 \AA{} UO$_2$ (111) film deposited on YSZ. The film was then irradiated with $^{40}$Ar$^+$ ions for various times and then annealed at various temperatures. On the basis of the measured spectral parameters one can conclude that the annealed film with $x$ = 0.12 contains mostly the U$^{4+}$ and U$^{5+}$ ions with some small amount of U$^{6+}$ ions also present. Embedding Xe into UO$_2$ films has also been reported by Usov {\it et al.} \cite{Usov2014}.

\subsubsection{Use of films to assure high-quality surfaces} \label{s:UO2surfacescience}

As mentioned above, in the case of experiments that are very surface sensitive, it sometimes is easier to use a thin film than go to the lengths of polishing and annealing a single crystal surface. It is known from a number of studies (e$.$g$.$ \cite{Watson2000}) that when exposed to air UO$_2$ acquires a layer of $\sim$ 30 \AA{} where the surface is UO$_{2+x}$, and this is quite independent of whether the surface is polar or non-polar. No doubt a careful examination of such effects would show a similar characteristic as those measured by Stubbs {\it et al.} \cite{Stubbs2015,Stubbs2017}; however, if the measuring technique is extremely surface sensitive, then clearly the first few layers of a sample exposed to air do not represent stoichiometric UO$_2$. Such effects can be minimised by having the sample under vacuum, such as in photoemission experiments, but may not be completely eliminated.

\begin{figure}[htb]
\centering
\includegraphics[width=0.55\linewidth]{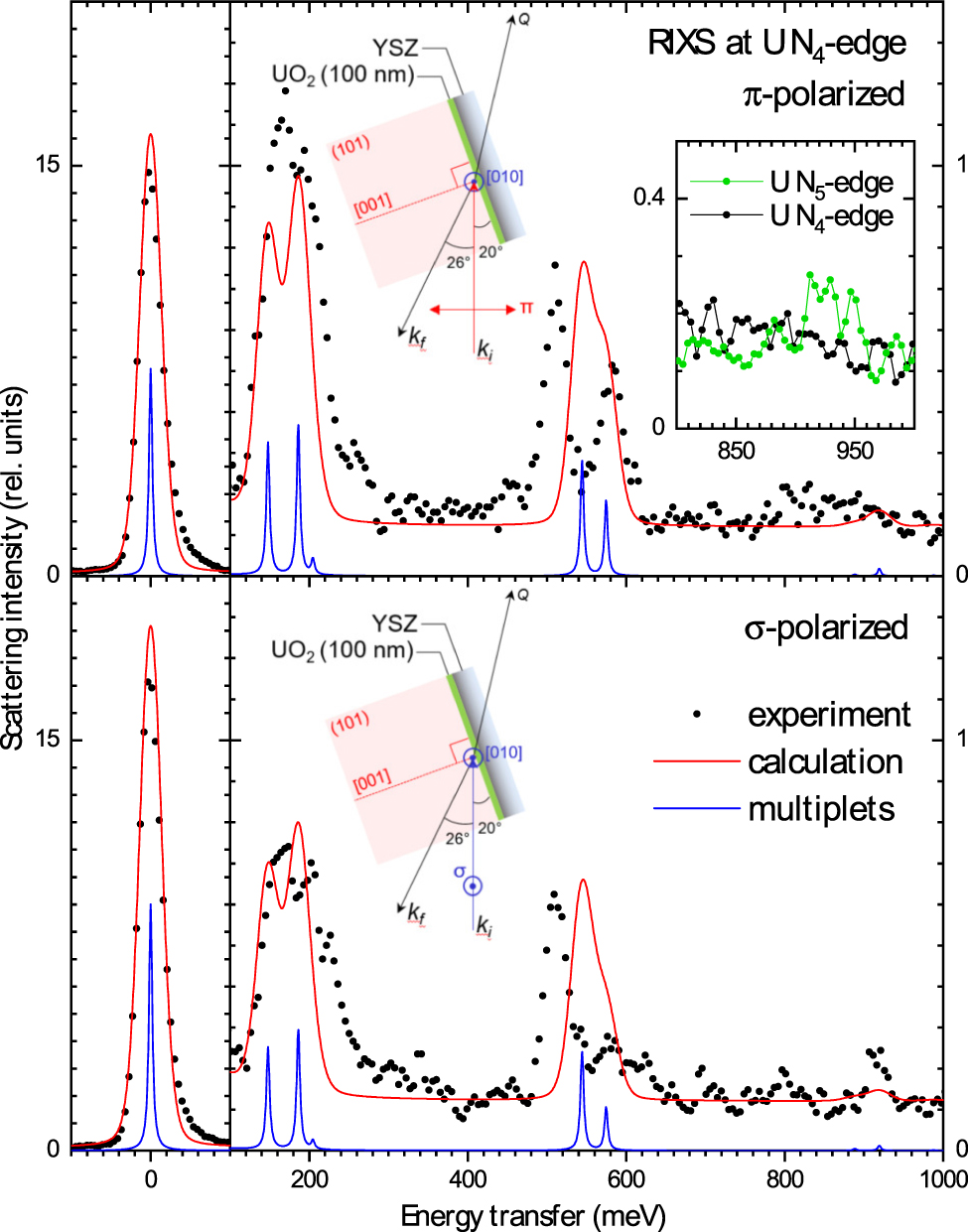}
\caption{RIXS spectra of UO$_2$ at the U $N_4$-edge (dots). Data were taken at 20$^{\circ}$ incident and 154$^{\circ}$ scattering angle with photon energy of 778 eV for linear horizontal $\pi$ (top) and vertical $\sigma$ polarization (bottom) as shown in the insets. The sample temperature was T = 15 K. The red lines show the calculated broadened RIXS spectrum for the U$^{4+}$ 5$f^2$ configuration for the CF parameters. The spectra recorded at the U $N_5$-edge in the energy transfer range 0.8–1.0 eV are shown in the inset of the upper panel ($\pi$-polarization channel).  Spectra are calculated with a Lorentzian width of 5 meV (FWHM) and convoluted with a Gaussian of 30 meV (FWHM). The blue lines show the underlying multiplet peaks (with no line broadening) of the CF excitations.  Taken from Ref. \cite{Lander2021}. }
\label{fig5_4_4}
\end{figure}

Thin epitaxial films give a simple method to eliminate such effects. The films are prepared at high temperature in vacuum, so that if they are removed in a “vacuum suitcase” they may be loaded into another vacuum chamber without exposure to air. This method was used in some recent experiments \cite{Lander2021} using soft X-ray spectroscopy, as shown in Fig. \ref{fig5_4_4}. The incident X-ray energy was tuned to the $N_{4,5}$ edges of uranium (779 and 737 eV, respectively), which have a wavelength of $\sim$ 16 \AA{}. At such an energy the penetration depth of the X-ray beam is certainly only a few 10’s of \AA{} at best, and the presence of any non-stoichiometric UO$_{2+x}$ and/or roughness at the UO$_2$ surface may strongly attenuate the outgoing X-rays. The technique, known as Resonant Inelastic X-ray Scattering (RIXS) is able to resolve the high-energy multiplets, a question that has been discussed about UO$_2$ (and other actinide systems) for the last 50 years. This work has now been followed by experiments on epitaxial films of U$_3$O$_8$ and UN \cite{Bright2023}. More information on this (and other) X-ray techniques may be found in Caciuffo {\it et al.} (2022) \cite{Caciuffo2022}.

\subsubsection{Studies of higher oxides} \label{s:higheroxides}
The UO$_2$ – UO$_3$ phase diagram consists of a number of different phases. Good summaries of the defect structures that appear up to UO$_{2+x}$, where $x$ $\sim$ 0.20, complementing the pioneering 1963 paper by B. T. M. Willis (1963) \cite{Willis1963}, are those of Garrido {\it et al.} (2006) \cite{Garrido2006}, Rousseau {\it et al.} (2006) \cite{Rousseau2006}, Wang {\it et al.} (2014) \cite{Wang2014}, and J. M. Elorrieta {\it et al.} (2016) \cite{Elorrieta2016}.

Distinct phases exist in UO$_{2+x}$ at $x$ = 0.25 (U$_4$O$_9$), $x$ = 0.33 (U$_3$O$_7$), $x$ = 0.67 (U$_3$O$_8$), and $x$ = 1 (UO$_3$), with some having more than one allotrope. There is a copious literature on these systems, some of it going back to the 1960's . In terms of films, as shown in Table II. Only a few have been grown as epitaxial films. Including UO$_2$, these are U$_3$O$_8$ and UO$_3$ \cite{Strehle2012,Enriquez2020,Kruk2021}.

\vspace{11pt}

{\bf(i) U$_3$O$_8$}

U$_3$O$_8$ is particularly interesting as it is known to have a mixture of U$^{5+}$ and U$^{6+}$ with twice as much of the former compared to the latter. The structure was first solved in 1964 by Loopstra \cite{Loopstra1964}. Recently, two papers reporting studies of polycrystalline U$_3$O$_8$ have been reported, and these raise a number of questions that may be answered with using epitaxial films. No bulk crystals exist, which is true for all $x$ $>$ 0 in the U–O phase diagram, so epitaxial films represent the only single-crystal samples available. We have already reported the electronic structure of U$_3$O$_8$ in \cite{Bright2023}, and this (and associated XPS core-hole spectroscopy) is consistent with the model of U$^{5+}$/U$^{6+}$ mentioned above \cite{Leinders2017}.

Enriquez {\it et al.} (2020) \cite{Enriquez2020} have used thin films to measure the optical band gaps and report UO$_2$ films have a direct band gap of 2.61 eV, whereas epitaxial $\alpha$-U$_3$O$_8$ and $\alpha$-UO$_3$ films exhibit indirect band gaps of 1.89 and 2.26 eV, respectively. This value for UO$_2$ seems somewhat higher than the accepted value of $\sim$ 2.1 eV, but we agree that the indirect band gap for U$_3$O$_8$ is $\sim$ 1.8 eV \cite{Bright2023}.

\vspace{11pt}

{\bf(ii)	Transition UO$_2$ to U$_3$O$_8$}

Another key question is how the transition occurs between UO$_2$ and U$_3$O$_8$. The paper by Allen \& Holmes (1995) \cite{Allen1995} proposed that the transformation started from the (111) face of UO$_2$ and proceeded in a number of steps, so that the short $c$ (4.146 \AA{} at RT) of the orthorhombic structure of U$_3$O$_8$ resembles the (111) interplanar spacing of cubic UO$_2$ (3.138 \AA{}) after suitable relaxation.

\begin{figure}[htb]
\centering
\includegraphics[width=\linewidth]{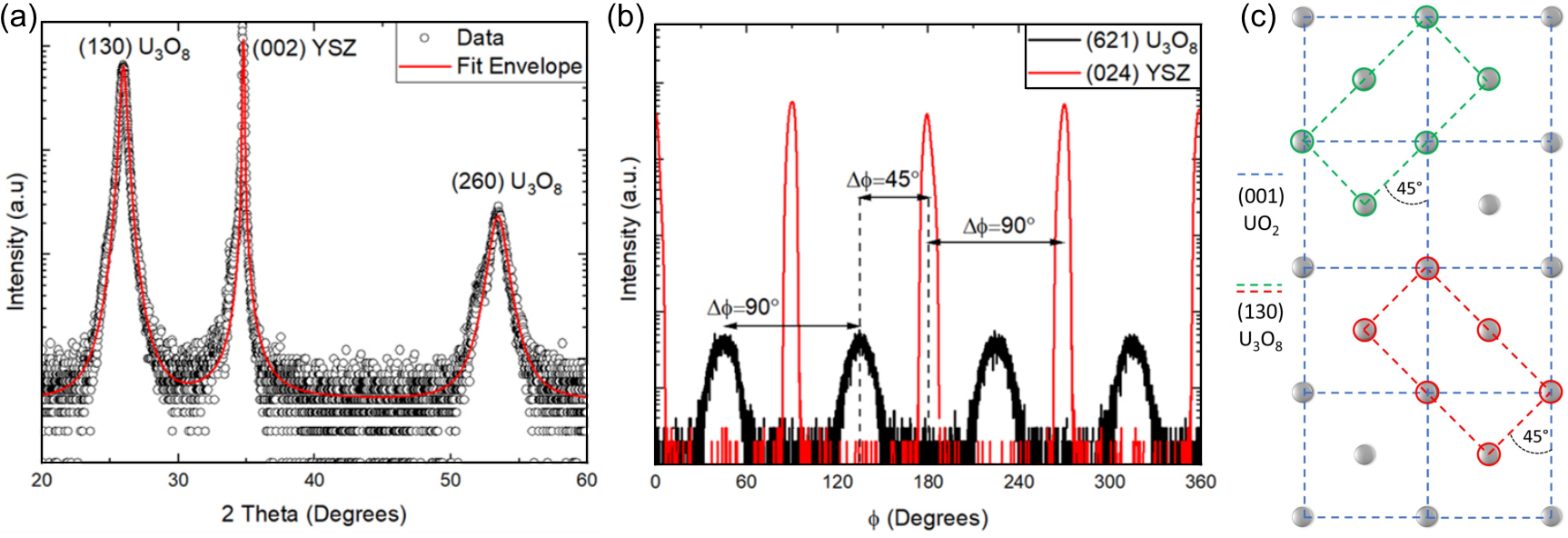}
\caption{(a) Specular scan from a 500 \AA{} UO$_2$ (001) film after full oxidation in an O$_2$ atmosphere showing the predominant reflections are (130) and the second order (260) reflection of U$_3$O$_8$. (b) $\phi$-scan showing that the reflections from U$_3$O$_8$ occur at 90$^{\circ}$ and are shifted by 45$^{\circ}$ from the substrate cube axes. (c) Atoms of uranium are represented by the gray spheres. Blue dash lines show the cubic structure of the UO$_2$ (001) phase. The red and green lines represent possible arrangement of the (130) plane of U$_3$O$_8$ on top of UO$_2$. This shows the possible domains existing in the U$_3$O$_8$ highly textured sample. Taken from \cite{Wasikthesis}.}
\label{fig5_4_5}
\end{figure}

Experiments by J. Wasik (University of Bristol thesis, 2021) \cite{Wasikthesis} show that this idea is incorrect. Instead, oxidation starts from the (100) face of cubic UO$_2$ and distortions are made to produce a (130) plane of U$_3$O$_8$, as shown in Fig.~ \ref{fig5_4_5}. The UO$_2$ films were grown on YSZ for the experiments reported in this thesis because all three principal orientations can be grown on YSZ \cite{Strehle2012}. A further point from these experiments is that no similar observation could be made starting with either (110) or (111) UO$_2$ thin films. In both cases the oxidation resulted in the product peeling off the substrate, and no phases could be identified.

The process of going from one epitaxial symmetry to another is called “topotaxy” \cite{Shannon1964} and quite unusual, although important in semiconductor technology \cite{Mairoser2015}.  Note that there are reports of preparing polycrystalline U$_3$O$_8$ films in which the authors note a strong texture with the (130) diffraction line being the preferred direction.

\section{Uranium Hydrides, Nitrides, and Silicides} \label{s:Uides}

\subsection{Introduction} \label{s:Uidesintro}

Along with the element (Sec. \ref{s:metals}), and the oxides (Sec. \ref{s:oxides}), the most work on bulk samples incorporating uranium has been done on U compounds, especially those such as UPt$_3$ or URuSi$_2$, which are representative of the well-known heavy-fermion superconductors. For the moment, except for the notable exception of the early work on UPd$_2$Al$_3$ and UNi$_2$Al$_3$ at Mainz \cite{Huth1994,Jourdan1999,Dressel2002,Jourdan2004,Foerster2007,Bernhoeft1998} starting in the mid 1990's, we are unaware of any work with epitaxial films on U heavy-fermions. This will certainly change, and we will discuss some of the possibilities opened up by fabricating such materials as epitaxial films in the Conclusions (Sec. \ref{s:conclusions}).

However, the fabrication and exploration of thin films of the alloys, hydrides, nitrides, and silicides already represents a large and important field. Both nitrides and silicides have been discussed in terms of advanced fuels, primarily because both have much higher thermal conductivities than the presently used UO$_2$. The nitrides, particularly, UN, have been the object of much basic research since the 1960's.

A great deal of work has been reported on the uranium hydrides. As discussed later, UH$_3$ is a ferromagnet, and has been of considerable interest over the years. Uranium-hydrides have also been thoroughly investigated as the reaction between these two elements is very strong, exothermic, and constitutes a considerable safety challenge if, for example, hydrogen is produced inside a vessel containing U metal. We shall discuss an experiment using bilayers of U metal and oxide that starts to address these reactions.

\subsection{Growth of thin films of hydrides, nitrides, and silicides} \label{s:Uidesgrowth}
The growth of U-H and U-N phases has been achieved by reactive growth with the presence of the relevant gas in the chamber during U deposition. The U-Si, as well as a range of other materials, have been prepared by co-deposition. Table \ref{tab:comp} summarises the various thin films in these categories that have been grown to date.

\begin{landscape}
\begin{table}[]
\centering
\caption{Uranium compounds thin films (excluding oxides) that have been produced, with references of their first mention in publication.}
\label{tab:comp}
\begin{tabular}{|l|l|l|l|l|} \hline
Material         & Form            & Substrate                                        & Deposition method                & Reference          \\ \hline
UAs              & amorphous       & glass                                            & mag. co-sputtering          & \cite{mcguire1992}     \\
USb              & amorphous       & glass                                            & mag. co-sputtering          & \cite{Gambino1991}        \\
UBi              & amorphous       & glass                                            & mag. co-sputtering          & \cite{Gambino1991}        \\
USbMn            & amorphous       & glass                                            & mag. co-sputtering          & \cite{Gambino1991}        \\
USbCo            & amorphous       & glass                                            & mag. co-sputtering          & \cite{Gambino1991}        \\
UN               & poly            & glass                                            & reactive DC sputtering      & \cite{Black2001}          \\
U$_2$N$_{3+x}$   & poly            & glass                                            & reactive DC sputtering      & \cite{Black2001}          \\
UC$_2$           & \hkl(001)       & \hkl(001) YSZ                                    & polymer-assisted deposition & \cite{Scott2014}          \\
UN$_2$           & \hkl(111)       & \hkl(001) LaAlO$_3$                              & polymer-assisted deposition & \cite{Scott2014}          \\
UN               & \hkl(001)       & \hkl(1-102) Al$_2$O$_3$ with \hkl(001) Nb buffer & reactive DC mag. sputtering & \cite{Bright2018} \\
U$_2$N$_3$       & \hkl(001)       & \hkl(001) CaF$_2$                                & reactive DC mag. sputtering & \cite{Bright2018} \\
$\beta$-UH$_3$   & poly            & Si                                               & reactive sputter deposition & \cite{Gouder2004}                    \\
UH$_2$           & poly            & fused silica                                     & reactive sputter deposition & \cite{Havela2018}                    \\
UPd$_{2}$Al$_{3}$ & \hkl(001)      & \hkl(1-102) and \hkl(11-20) Al$_2$O$_3$            & MBE                       & \cite{Huth1993}          \\
UNi$_{2}$Al$_{3}$ & \hkl(100)      &  \hkl(001) MgAl$_{2}$O$_{4}$, \hkl(112) YAlO$_{3}$ & MBE                       & \cite{Jourdan2004} \\
U-Si see Table \ref{tab:silicides}  &                &                                                  & DC mag. co-sputtering       & \cite{Harding2023}     \\
\hline
\end{tabular}
\end{table}
\end{landscape}

\subsection{Science with thin films of hydrides, nitrides, and silicides} \label{s:Uidesscience}

\subsubsection{Thin films of uranium hydrides} \label{s:UH}

The hydrides of uranium have been of interest for a long time. The most interesting property of UH$_3$ (which exists in two structural forms) is that it becomes ferromagnetic at $\sim$174\,K. This was first discovered in Wroclaw, Poland in 1952 \cite{Troc1995}. This was particularly interesting at the time as it was the first material to be found ferromagnetic with two elements that were themselves non-magnetic. Of course, there were originally doubts whether some impurity was playing a role, but the later work \cite{Lin1956} showed that the ferromagnetism was intrinsic with a moment of $\sim$ 1.5 $\mu_{\mathrm{B}}$/U atom. Since then, much work has been done on this compound. The material was first prepared as thin films in 2004 \cite{Gouder2004} by depositing at room temperature on Si(111) wafers. Later, many more samples were produced, including alloys incorporating Zr and Mo, as discussed in the review articles published recently by L. Havela and  collaborators \cite{Havela2020, Havela2023}. The main reason for the thin films is that the properties of interest are deduced from the photoemission spectra, and these are sensitive just to the top 20 - 50 \AA{}. The films must not be heated above $\sim$ 350 K as the compound decomposes. If exposed to air UH$_3$ catches fire, so it must be handled cautiously! Various attempts were made at Oxford, and then later at Bristol, to produce epitaxial films, but they have been unsuccessful to date. Epitaxial growth usually requires high temperatures to allow the atoms to have considerable energy, but this is not possible in the case of UH$_3$, as mentioned above.\\

{\bf (i) Stabilization of UH$_2$ }
\begin{figure}[htb]
\centering
\includegraphics[width=0.55\linewidth]{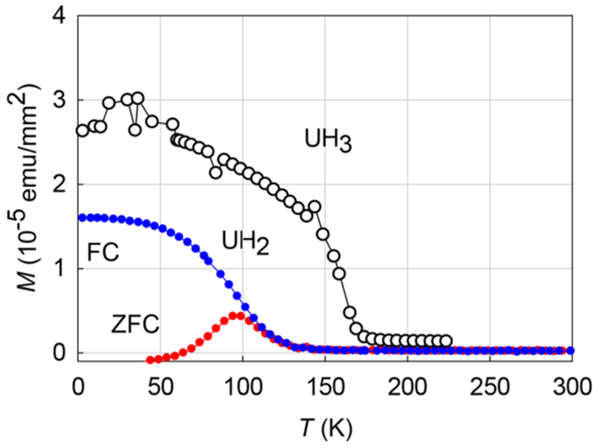}
\caption{Temperature dependence of the magnetization of a UH$_2$ film in the field-cooled (FC) and zero-field-cooled (ZFC) mode. The data are compared with the UH$_3$ film (FC mode only displayed)
Taken from Ref. \cite{Havela2018}.} \label{fig6_1}
\end{figure}

An interesting development \cite{Havela2018} that occurred in the UH$_3$ saga took place when the authors used Si(100) substrate with a = 5.43 \AA{}. This value is close to the lattice parameters of the known dihydrides NpH$_{2}$ ($a$ = 5.34 \AA{}) and PuH$_2$ ($a$ = 5.36 \AA{}). The idea was then to produce UH$_2$, which has not been reported as a stable compound. By cooling the substrate to $T = 177$ K, they managed to produce a film ($\sim$ 4000 \AA{}) of UH$_2$ with a lattice parameter $a$ = 5.36 \AA{}. Surprisingly, the film appears to be polycrystalline with a random domain distribution with domain sizes $\sim$ $500$ - $1000$ \AA{}. The sample was capped with 30 \AA{} of Mo. The absence of any preferred orientation suggests that the substrate-film interaction is relatively small, but certainly at these temperatures the chances of attaining epitaxy would be small. UH$_2$ exhibited a somewhat lower $T_{\mathrm{C}}$ than UH$_3$, but, like UH$_3$, it shows relatively wide hysteresis loops, suggesting a strongly anisotropic ferromagnet, despite the cubic structure \cite{Tereshina2023}.\\

{\bf (ii) Experiments with hydrogen on bilayers of U and UO$_2$ }

As discussed earlier, UH$_3$ is a dangerous material. When U metal is stored in the presence of either moisture or organic material in sealed containers hydrogen can be produced over long periods \cite{Harker2006}. The hydrogen reacts with the U producing pyrophoric finely divided radioactive powder in the container. A major safety incident can occur if the containers are then opened and a large amount of UH$_3$ is present \cite{Orr2016}. The process of hydrogenation of U is obviously a complex reaction, and will be dependent on the pressure of hydrogen, and the form of the uranium and its temperature. Because we have a method to make epitaxial films of U metal in its $\alpha$ form (see Sec. \ref{s:metals})  an experiment was undertaken to use such a film coated with a layer of UO$_2$ and then allow the hydrogen to react in-situ at the synchrotron beamlime to monitor the diffraction pattern as a function of time and temperature \cite{Darnbrough2018}. The UO$_2$ surface layer does not make any epitaxy with the underlying U, but there will be a preferred orientation with the (111) reflection dominant. Studies were performed on a nominal 1000 \AA{} metal layer with a deposited UO$_2$ layer of $\sim$ 200\,\AA{}, and also a sample left to oxidise in air over two days, forming a layer of $\sim$ 600 \AA{} oxide.
\begin{figure}[htb]
\centering
\includegraphics[width=0.8\linewidth]{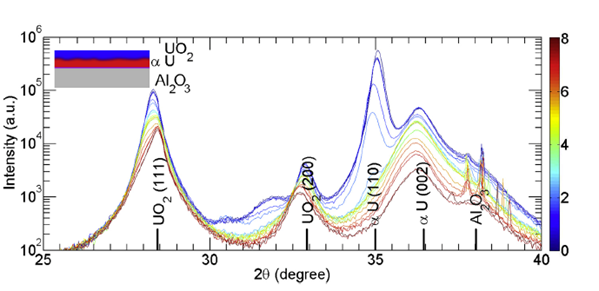}
\caption{For the air-grown oxide sample. A plot of intensity vs $2\theta$ with the line colour representing exposure to hydrogen dose of the high-angle X-ray diffraction taken with increasing cumulative dose. The colour coding indicated on the right-hand side corresponds to the total accumulative dose (min.bar). Before exposure to air the nominal thickness of U metal was $\sim$1000 \AA{}. After reaction with air, we estimate that the U metal reduced to 800 \AA{} $ $, with some $\sim$600 \AA{}$ $ of oxide. The oxide/metal interface is not smooth. The inset in the upper left is a schematic of the sample. Taken from Ref. \cite{Darnbrough2018} } \label{fig6_2}
\end{figure}
A number of points are clear from the diffraction patterns (Fig. \ref{fig6_2}) without further analysis.\\

(a) The main orientation of the U epitaxial layer has the (110) orientation, but there is a $\sim$ 10 \% formation of (002) grains, which are significantly smaller in size (probably of the order of 200 \AA{}) than those oriented in the (110) direction, which clearly extend over the whole thickness of the U film. However, the (110) grains are consumed much faster with hydrogen exposure – a property that was found throughout the experiment and indicating a significant anisotropy in the consumption of the metal.

(b) The peaks just below the position of the UO$_2$ (200) are coming from higher oxides in the case of oxidizing by air, and these are rapidly removed by exposure to hydrogen – as expected.

(c) Both UO$_2$ peaks decrease in intensity; they also move in opposite directions. The UO$_2$ (111) $d\mbox{-spacing}$ decreases and the UO$_2$ (200) $d\mbox{-spacing}$ increases with hydrogen doping. Of course, these observations are coming from different particles since the experiment is sampling only those with their scattering vectors exactly along the specular direction. However, this implies a directional effect of the hydrogen passing through UO$_2$, previously it was thought the UO$_2$ was not changed by H-exposure, but that is clearly incorrect. The change of intensity is not easy to explain, but may be related to the change of particle size.

(d) One mystery was the absence of any intensity from the UH$_3$ that must be forming with the clear consumption of U metal. The main diffraction line one would expect to observe would be at $2\theta$ $\sim$ 30$^{\circ}$, and there is no sign of a peak at this value at the end of the H exposure. This mystery was partly resolved by EELS work on a separate sample where the UH$_3$ was identified with defects and grain boundaries, and was almost certainly either nanocrystalline or amorphous.

These experiments have raised a number of questions about what is clearly a complicated process. The simplest method of hydrogenation would be that the hydrogen arrives from the UO$_2$ layer and starts consumption of the U from the top towards the bottom of the U film. However, such a process would lead to the particles with (110) getting smaller, and hence the peak widths increasing. This is not observed, and the EELS experiments also rule this out. A more likely model involves grain-boundary corrosion with the interface moving laterally into the grains. A consequence of the consumption of U is a proportional increase in the $d$-spacing of U(110), and to a lesser extent the U(002), in the direction of film thickness.

The experiment described here \cite{Darnbrough2018} was a first attempt at a complicated problem. More work needs to be done, including, for example, reflectivity studies and using better samples for the (110) and (002) $\alpha$-U orientations. In addition, neutrons could be very effective in this study, especially for locating the hydrogen as there is a strong contrast with neutrons between hydrogen and deuterium.

\subsubsection{Magnetism and electronic structure of uranium nitride thin films} \label{s:UNmag}

Uranium mononitride (UN) has been of interest for many years both to applied projects, as well as fundamental research.
On the applied side UN there is increased interest in the last two decades \cite{Zinkle2014}, in using UN as an advanced technology nuclear fuel to replace UO$_2$.
Compared to UO$_2$, UN has a higher U density (thus enabling a lower enrichment to be used), has a better thermal conductivity, and an equally high melting point (the last two affecting safety).
However, it has two disadvantages, a high reactivity with water and oxygen above 200 $^\circ$C, and the large cross section of $^{14}$N implies that there will a substantial amount of $^{14}$C produced.

On the fundamental side, UN is part of a large group of actinide pnictides that have the simple  {\it fcc} rocksalt structure, and many experiments on these materials have been reported since the 1960s.
There is still a debate over the electronic structure of UN.
A recent review \cite{Troc2016} advances the case that UN is a mixed system, with the 5$f$ electrons partly localized and partly itinerant.
This is not in agreement with earlier work using angular-resolved photoemission \cite{Fujimori2012} or with the results from neutron inelastic scattering \cite{Holden1984}, where broad excitations and no crystal-field levels were reported; both these experiments suggest an itinerant model would be more appropriate.
UN is known to be antiferromagnetic at 53 K from the work of Curry in 1965 \cite{Curry1965} with an ordered moment of only 0.75(10) $\mu_{\mathrm{B}}$.
The effective magnetic moment above T$_{\mathrm{N}}$ is in the range $\sim$ 2.7 $\mu_{\mathrm{B}}$.
Recent resonant inelastic X-ray scattering (RIXS) experiments show that UN cannot be described in terms of a localized $5f^3$ configuration, and that a band description is certainly more appropriate, in agreement with the ARPES and neutron experiments \cite{Bright2023}.

Thin films were first reported from ITU, Karlsruhe, in a paper by Black {\it et al.} \cite{Black2001}.
Using reactive sputter deposition onto glass, they showed that the stoichiometry of the films deposited, measured by XPS, could be varied by changing the N$_2$ partial pressure.
Structural analysis of these films showed  preferred orientation in the \hkl{111} direction with an average grain size of 170 \AA{} for the films deposited at room temperature, and somewhat larger for films deposited at 400 $^\circ$C \cite{Rafaja2005}.
Other properties also changed with deposition temperature, with the strength of preferred orientation, the residual stress and the density of structural defects all decreasing with increasing temperature.
The magnetic studies presented some puzzling results, and there was no clear sign of the AF transition in susceptibility measurements.
The long-range AF behaviour of crystalline UN is replaced by a ferromagnetic-cluster glass behaviour resulting from a defected antiferromagnetism in the films deposited at higher temperatures, with the highly disturbed thin films (low temperature deposition) exhibiting weak Pauli paramagnetism.
Similar experiments were done on US films (US has the same crystal structure as UN, and in the bulk is a ferromagnet at 177 K), and some suppression of the $T_{\mathrm{C}}$ was observed \cite{Havela2006}, but less than observed for UN.

The first {\it epitaxial} single-crystal UN thin films, fabricated by E. Lawrence Bright {\it et al.} \cite{Bright2018}, used a \hkl(001) Nb buffer on a (1-102) Al$_2$O$_3$ substrate, with the Nb acting as a physical and chemical buffer to stabilise (001) UN.

These UN films were used for an attempt to find the crystallographic distortion that has been controversial at the AF ordering temperature $T_{\mathrm{N}}$. In the original AF model of UN, Curry assumed that the ordering was of type-I with ferromagnetic layers of atoms arranged in a + – + – arrangement, the so-called 1{\bf k} arrangement, with the ferromagnet moments perpendicular to the layers.
Since the overall symmetry is cubic, this implies that any cube axis can be the direction of the moments, so that there are three clearly different spatial domains. Each domain will have tetragonal symmetry when the U moment orders, so there should be a magnetostrictive distortion at T$_{\mathrm{N}}$. This should result in a clear splitting of the reflections from the different domains. Marples {\it et al.} \cite{Marples1975} claimed to have found such a distortion, but a careful examination of their paper shows that they did not observe a finite splitting of the peaks at high angle, but simply a broadening of the peaks. Whereas this might indicate a distortion, it could also be a strain effect. In a re-examination of this effect the experiment \cite{Bright2019a} showed that the distortion, if present, is much smaller than suggested by Ref. \cite{Marples1975}, and more in line with the experiments reported by Knott {\it et al.} \cite{Knott1980}. However, strain effects were observed in the experiment. A 700 \AA{} film was used and the tensile strain was measured as $+ 20 \times 10^{-6}$. We show below in Fig. \ref{fig6_4} the nominal change of the lattice parameter at the magnetic ordering temperature. The AF ordering temperature ($T_{\mathrm{N}}$) is slightly lower in the film (46 K) than reported in the bulk (53 K).

\begin{figure}[htb]
\centering
\includegraphics[width=0.45\linewidth]{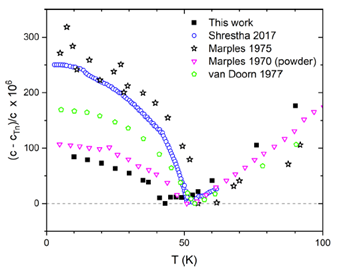}
\caption{Relative variation in the lattice parameter as a function of temperature around $T_{\mathrm{N}}$. Measurements have been made with X-rays by Marples and the work reported, and by strain gauge techniques by Shrestha and van Doorn. Taken from Ref. \cite{Bright2019a}. } \label{fig6_4}
\end{figure}

The point of this figure is to show that although the expansion of the lattice parameter at low temperature is clearly present, the magnitude of this effect is sample dependent and can vary by almost a factor of three between different samples. This expansion is certainly a property of UN, but its magnitude is determined by the macro-strain properties of the individual samples. Marples {\it et al.} \cite{Marples1975} proposed the distortion at $T_{\mathrm{N}}$ gave a strain such that $2(c - a)/(c +a) = - 650 \times 10^{-6}$. Lawrence Bright {\it et al.} \cite{Bright2019a} have lowered that to $< |200 \times 10^{-6}|$ in agreement with Ref. \cite{Knott1980}. With the absence of such a distortion there remains some doubt over the magnetic structure of UN. An alternative interpretation would be that the structure is a 3{\bf k} type-I structure as found in USb \cite{Magnani2010}. In such a system the symmetry below $T_{\mathrm{N}}$ is cubic. Further work on this is warranted; however, neutron experiments such as Ref. \cite{Magnani2010} require large single crystals ($\sim$ 1 g) and cannot be performed on thin films.\\

(001) U$_2$N$_3$ was stabilised on (001) CaF$_2$.
The second experiment with synchrotron radiation was performed on the U$_2$N$_3$ epitaxial film. The crystal structure is known to be the bixbyite type of body-centered cubic, isostructural with Mn$_2$O$_3$ and also rare-earth systems such as Gd$_2$O$_3$. Earlier work on U$_2$N$_3$ was reported by Troc \cite{Troc1975} and showed that stoichiometric U$_2$N$_3$ has a lattice parameter of 10.70 \AA{}, but that with additional nitrogen this reduces to $\sim$ 10.60 \AA{} by about UN$_{1.80}$. At the same time the stoichiometric material (UN$_{1.50}$) orders antiferromagnetically at $\sim$ 90 K, and with added nitrogen the $T_{\mathrm{N}}$ reduces so that by UN$_{1.80}$ there is no ordering. There is no report of a successful neutron experiment finding the magnetic structure of U$_2$N$_3$, so that aspect is unknown.

There was some strain found in the U$_2$N$_3$ epitaxial film with the (001) orientation growth on CaF$_2$ substrates. The growth direction parameter was found as 10.80(1) \AA{}, whereas the in-plane parameters were 10.60(2) \AA{} (i.e. strain = + 1.8 \%). The atomic volume corresponds to a lattice parameter of 10.67 \AA{}, suggesting the films are close to stoichiometry. When the X-ray energy was tuned to 3.726 keV, which corresponds to the U $M_4$ resonance, extra peaks were found at the non-{\it bcc} Bragg peaks for temperatures below $T_{\mathrm{N}}$ = 73.5 K. This gives a simple {\bf q} = 1 AF wave-vector for the magnetic structure, as was found for the isostructural Yb$_2$O$_3$ \cite{Moon1968}. Determining the magnetic configuration is considerably more difficult. In the case of Yb$_2$O$_3$ the configuration is non-collinear, but the T$_{\mathrm{N}}$ in that case is 2.3 K, so the interactions are certainly stronger in U$_2$N$_3$ than the rare-earth materials, and probably involve more direct exchange interactions. In principle, the resonant magnetic intensities are related to the arrangement of moments, but the main problem is making reliable absorption corrections for the off-specular reflections when using a film of 2000 \AA{}. There are two independent sites for U atoms in U$_2$N$_3$, and there is interest in knowing the magnetic configuration. An attempt was made with neutron diffraction at the WISH instrument on the ISIS spallation source, but although both a film and a 1 g polycrystalline sample were used, no magnetic scattering was observed. This suggests the U moments are below $\sim$ 0.5 $\mu_{\mathrm{B}}$.

As discussed above, the crystal structure of UN is the simple rocksalt structure, that of U$_2$N$_3$ being the bixbyite structure, with two independent sites for the uranium atom (U1 and U2), whereas only one exists in UN. A number of methods, including photoemission experiments \cite{Bright2018}, have been used to estimate the valency of these materials, and for UN this is $\sim$ 3+, i.e. U(III), but for U$_2$N$_3$ the valency is higher. Such methods are not site selective, so leave open the question of the valency at each individual site. This is important as the U(VI) valent state is highly soluble in water, so if that is present in at least one of the sites of U$_2$N$_3$, this could explain its high corrosion rates discussed below \cite{Bright2019}.

The magnetic properties give one clue to the valency; for example, U(VI) has no 5$f$ electrons so cannot be magnetic. In an effort to extract further information on the valency and bonding of the two separate uranium sites 'Diffraction absorption experiments' were performed at the U $M_4$ edge on a number of Bragg reflections of the  {\it bcc} 2000 \AA{} U$_2$N$_3$ film.

\begin{figure}[htb]
\centering
\includegraphics[width=0.45\linewidth]{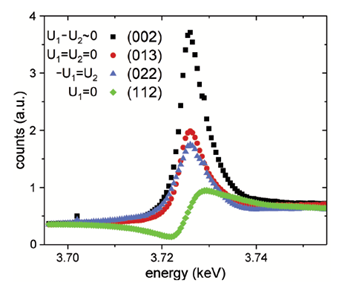}
\caption{Energy profiles of various Bragg reflections from the  {\it bcc} U$_2$N$_3$ film. The profiles are independent of temperature. A normal ``energy dispersive'' curve is shown in green from the (112) reflection. The other profiles represent reflections in which the strong (spherical) Thomson scattering from the two uranium sites cancels, or almost cancels. They represent so-called anisotropic resonant X-ray scattering and show that there is an aspherical charge density associated with the U$_2$ sites. Such a charge density is almost certainly due to covalency between the uranium 5$f$ electrons and the 2$p$ states of nitrogen. Adapted taken from \cite{Bright2019a}. } \label{fig6_5}
\end{figure}

The reflections have different contributions from the two independent uranium atoms, as the atomic sites have different symmetries. For strong Bragg reflections, in which the scattering from both the U$_1$ and U$_2$ atoms are in-phase, or one set is absent, the expected result is a dispersive curve that reflects the combined effect of both the real (f$_o$ + f') and imaginary (f'') parts of the uranium scattering factor. Such a curve is shown in the green curve in Fig. \ref{fig6_5} for the (112) reflection, in which only the U$_2$ atoms participate. (Scattering from nitrogen is neglected, as it is far weaker than that from uranium; in addition, there is no edge sensitive to nitrogen in the energy range covered).

However, for reflections in which the strong Thomson scattering (from the 86 core electrons of uranium) is reduced by the cancellation between the two separate uranium atoms, a very interesting profile is shown in Fig. \ref{fig6_5} that is precisely the energy profile at the $M_4$ edge of the imaginary part (f'') from the U atoms. This profile reflects the fact that around one of the U sites is an aspherical charge density, which involves the U 5$f$ electrons. For example, for the (013) reflection, which is forbidden and has no contribution from the Thomson (spherical) charge density, this aspherical part is the only contribution to the scattering intensity. Similarly, for the (002) and (022), in which the strong spherical charge density contributions almost cancel, the aspherical part is also observed. From the pattern of the intensities, it becomes clear that any aspherical contribution from the U$_1$ sites must be small, suggesting that these sites may possibly have the U(VI) valency, in which there are no occupied 5$f$ states.

This effect has been observed before, mainly at the {\it K} edge of the transition metals \cite{Collins2001}. However, at the {\it K} edge with the $d$ transition metals there is the possibility of both dipole and quadrupolar transitions, making the identification of the underlying physics complicated. For the U $M_4$ edge this ambiguity is removed; the transition is definitely of dipole symmetry illuminating an aspherical contribution from 5$f$ states. The local non-centrosymmetric coordination of this distribution around the U nucleus then couples to the imaginary scattering factor (f'') giving rise to scattered intensity, with a distinctive energy profile, at the Bragg position \cite{Kokubun2012}. In the case of U$_2$N$_3$ the effects are temperature independent, and so not related to the magnetic order at $\sim$75\,K. They are almost certainly induced by covalency, probably mixing between the U 5$f$ states and the nitrogen 2$p$ states.

In conclusion, these experiments strongly suggest that the U1 site may have a significantly higher valency, quite possibly U(VI), and this is responsible for the rapid corrosion rates of U$_2$N$_3$. In addition, these experiments have opened the way for more quantitative modelling in such systems, based on the observation of an aspherical 5$f$ charge distribution around the U2 nucleus.

\subsubsection{Reactivity studies of UN } \label{s:UNcorr}

\begin{figure}[htb]
\centering
\includegraphics[width=0.45\linewidth]{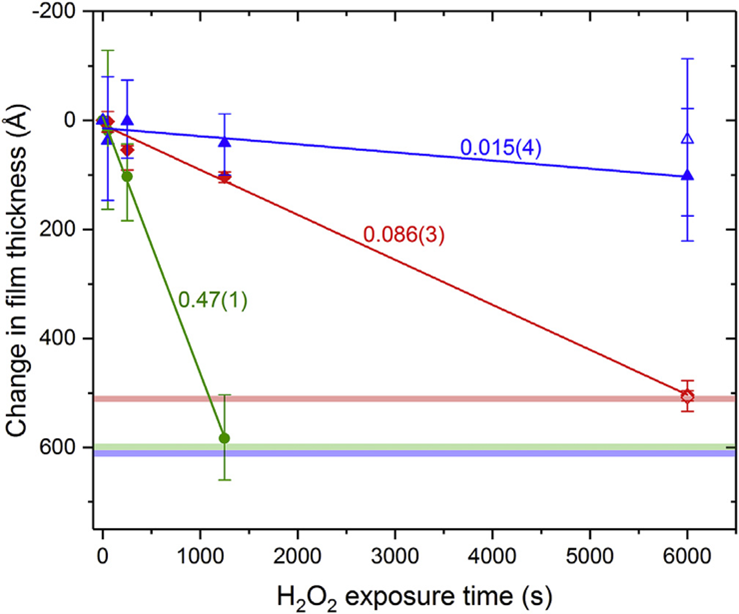}
\caption{Change in sample thickness of UO$_2$ (red diamond), UN (blue triangle), and U$_2$N$_3$ (green circle) as a function of exposure time to H$_2$O$_2$. Closed points show single exposures and open points show the cumulative time after 4 exposures of 1250 s. The pale lines denote the initial total sample thickness and the solid line shows a linear fit to the data, labelled with the gradient. Taken from Ref. \cite{Bright2019}. } \label{fig5_2_3a}
\end{figure}

Thin films have become relatively popular for investigating the oxidation and corrosion behaviour across the uranium-nitrogen phase diagram, with several studies coming from Bristol \cite{Bright2019,Bright2022}, and from the Science and Technology on Surface Physics and Chemistry Laboratory, Mianyang, China \cite{Liu2013,Lu2016,Luo2017,Luo2018}. Dissolution experiments comparing films of both UN and U$_2$N$_3$ of $\sim$ 600 \AA{} with similar UO$_2$ films in a 0.1 M H$_2$O$_2$ solution showed surprising results, shown in Fig. \ref{fig5_2_3a}. Reflectivity measurements allowed a measurement of the film thickness, and hence the dissolution rate \cite{Bright2019}. The dissolution results were unexpected, being equivalent to 0.033(1), 0.010(2), and 0.19(3) mg/cm$^2$/hr for UO$_2$, UN, and U$_2$N$_3$, respectively. Although in the literature the UN dissolution rate in water is actually greater than that for UO$_2$, when the effect of radiolysis is simulated using H$_2$O$_2$, the results are different.

Studies from Mianyang have looked at surface oxidation thin films with different N/U ratios produced by reactive RF magnetron sputtering. As UN does not accommodate stoichiometry changes, this produced mixed phase films. Auger electron spectroscopy (AES) of U, UN$_{0.23}$ (composed of U and UN), UN$_{0.68}$ (mainly UN) films before and after oxygen exposure found that an oxide layer of UO$_2$ formed on the surface \cite{Lu2016}. However, later work using XPS revealed that the oxidation of the UN and metallic U phases in the films is not a simple combination of two independent oxidation behaviors, but an interactive association \cite{Luo2018}.

Oxidation studies on UN$_{1.66}$ films found that UN$_x$O$_y$ oxides formed, both when investigated with AES \cite{Lu2016} and XPS \cite{Luo2018}. Further investigations on a UN$_{1.85}$ film found a three-layered oxidation surface structure, composed of uranium oxides (U$_4$O$_9$,UO$_{2+x}$ and UO$_2$), a U-N-O ternary compound layer, and an N-rich uranium nitride UN$_{1.85+x}$ layer. This layered structure is proposed to be responsible for the measured long-term stability of the surface oxide layer \cite{Luo2020}.

\begin{figure}[htb]
\centering
\includegraphics[width=0.85\linewidth]{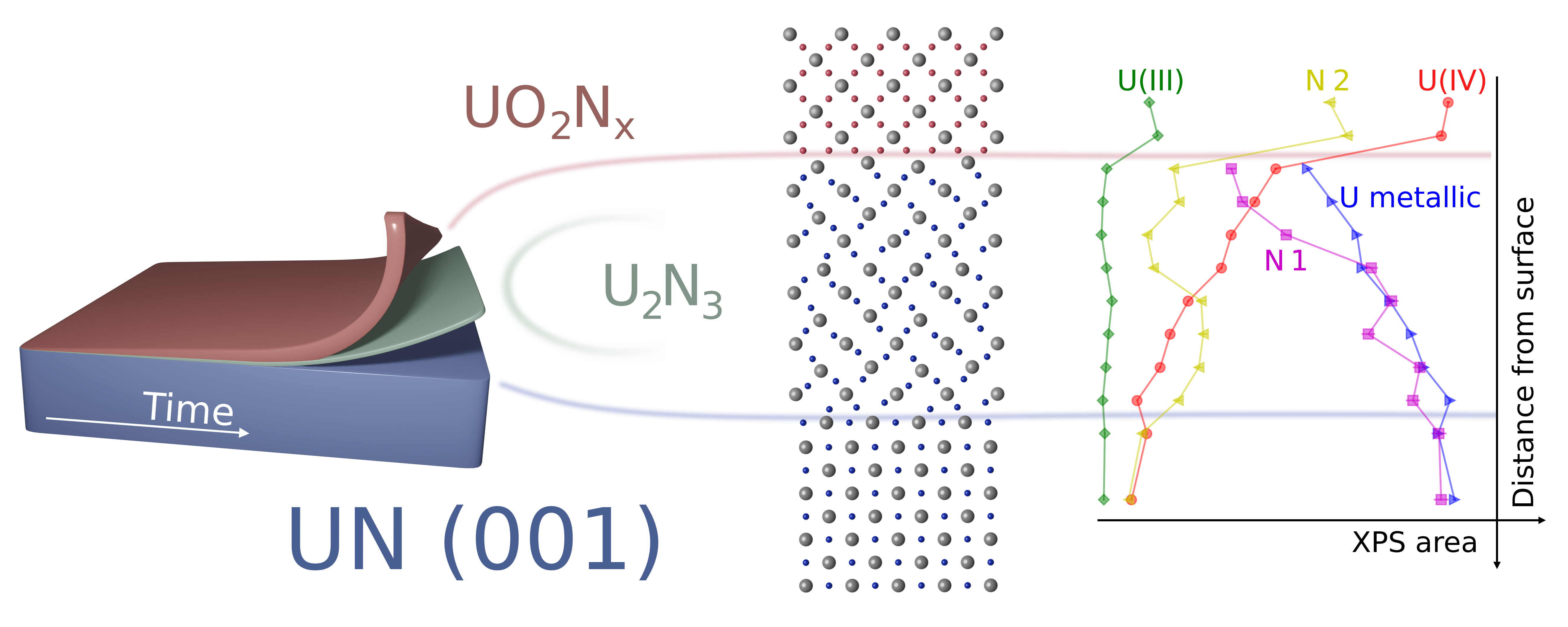}
\caption{Graphical abstract showing a summary of the study on the oxidation of a UN (001) surface showing (centre) the topotactic relationship between the film and surface layers, and . Figure taken from \cite{Bright2022}.} \label{fig5_2_3b}
\end{figure}

Epitaxial films grown by Lawrence Bright at Bristol \cite{Bright2018} have been used to investigate the oxidation of a UN (001) surface \cite{Bright2022}. Using such single phase films allowed the reaction of UN to be investigated without the influence of a secondary metallic U phase. XRR measurements of the thickness of the surface layers that formed on exposure to air showed that the surface passivated. The chemistry of these layers was investigated with a XPS depth profile, identifying a surface UO$_{2+x}$N$_y$ layer and U$_2$N$_3$ intermediate layer, not dissimilar to the oxidised surface of U$_2$N$_{3+x}$ described above \cite{Luo2020}. The epitaxial nature of the sample, producing a single (001) oriented UN single crystal surface, provided further insight into the reaction: XRD measurements of the oxidised surface showed a topotactic relationship (see Fig. \ref{fig5_2_3b}) between the film and surface oxide, which is proposed to play a critical role in the passivation mechanism.

\subsubsection{Thin films of uranium silicides}	\label{s:USi}

Uranium silicide phases have been of interest since the 1940's.
Work conducted by A.R. Kaufmann, B.D. Cullity, and G. Bitsianes, as reported by W.H. Zachariasen in $1948$ \cite{Zachariasen1949}, described the crystal structure of a tri-silicide phase (USi$_3$), and proposed additional uranium silicide phases: USi$_2$, U$_2$Si$_3$, USi, U$_5$Si$_3$, and U$_{10}$Si$_3$.
Zachariasen presented in the $1948$ paper \cite{Zachariasen1949} the uranium disilicide phase, USi$_2$, which was found to be isomorphous with ThSi$_2$ and PuSi$_2$, all with  body-centred tetragonal structures, and I$4$/amd space groups.
The uranium-silicon binary phase diagram, provided by Okomoto {\it et al.} \cite{Okamoto2013} and Middleburgh {\it et al.} \cite{Middleburgh2016}, indicates there are around seven stoichiometric phases which all exist as line-compounds.
The nature of these line-compounds suggest the fabrication of individual phases in the bulk is challenging. Middleburgh {\it et al.} \cite{Middleburgh2016}, suggests that the U$_3$Si$_2$ phase could not incorporate additional uranium into the lattice without forming mixed-phases.  It is this factor that results in the separation and production of multiple phases if the stoichiometric U:Si ratio is not satisfied. As a result, engineering of U-Si phases is particularly difficult for bulk investigations.

The Reduced Enrichment for Research and Test Reactors (RERTR) Program, initiated by the Department of Energy (US-DOE), suggested the use of U-Si phases in order to implement low-enriched uranium (LEU) fuel compounds within research reactors to prevent proliferation \cite{Snelgrove1996}. 

With being highlighted as nuclear fuel compounds, the U-Si phases have gained attention with regards to their fundamental behaviours. Remschnig {\it et al.}, \cite{Remschnig1992} investigated the structural and magnetic behaviour of the binary U-Si phases, probing: U$_{3}$Si, U$_{3}$Si$_{2}$, USi, U$_{3}$Si$_{5}$, USi$_{1.88}$, and USi$_{3}$. Bulk single crystals of U$_{3}$Si$_{2}$, USi, U$_{3}$Si$_{5}$, and USi$_{1.88}$ were extracted from arc melted samples within this study. Additional investigations conducted by Antonio {\it et al.}, \cite{Antonio2018} probed the thermal and transport properties of the primed ATF candidate, U$_{3}$Si$_{2}$. Understanding the thermal behaviour of fuel candidates is integral to its consideration as a commercial fuel and eventual implementation into the nuclear fuel cycle. Here, the heat capacity, electrical resistivity, Seebeck and Hall effects, and thermal conductivity were probed in a temperature range of $2 $–$ 300$ K in magnetic fields up to $9$ T. The U$_{3}$Si$_{2}$ samples used in this investigation were engineered via the arc-melting of elemental U and Si powders. Impurities of USi and UO$_{2}$ were observed using XRD during sample characterisation. Low temperature thermal investigations on the U-Si phases are complementary to the high-temperature studies conducted by White {\it et al.}, \cite{White2015} on bulk U$_{3}$Si$_{2}$ sintered samples. A major roadblock in investigating uranium silicide phases is sample fabrication.
The production of U-Si materials often results in the formation of multi-phased systems \cite{Kardoulaki2021,Harp2015}. As a result, it can be difficult to identify and attribute structural or chemical behaviours to a particular phase.

The adaptability of engineering U-based surfaces on substrates suggests that producing uranium-silicon phases in this form is suitable for both applied and fundamental investigations. An early investigation, conducted by S. Fujimori in $1988$ \cite{Fujimori1998} presented the electronic structure, probed using X-ray photoelectron spectroscopy, of uranium deposited upon a [$100$]-oriented silicon surfaces. The study aimed to understand the interactions between the two elements, and to probe the possibility of the epitaxial growth of uranium silicides upon the [$100$]-silicon surface. The uranium layers were deposited using an `MBE-like' technique, and characterisation of the surfaces was conducted using X-ray photoelectron spectroscopy (XPS). The work presented in \cite{Fujimori1998} did not conclude if the annealing of uranium deposited on [100]-oriented Si surfaces resulted in the formation of uranium silicide phases.

A second paper, produced by Fujimori in $2000$ \cite{Fujimori2000}, showed the deposition of U metal onto a prepared ($111$)-oriented Si surface, as a different way of controlling the $5${\it f} electrons when compared to bulk studies, and to further understand the possibility to produce epitaxial uranium silicide phases. Similar to the initial paper, the uranium surfaces were deposited using a method which is described as `MBE-like', and were subsequently characterised using in-situ XPS. Valence band spectra collected from the U surfaces suggested structural disorder at the interface between substrate and film. This was noted with a broadening of the Si {\it sp} band states at $3$, $7.5$, and $10$ eV. Additionally, the Si and U atomic cross-sections provided by Yeh {\it et al.} \cite{Yeh1985}, vary with $0.13$ Mb for U $5${\it f}, $0.01$ Mb for Si $3${\it s}, and $0.0017$ Mb for Si $3${\it p}, suggesting that small amounts of U deposited upon the (111)-Si may dominate the valence band spectra.

A large body of work conducted by Harding {\it et al.} \cite{Harding2023} in $2023$ showed the epitaxial stabilisation of four uranium silicide phases as epitaxial thin films: U$_{3}$Si, U$_{3}$Si$_{5}$, $\alpha$-USi$_{2}$, and USi$_{3}$, with poly-crystalline U$_{3}$Si$_{2}$. The U-Si phases presented in \cite{Harding2023} were all synthesised using DC magnetron sputtering where a co-sputtering technique was implemented allowing for material from U and Si targets to be deposited simultaneously under UHV conditions. The technique allowed for control over the relative U and Si contents, resulting in the formation of U-Si phases that extend over the entire binary phase diagram.

These phases were structurally found to be epitaxial with their respective single-crystal substrates using X-ray diffraction techniques (Table \ref{tab:silicides}). A deeper study into the chemical bonding of U-Si phases was also presented in this work. Using XPS, Harding {\it et al.} \cite{Harding2023} presented the metallic nature of these U-Si thin films with clear asymmetry noted in the U-$4${\it f} core level spectra.
Using area ratios between the U-$4${\it f} and Si-$2${\it s}, the U-Si thin films were found to be stoichiometric within error, with the exception of the $\alpha$-USi$_{2}$ phase.
The uranium disilicide, presented in Table \ref{tab:silicides}, has a U:Si ratio representative of a uranium monosilicide phase.
From the structural characterisation presented in \cite{Harding2023}, the data suggested the formation of the tetragonal $\alpha$-USi$_{2}$ phase similar to the data presented by Sasa {\it et al.}. \cite{Sasa1976} in their $1976$ manuscript.

The work conducted on the uranium silicide phases has demonstrated the ability to control stoichiometry using epitaxial lattice matching.
The understanding of the U-Si epitaxial system, as presented by Harding {\it et al.} \cite{Harding2023} and trialled by Fujimori \cite{Fujimori1998, Fujimori2000}, can form the basis of using epitaxial stabilisation to navigate other phase diagrams.

\begin{landscape}
\begin{table}[]
\centering
\caption{Structural and chemical parameters extracted from uranium silicide thin films designed by Harding {\it et al.}. All samples deposited using DC magnetron sputtering.}
\label{tab:silicides}
\begin{tabular}{p{1.5cm}|p{2cm} p{1cm} p{2cm} p{1.75cm} p{1.25cm} p{4.5cm} p{1.5cm}}
\hline
Material            & U:Si & Form        & Structure    & Space Group   & Domains & Substrate                 & Reference \\ \hline
U$_{3}$Si           & $3.2\,\pm\,0.3$    & \hkl(001)   & Tetragonal   & I$4$/mcm      & $2$     & \hkl(001) CaF$_{2}$       & \cite{Harding2023} \\
U$_{3}$Si$_{2}$     & $1.5\,\pm\,0.2$    & Poly        & Tetragonal   & P$4$/mbm      & n/a     & \hkl(001) CaF$_{2}$       & \cite{Harding2023}   \\
U$_{3}$Si$_{5}$     & $0.6\,\pm\,0.1$    & \hkl(100)   & Hexagonal    & P$6$/mmm      & $2$     & \hkl(001) CaF$_{2}$       & \cite{Harding2023}    \\
$\alpha$-USi$_{2}$  & $1.1\,\pm\,0.1$    & \hkl(001)   & Tetragonal   & I $41$/amd    & $1$     & \hkl(001) MgO, \hkl(001) CaF$_{2}$  & \cite{Harding2023}  \\
USi$_{3}$           & $0.33\,\pm\,0.03$  & \hkl(001)   & Cubic        & Pm-$3$m       & $1$     & \hkl(001) CaF$_{2}$       & \cite{Harding2023} \\
\hline
\end{tabular}
\end{table}
\end{landscape}

\section{Conclusions and future prospects} \label{s:conclusions}

We have attempted in this review to give an account of several decades of work on U-based thin film. Various efforts were made in the period 1960 – 2000, most of which are discussed in this review, but none managed to continue over a long enough period to build up a substantial body of work that encouraged other Laboratories to start a significant effort. It is important to distinguish between attempts to produce thin samples to reduce the radioactive inventory, which have been widespread over the years, and thin films on chosen substrates in an effort to make epitaxial (or at least strongly textured) thin films. Thomas Gouder and his collaborators at the European Commission’s Laboratory in Karlsruhe, Germany, have been involved primarily in the first effort discussed above, and have pushed beyond U into Pu, Np, Am, and even Cm.

This changed with our own program, first starting at Oxford University in $\sim$ 2002 and then transferred in 2011 to Bristol University, and also the program at Los Alamos National Laboratory, which started at about the same time \cite{Burrell2007,Scott2014}. Both these programs have concentrated on uranium, and aimed to prepare epitaxial films. As discussed in Sec. \ref{s:UO2_PuO2_Photoemission}, the epitaxial films of PuO$_2$ \cite{Scott2014,Kruk2021} were also made at LANL by the polymer assisted deposition (PAD) technique, but all other samples have been with U. More recently, an effort using pulsed laser deposition (PLD) has also been mounted at LANL \cite{Enriquez2020,Sharma2022}.

In discussing these potential advances in actinide materials, we need to be aware that these materials are radioactive, and not familiar to the general public, except in connection with nuclear fuel (especially irradiated) or nuclear weapons. However, uranium is a natural element found in all soil, and also in the human body to the extent of between 50 – 100 $\mu$g for an adult. Thorium is similar. To put this in perspective, a thin film of UO$_2$ of $5 \times 5$ mm$^2$ with a thickness of 1000 \AA{} has a mass of uranium of $\sim$ 25 $\mu$g and an activity of 0.3 Bq. A banana has an activity of $\sim$ 15 Bq. Working with these materials we usually cover them with a 50 \AA{} layer of Nb, so no radioactive particles can escape, and any potential device would be suitably encapsulated. It is clear that the use of thin films of thorium or uranium (which would probably have a thickness less than 1000 \AA{}) does not represent any kind of hazard. Of course, working with heavier actinides than uranium, e. g. plutonium, is a different matter, and they could be only prepared and used in specialized laboratories.

Thin-film methods, capable of producing high-quality single crystal films with thicknesses ranging from 100-5000 \AA{}, have played an increasing role in the study of actinide containing materials as they are, by their very nature, extremely low mass samples naturally reducing radioactive risks. The recent and dramatic increase in X-ray flux available at many large synchrotron facilities, plus the significant development in grazing- or low-incidence angle scattering techniques, has mitigated the experimental difficulties previously associated with such extremely low scattering volume samples; opening the door to wealth of experimental opportunities, some of which are discussed in this review. A wider perspective on synchrotron use with actinides can be found in \cite{Caciuffo2022}.

As well as providing a safer method for actinide studies, the use of thin films, in particular epitaxial thin films, has numerous other benefits over bulk crystals and we expect these to play an increasing role in the future of actinide research. Lattice matching provides an additional dimension to the synthesis phase space, through which various crystal film orientations can be synthesized. Strain – stretching or compressing particular crystallographic directions - can be used as a tuning parameter for various physical phenomena, metastable phases can be stabilized at ambient conditions, and some phases not present in the bulk phase diagram can even by synthesized in thin film form. We report on early experiments with elemental uranium in this context (see Sec. \ref{s:alphacdw} et seq.). The work by Sharma {\it et al.} \cite{Sharma2022} with UO$_2$ is another, see Section \ref{s:UO2paramagstrain}.

Such films are naturally suited to almost all forms of transport measurements, in particular in-plane directionally dependant studies. Interface and proximity effects can be explored systematically through the fabrication of \AA{}ngstrom scale precision designed multi-layers and heterostructures, and such methods naturally extend to the exploration of exotic devices with actinide containing functional layers for possible advanced devices. The theoretical paper by Dennett {\it et al.} \cite{Dennett2022} discusses such a project.

An important step was made at the University of Illinois with the discovery \cite{Strehle2012} that epitaxial UO$_2$ could be deposited by sputtering U onto yttrium-stabilised zirconia (YSZ). There has been an effort also at Charles University, Prague, especially on the hydrides, and they discovered that UH$_2$ (with the cubic fluorite structure) could be stabilized as a thin film (see Sec. \ref{s:UH}). Some experiments have been reported from facilities in China, especially at the Surface Physics and Chemistry Laboratory, Jiangyou. Recent work from them suggests that cubic UN$_2$ with the CaF$_2$ structure can be produced as a thin film \cite{Luo2020}.

More recently, Idaho National Laboratory (INL) have announced that they are starting a thin-film program on the actinides using molecular-beam epitaxy. They have also published an review of possible work with actinide thin films \cite{Vallejo2022}. This is a most useful exercise in setting the stage for further work, and contains an important source of the literature on this subject. We note however, that the “mineralization” technique discussed in that review is used to prepare single crystals, and does not involve any substrate, so does not belong in the category of “thin-film samples” as we have discussed in this review. Furthermore, the methods mentioned on transuranium systems do not explore the possibility of producing epitaxial (i$.$e$.$ single crystal) samples, so the number of {\it bona fide} “thin film samples” produced is smaller than it might first appear.

In covering this wide field of endeavor, it is useful to make a distinction between different samples on the following basis. {\it First}, (category A) one can imagine that epitaxial samples can be made that allow the study of essentially {\it bulk} properties. A good example in this category are epitaxial samples of the {\it bcc} alloys with U-Mo (Sec. \ref{s:corrdisor}). No single crystals of these can be made in the bulk, although there are numerous studies of the alloy system. The production of epitaxial samples thus allowed bulk properties, such as the phonon-dispersion curves, to be examined. In this study important diffuse scattering was observed \cite{Chaney2021} and led to the discovery of a new type of correlated disorder. A second example would be U$_2$N$_3$ (Sec. \ref{s:UNmag}), where, again, single crystals of the bulk material have not been produced, so the synchrotron experiments discovered some new effects in this system \cite{Bright2019a} that were not evident earlier. Yet another example is the study of the phonons in radiation damaged UO$_2$ (Sec. \ref{s:UO2phonons}); here the epitaxial films allow homogeneous damage by irradiation in beams of charged particles to the films, which can then be examined by synchrotron-radiation techniques \cite{Rennie2018b}.

The {\it second} type of investigation (category B) is based on creating epitaxial samples that have new properties, some of which may be a consequence of the interaction of the substrate. An excellent example of this is all the work performed so far on the elemental metal $\alpha$-uranium (Sec. \ref{s:alphacdw}), where the properties of the charge-density wave can be manipulated by different strains of the $\alpha$-U $\bm{a}\mbox{-axis}$ caused by depositing on different substrates. This has added considerably to our understanding of the metal’s exceptional properties, and led to a further effort to prepare {\it hcp}-U, which does not exist in the bulk (Sec. \ref{s:hcpgrowth}). In this category, we must also include dissolution experiments (Sec. \ref{s:UO2diss}), initially performed on thin ($\sim$ 100 \AA{}) films of UO$_2$. Of course, this effect is present at any UO$_2$ surface, but if one uses a bulk crystal then the effect at the surface will be swamped by the response of the bulk crystal, and it will be essentially impossible to measure such an effect. A third example is the work on bilayers of U/Fe (Sec. \ref{s:Ubilayers}) and multilayers including U (Sec. \ref{s:U_MLs}).

Quoting this third example, as well as the ideas proposed by Dennett {\it et al.} \cite{Dennett2022}, brings us to the question of the {\it interface}. As pointed out in this latter reference, the interface is crucial to the operation of any heterostructure. So far, our experiences with the interfaces in U systems have been somewhat mixed.  This was also an issue of the early attempts (at IBM) to make memory systems from multilayers of amorphous UAs (which is a ferromagnet at T$_c$ $\sim$ 140 K), and elemental cobalt \cite{Fumagalli1993}. A close examination of one of these samples \cite{Kernavanois2004} showed a poor interface and mixing of the two individual elements over a region of $\sim$ 10 \AA{} between the two components. Following these efforts, a series of multilayers were made at Oxford University (Sec. \ref{s:U_MLs}) of the ferromagnet elements and U metal. In this work, the interfaces involving the ferromagnetic 3$d$ elements, Fe, Co and Ni, were poor, with considerable inter-diffusion over a region of at least 15 \AA{} across the nominal interfaces \cite{Springell2008b}, but with U/Gd, the interfaces were extremely sharp and there appeared to be no interdiffusion \cite{Springell2010}.  Similarly, the interfaces of the sample with permalloy/U \cite{Singh2015} and U/Fe \cite{Gilroy2021} are probably also poor. A considerable effort needs to be made to understand how to make these interfaces better, perhaps by using an alloy of U, or even a compound, or depositing an intermediate thin layer. Dennett {\it et al.} \cite{Dennett2022} have suggested UN$_2$ and GaAs (created as a superlattice) might be a useful device, but at this stage no work has been done on trying to make such a compound; all we know is that, if it exists, the interfacial strain would be $\sim$ 1\%. In fact, no superlattices of any material containing U have ever been produced! The closest is the work on U/Gd multilayers, \cite{Springell2010}, but even there, the {\it hcp}-U was not ordered in-plane, as there is a large misfit between the in-plane parameters of U ($\sim$ 3 \AA{}) and those of Gd (3.64 \AA{}). Much remains to be done, but our progress over the last two decades gives rise to optimism that these challenges can be met.

There are undoubtedly many compounds containing U (Category A) that can be fabricated as thin films. An excellent example is the early work (in the 1990s) done at the Universities of first Darmstadt and then Mainz by Adrian and his colleagues. This work produced epitaxial films of the heavy-fermion compounds UPd$_2$Al$_3$ and UNi$_2$Al$_3$ \cite{Huth1994,Jourdan1999,Dressel2002,Jourdan2004,Foerster2007,Bernhoeft1998} deposited on LAO with the molecular-beam technique. No effort, so far, has been made at either Oxford/Bristol or Los Alamos National Laboratory to prepare any heavy-fermion materials, so this is a completely open field. For example, a great deal could be learnt about actinide electronic structure by having thin epitaxial films of the isostructural UCoGa$_5$, NpCoGa$_5$, and PuCoGa$_5$. The 5$f$ states in these three systems show paramagnetic behavior in the first compound, antiferromagnetic behavior (at 47 K) in the second, and superconducting behavior (at 18 K) in the Pu compound. One could imagine a series of important experiments on such samples, including angular-resolved photoemission, that would shed further light on the electronic structure of the 5$f$ states, and what features of that electronic structure are responsible for such diverse behaviour in three isostructural compounds.

Theoretical studies have proposed a number of interesting properties in U systems where single crystals would certainly be welcome for further studies and comparisons with theory. Examples include a topological insulator-to Weyl semimetal transition in the system UNiSn \cite{Ivanov2019}. Searching for two-dimensional actinide systems (such as can be stabilized in thin films) Lopez-Bezanilla \cite{Lopez2020} has identified the system UB$_4$ as potentially of considerable interest as the theory shows that the dispersion in the band states (resulting in Dirac cones) is driven by the hybridization of the uranium 5$f$ orbitals with the {\it p}$_z$ orbitals of the B atoms. Another example concerns possible magnetism in U/W(110) superlattices \cite{Zarshenas2012}. Of course, magnetism in some form has already been predicted at pure U surfaces \cite{Stojic2003}, and this would be interesting, but difficult, to detect. Most of the theoretical techniques used for these predictions have used pure density functional theory (DFT), and it is known that these methods over-estimate the tendency for spin polarization in the actinides. Isaacs and Marianetti \cite{Isaacs2020} have argued that a combination of DFT and dynamical mean-field theory (DMFT) are needed to address electronic structure of correlated electron materials \cite{Kotliar2006}. However, such materials need to be prepared and examined before we can draw any definitive conclusions.

In guessing or proposing systems in Category B, the problem is the challenge of the unknown. Indeed, heterostructures may well be of great interest in the actinides. We have started down this path with work reported with U, and also an effort on producing exchange bias with UO$_2$ films in Sec. \ref{s:UO2exchangebias}, but difficulties are encountered in both cases in fabricating good interfaces. Ultimately, of course, we should be able to fabricate “superlattices” where the interfaces are of high quality and the strains across them are small by choosing the correct materials. Only then will we be able to answer questions posed by theories, such as that on Pb-Pu superlattices by Rudin \cite{Rudin2007}.

All of these theoretical predictions depend crucially on the unusual behavior of the 5$f$ electrons in the actinides. Situated far away from the nucleus, they are not shielded from the neighboring electrons, as is the case of the tightly packed 4$f$ electrons in the lanthanide series, so that the 5$f$ electrons readily interact with electrons of neighboring atoms. This gives them, for example, the property that actinide ions can exist in multiple valence states, depending on their environment. In addition, the 5$f$ electrons carry a large orbital moment. Some properties involving orbital moments are proportional to a higher power ($Z^4$) of the atomic number, giving the actinide series an obvious advantage. The ability to interact with electrons from neighboring atoms is understood in the word “hybridization”, although this can take many different forms, and requires advanced theoretical methods to describe the process. In some sense, these properties have made the actinides complicated to work with; now, hopefully, with new theory and the capabilities of fabricating thin films, heterostructures, and superlattices, we are on the verge of making these peculiar properties useful in technology!

Although a great deal of the initial work on U-based thin films was devoted to fundamental physics and the exploration of correlated electronic states, it is clear that the last decade has seen a significant shift in activity towards the investigation of applied nuclear materials, as more research groups realise the strength of this approach. Working swiftly on large bulk sample sets of active materials is difficult and this can be overcome, using thin film deposition. The controlled engineering of samples is also a contributing factor, where we are able to reduce complex systems to simpler, purer experimental analogues, where variables can be carefully controlled in order to precisely determine mechanisms of degradation, in structure, chemistry, and thermal transport etc., all important aspects for nuclear materials research.

Some of the most important aspects, from a storage and operation perspective, for nuclear fuels research concerns the interaction of the fuel with air and water. Ambient oxidation processes and dissolution in long-term storage can be slow and yield only small changes that are difficult to detect with bulk samples. By using thin film analogues of nuclear fuels, with modern, sophisticated diffraction techniques it is possible to measure surface changes on the order of \AA ngstroms. Research has already begun, using thin films to provide important corrosion rate data to model spent nuclear UO$_2$ fuels in radiolytic environments, the ambient degradation and oxidation of candidate advanced technology fuels, and the corrosion, oxidation and hydriding of stored metallic waste forms. This could be extremely useful in the study of spent fuel behaviour in long term repository conditions; not just for the UO$_2$ fuel that is already stored by many countries across the globe, but also for possible advanced technology fuels (UN, U$_3$Si$_2$ for example) and more complex composites (UCO in TRISO), proposed in the next generations of small and advanced modular reactors.

Looking to the future, the advancement of high resolution physical and chemical characterisation techniques, including laboratory-based equipment, synchrotron and neutron beamline experiments, and the more frequent application of these techniques to U-based thin film systems will yield ever more detailed understanding of these complex systems. This will not only help to underpin our theoretical framework for important nuclear materials, but in some instances could provide crucial data for the utilisation of uranium in device technologies.

\section{Acknowledgments}
We would like to acknowledge the funding and support from the Engineering and Physical Sciences Research Council (EPSRC), UK. Recent grants in advanced fuels (ATLANTIC, EP/S011935/1) and nuclear waste and decommissioning (TRANSCEND, EP/S01019X/1) have opened up new research avenues and provided PDRA and PhD student support, which has been invaluable to the continuation of this field. More directly, we thank the EPSRC for the recent award of a new deposition and surface characterisation facility (EP/V035495/1), becoming a national nuclear user facility (FaRMS: https://www.nnuf.ac.uk/farms).

We would like to thank our many industry supporters, particularly Dave Goddard and Rob Burrows of the National Nuclear Laboratory, and Dave Geeson and Norman Godfrey of the AWE for provision of advice and materials over the years. We would also like to acknowledge our international collaborators at the JAEA, INL, CEA and ESRF who have helped enrich and expand the scope of this growing field of research.

We would like to give particular credit to Bill Stirling, Mike Wells, Stan Zochowski, Mike Thomas, Sean Langridge, and the late Roger Cowley for their interest and support over many years. Early funding for this program was obtained from the European Commission's Joint Research Center, Karlsruhe, Germany, and we thank Jean Rebizant for organizing this valuable assistance. Constructive comments on the manuscript were made by Ladia Havela from Prague, and we thank him for these. We also acknowledge direct and indirect input from several generations of PhD students, in particular Rebecca Nicholls.

\bibliographystyle{tfq}
\bibliography{final_bib}

\end{document}